\documentclass[preprint]{raa_twocolumn}

\usepackage{graphicx,times}
\usepackage{natbib}
\usepackage{amssymb,amsmath}
\bibpunct{(}{)}{;}{a}{}{,}

\usepackage[pagebackref=true]{hyperref}

\begin{document}

   \title{Mapping Dust Attenuation at Kiloparsec Scales. II. Attenuation Curves from Near-Ultraviolet to Near-Infrared}

 \volnopage{ {\bf 2025} Vol.\ {\bf X} No. {\bf XX}, 000--000}
   \setcounter{page}{1}

   \author{Ruonan Guo
   \inst{1}, Cheng Li\inst{1}, Shuang Zhou\inst{3}, Niu Li
      \inst{1,2}, Tao Jing\inst{1}, Zhuo Cheng\inst{1,4}
   }

   \institute{Department of Astronomy, Tsinghua University, Beijing 100084, China; {\it grn19@mails.tsinghua.edu.cn; cli2015@tsinghua.edu.cn}\\
        \and
             National Astronomical Observatories, Chinese Academy of Sciences, Beijing 100101, China \\
	    \and
             INAF–Osservatorio Astronomico di Brera, via Brera, 28, 20159 Milano, Italy \\
        \and 
             Department of Physics, The Chinese University of Hong Kong, Sha Tin, NT, Hong Kong, China\\
\vs \no
   {\small Received 2025 Month Day; accepted 2025 Month Day}
}

\abstract{This is the second paper in a series that utilize integral field spectroscopy from MaNGA, NUV imaging from Swift/UVOT and NIR imaging from 2MASS to investigate dust attenuation properties on kpc scales in nearby galaxies. We apply the method developed in Paper I \citep{2023ApJ...957...75Z} to the updated SWiM\_v4.2 catalog, and measure the optical attenuation curve and the attenuation in three NUV bands for 2487 spaxels selected from 91 galaxies with S/N$>20$ and $A_V>0.25$. We classify all spaxels into two subsets: star-forming (SF) regions and non-SF regions. We explore the correlations of optical opacity ($A_V$) and the optical and NUV slopes of the attenuation curves ($A_B/A_V$ and $A_{\tt w2}/A_{\tt w1}$) with a broad range of stellar population and emission-line properties, including specific surface brightness of H$\alpha$ emission ($\Sigma_{\text{H}\alpha}/\Sigma_\ast$), stellar age, stellar and gas-phase metallicity, and diagnostics of recent star formation history. Overall, when comparing SF and non-SF regions, we find that $A_V$ and $A_B/A_V$ exhibit similar correlations with all the stellar population and emission-line properties considered, while the NUV slopes in SF regions tend to be flatter than those in non-SF regions. The NUV slope $A_{\tt w2}/A_{\tt w1}$ exhibits an anti-correlation with $\Sigma_{\text{H}\alpha}/\Sigma_\ast$, a trend that is primarily driven by the positive correlation between $A_{\tt w2}/A_{\tt w1}$ and $\Sigma_\ast$. The NUV slope flattens in SF regions that contain young stellar populations and have experienced recent star formation, but it shows no obvious dependence on stellar or gas-phase metallicity. The spatially resolved dust attenuation properties exhibit no clear correlations with the inclination of host galaxies or the galactocentric distance of the regions. This finding reinforces the conclusion from Paper I that dust attenuation is primarily regulated by local processes on kpc scales or smaller, rather than by global processes at galactic scales.
}

   \authorrunning{Guo, Li, Zhou et al. }            
   \titlerunning{Mapping Dust Attenuation at Kiloparsec Scales (II)}  
   \maketitle

%
\section{Introduction}           
\label{sect:intro}

Dust is ubiquitously distributed throughout the interstellar medium, absorbing and scattering stellar and nebular radiation at wavelengths shorter than the near-infrared (NIR) while re-emitting it in the mid-infrared and far-infrared \citep[e.g.][]{2008MNRAS.388.1595D,2018ARA&A..56..673G}. Consequently, despite constituting only a small fraction of the baryonic mass \citep{2014A&A...563A..31R,2018MNRAS.475.2891D}, interstellar dust plays a crucial role in shaping a galaxy's observed spectral energy distribution (SED). The wavelength-dependent extinction of starlight along different sightlines---referred to as {\em dust extinction curves}---is determined by dust absorption and scattering and depends on the chemical composition  and size distribution of dust grains \citep{2001ApJ...548..296W}. Extinction curves can be directly measured only for the Milky Way and a few nearby galaxies \citep[e.g.][]{1984A&A...132..389P,1985ApJ...288..558C,1986ApJ...307..286F,1989ApJ...345..245C,1998ApJ...500..816G,1999PASP..111...63F,2003ApJ...594..279G,2015ApJ...815...14C}. For more distant galaxies, the observed wavelength-dependent dimming of light---described as {\em dust attenuation curves}---accounts not only for absorption and scattering but also for additional effects such as light scattered into the line of sight, as well as the complex spatial distribution of dust relative to the stars \citep{2001PASP..113.1449C}. A fundamental task in extragalactic astrophysics is to measure dust attenuation curves and understand the underlying mechanisms that shape them. This is essential for accurately recovering the intrinsic SED and correctly measuring key physical properties of galaxies \citep{2013ARA&A..51..393C,2020ARA&A..58..529S}.

For a galaxy or a local region within a galaxy, the dust attenuation curve is obtained by comparing the observed spectrum or SED with its intrinsic, dust-free counterpart. Broadly, the intrinsic spectrum or SED can be obtained using two categories of  methods, either through empirical comparison with selected reference galaxies that are less attenuated \citep[e.g.][]{1994ApJ...429..582C,1994ApJ...429..172K,1997AJ....113..162C,2000ApJ...533..682C,2007ApJS..173..392J,2011MNRAS.417.1760W,2016ApJ...818...13B,2017ApJ...851...90B,2017ApJ...840..109B}, or by fitting the observed spectrum/SED with stellar population models \citep[e.g.][]{1971ApJS...22..445S,1972A&A....20..361F,1998AJ....115.1329S,2001ApJ...559..620P,2003MNRAS.341...33K,2005ApJ...619L..39S,2007ApJS..173..267S,2009A&A...507.1793N,2013ARA&A..51..393C,2017ApJ...837..170L,2019A&A...622A.103B,2019ascl.soft05025J,2019MNRAS.485.5256Z,2022A&A...663A..50B,2022MNRAS.514.5706J,2022ApJ...932...54N,2023ApJ...953...54B,2023ApJ...957...75Z,2025arXiv250115082H}. Over the past three decades, numerous observational studies have employed both categories of methods to measure attenuation curves and investigate their correlations with galactic properties on global scales, spanning both the local Universe and high redshifts (see \citealt{2020ARA&A..58..529S} for a comprehensive review). 

These studies have consistently shown that dust attenuation curves vary in three key aspects: the overall slope from the optical to the 
 ultraviolet (UV), which is typically approximated by a power law; the excess attenuation around 2175\AA, commonly referred to as the 2175\AA\ bump; and their relationship with nebular attenuation, often quantified using the Balmer decrement of emission lines. Early studies found that the attenuation curves of local starburst galaxies tend to have shallow slopes, a weak or absent 2175\AA\ bump, and an average stellar-to-nebular color excess ratio of $E(B-V)_{\text{star}}/E(B-V)_{\text{gas}}\approx 0.44$ \citep{1994ApJ...429..582C,2000ApJ...533..682C}. In contrast, quiescent galaxies, such as the Milky Way, typically exhibit steeper attenuation slopes and a prominent 2175\AA\ bump \citep[e.g.][]{1989ApJ...345..245C,1999PASP..111...63F}. It is now well established that attenuation curves span a broad range in both UV-through-optical slopes and the strength of the 2175\AA\ bump, encompassing curves with no 2175\AA\ bump and slopes shallower than the Calzetti curve, as well as those with strong 2175\AA\ bumps and slopes steeper than the Milky Way-type curves. The ratio of $E(B-V)_{\text{star}}/E(B-V)_{\text{gas}}$ is also found to span a wide range from 0.44 to $\gtrsim 1$ depending on the mass, star formation rate (SFR) and axis ratio of galaxies \citep[e.g.][]{2011MNRAS.417.1760W,2011ApJ...738..106W,2017ApJ...847...18Z}. In addition, the slopes of attenuation curves are found to strongly correlated with the optical opacity, with shallower slopes in galaxies of lower visual attenuation. As summarized in \citet{2020ARA&A..58..529S}, both the correlations between the slope and optical opacity and the variation of the 2175\AA\ bump may be explained as consequences of geometric and radiative transfer effects based on theoretical studies. 

Therefore, spatially resolved observations down to the scale of star-forming regions and covering both the optical and UV bands are essential for disentangling the effects of local and galaxy-wide processes on variations in dust attenuation curves. Over the past decade, integral field spectroscopy observations have facilitated numerous studies investigating spatially resolved dust attenuation in nearby galaxies at kiloparsec scales or smaller \citep[e.g.][]{2013ApJ...771...62K,2016ApJ...825...34J,2017MNRAS.467..239B,2019ApJ...872...63L,2020MNRAS.495.2305G,2020ApJ...896...38L,2020ApJ...888...88L,2020ApJ...893...94T,2021ApJ...917...72L,2021MNRAS.503.4748R,2023A&A...670A.125J,2024ApJ...975..234L,2024A&A...691A.201L}. For instance, in a series of studies on optical dust attenuation using data from the Mapping Nearby Galaxies at Apache Point Observatory survey \citep[MaNGA;][]{2015ApJ...798....7B}, \citet{2020ApJ...896...38L} developed a novel method to derive model-independent attenuation curves from optical spectra. This method was subsequently applied in \citet{2021ApJ...917...72L} to conduct a comprehensive analysis of the correlations between stellar and nebular attenuation and a broad range of stellar population and emission-line properties, and in \citet{2024ApJ...975..234L} to further examine the radial variations of these correlations. The measurements of optical dust attenuation properties as well as stellar population and emission line properties for the final sample of MaNGA are publicly released in \citet{2023ChPhB..32c9801L}. 
Their findings revealed that the value of $E(B-V)_{\text{star}}/E(B-V)_{\text{gas}}$ at kiloparsec scales spans an even larger range than previously observed at galactic scales. More importantly, their results suggest that stellar age is the primary driver of variations in $E(B-V)_{\text{star}}/E(B-V)_{\text{gas}}$, implying that both young and old stellar populations may play significant roles in shaping the properties of dust attenuation. Recently, a spatially resolved study of the multiband SED of the nearby galaxy NGC 253 further reinforced this finding, revealing a clear correlation between the difference in stellar and nebular attenuation and stellar age \citep{2025arXiv250115082H}.

In recent years, UV photometry has been incorporated alongside optical IFS data, enabling the inclusion of the UV slope in studies of spatially resolved dust attenuation. Such studies have been conducted in individual nearby galaxies, including M81 and Holmberg IX \citep{2011AJ....141..205H}, M82 \citep{2014MNRAS.440..150H,2015MNRAS.452.1412H}, the Small Magellanic Cloud \citep{2017MNRAS.466.4540H}, NGC 628 \citep{2019MNRAS.486..743D}, the Milky Way \citep{2021MNRAS.505..283F}, and NGC 253 \citep{2025arXiv250115082H}. Additionally, such analyses have been extended to samples of galaxies across the local Universe \citep[e.g.][]{2020MNRAS.494.4751M,2023ApJ...953...54B,2023MNRAS.526..904D,2023ApJ...957...75Z,2025PASA...42...22B}.
For instance, utilizing NUV photometry from Swift/UVOT and optical IFS data from MaNGA, as provided by the SwiM\_v3.1 catalog \citep{2020ApJS..251...11M}, in addition to NIR imaging from 2MASS, \citet[][hereafter Paper I]{2023ApJ...957...75Z} applied the technique of \citet{2020ApJ...896...38L} and the Bayesian spectral fitting code {\tt BIGS} developed in \citet{2019MNRAS.485.5256Z} to measure the 2175\AA\ bump and optical attenuation curve for a sample of kpc-sized regions in nearby galaxies. They found that previously established correlations—namely, the anti-correlation between the attenuation curve slope and optical opacity, as well as the decrease in 2175\AA\ bump strength with increasing star formation rate—persist at kpc scales. This finding strongly suggests that dust attenuation is primarily regulated by local processes on kpc scales or smaller, rather than by global processes at galactic scales.

Following Paper I, this is the second paper in a series studying dust attenuation at kiloparsec scales in nearby galaxies, utilizing NUV photometry from Swift/UVOT, optical IFS data from MaNGA, and NIR photometry from 2MASS. This paper extends the work of Paper I by employing the latest version of the SwiM catalog, SwiM\_v4.2 \citep{2023ApJS..268...63M}, which is nearly four times larger than the SwiM\_v3.1 catalog used in Paper I. Additionally, we classify all kpc-sized regions in the catalog into two subsamples: star-forming (SF) regions and non-SF regions. This classification allows us to investigate dust attenuation properties separately for regions dominated by young stellar populations and star formation-related processes, and for regions dominated by older populations and diffuse-ionized gas. Furthermore, while Paper I primarily examined the specific SFR, this work incorporates a broader range of stellar population and emission-line properties into the analysis, including stellar age, stellar and gas-phase metallicity, and diagnostics of recent star formation history. In this paper, we focus on optical opacity, characterized by the $V$-band attenuation ($A_V$), and the slopes of the attenuation curve in the optical and NUV. In the next paper (Paper III, Guo et al. in prep.), we will investigate the 2175\AA\ bump and its correlations with stellar population and emission-line properties. In the fourth paper (Paper IV, Guo et al. in prep.), we will use the attenuation curve measurements obtained in this study to constrain a two-component dust model, enabling us to derive the distributions of dust grain sizes and the fractional contributions of dust mass from silicate and graphite compositions.


This paper is organized as follows. In \autoref{sec:data}, we  describe the data used in this work, along with the methods for measuring stellar population and emission-line properties. We present our results in \autoref{sec:results}, discuss our findings in \autoref{sec:discussion}, and summarize our conclusions in \autoref{sec:summary}. Throughout this paper, we assume a standard $\Lambda$CDM cosmology with $\Omega_{m}=0.3$, $\Omega_{\Lambda}=0.7$, and $H_{0}=70~\text{km~s}^{-1}~\text{Mpc}^{-1}$.

\begin{figure*}[ht!]
  \centering
   \includegraphics[width=0.45\textwidth]{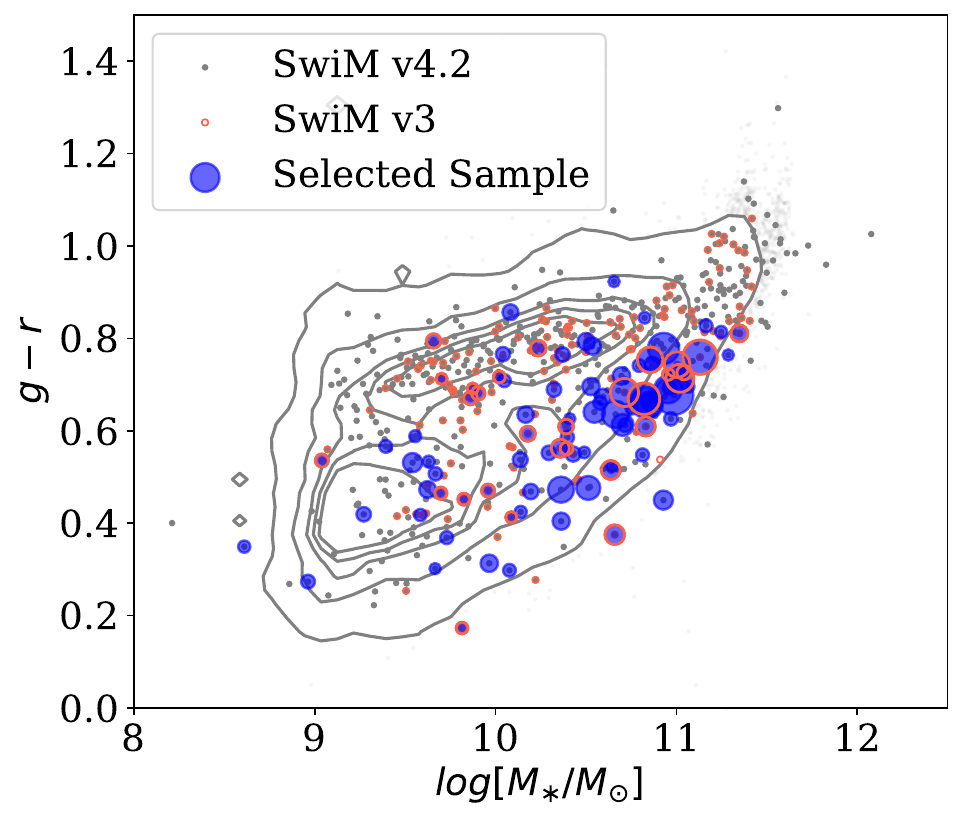}
   \includegraphics[width=0.45\textwidth]{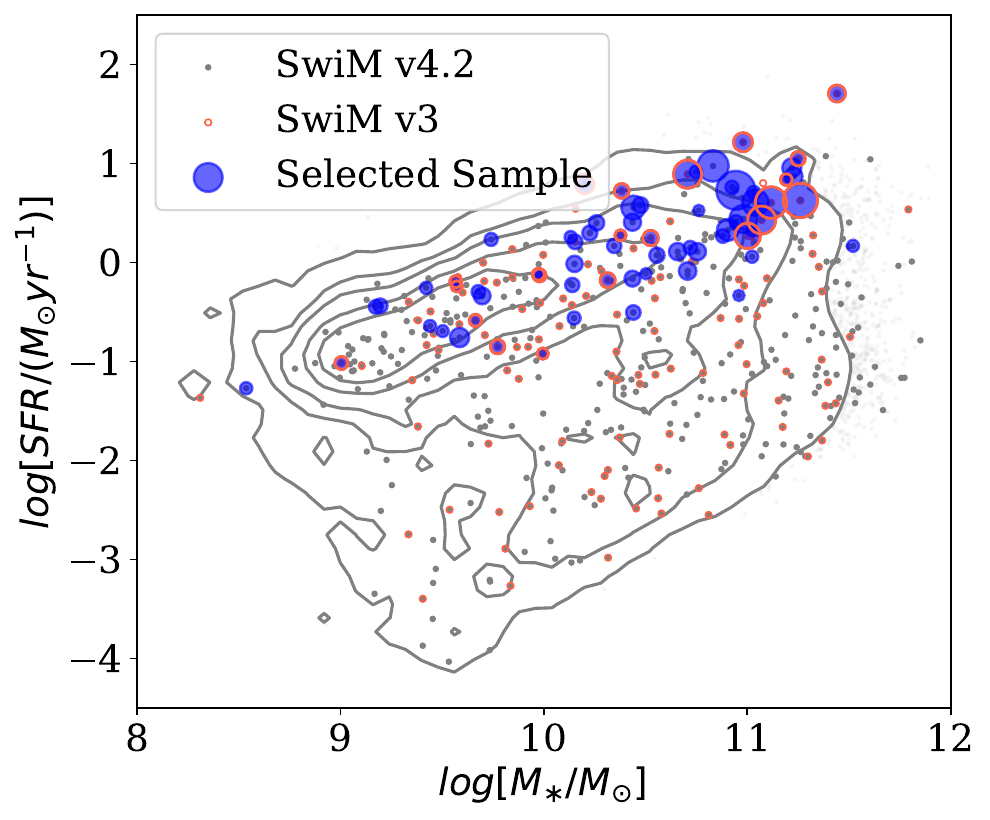}
  \caption{\emph{Left}: The global $g-r$ color versus stellar mass diagram for the SwiM\_v4.2 catalog (gray filled circles), compared with the full sample of MaNGA galaxies, shown as black contours. Red open circles represent galaxies included in the earlier SwiM\_v3.1 catalog, while blue filled circles indicate galaxies from SwiM\_v4.2 that contain at least one spaxel meeting our selection criteria. The size of the circles represents the relative number of selected spaxels per galaxy. \emph{Right}: The global star formation rate versus stellar mass diagram, with all symbols, lines, and colors consistent with those in the left panel.}
  \label{fig:global}
\end{figure*}

\section{Data and Methodology} \label{sec:data}

\subsection{SwiM\_v4.2, MaNGA and 2MASS} \label{subsec:swim}

We use the latest version of the Swift/UVOT+MaNGA (SwiM) Value-Added Catalog, SwiM\_v4.2, which is publicly available from the SDSS website\footnote{\url{https://www.sdss4.org/dr17/data_access/value-added-catalogs/?vac_id=swift-manga-value-added-catalog}} and described in detail in \citet{2023ApJS..268...63M}. The SwiM\_v4.2 catalog includes 559 galaxies, nearly four times more than the original SwiM\_3.1 catalog \citep{2020ApJS..251...11M} used in Paper I. By construction, the SwiM galaxies have optical integral field spectroscopy (IFS) from MaNGA \citep{2015ApJ...798....7B, 2016AJ....152..197Y}, as well as NUV photometry from Swift/UVOT \citep{2005SSRv..120...95R} in three bands: {\tt uvw2}, {\tt uvm2}, and {\tt uvw1}, centered at 1928\AA, 2246\AA, and 2600\AA, respectively. The {\tt uvm2} filter is centered near 2175\AA, thus allowing the investigation of the UV bump feature in dust attenuation curves.

The galaxies in SwiM\_v4.2 are selected by cross-matching the final sample of MaNGA with the UVOT data archive as of August 2021. MaNGA is one of the three major experiments of the SDSS-IV project \citep{2017AJ....154...28B}, accomplished over a period of six years from July 2014 through August 2020 \citep{2015ApJ...798....7B}. The full MaNGA sample consists of 10,010 unique galaxies selected from the NASA Sloan Atlas \citep[NSA;][]{2011AJ....142...31B}, covering a redshift range of $0.01<z<0.15$ with a median redshift of $z\sim0.03$ and a stellar mass range of $5\times10^{8}\text{M}_{\odot} \leq M{\ast} \leq 3\times10^{11}\text{M}_{\odot}$ \citep{2017AJ....154...86W}. The galaxies were observed with a typical exposure time of 3 hours, using 17 pluggable hexagonal-formatted Integral Field Units \citep[IFUs;][]{2015AJ....149...77D} that are fed to the two dual-channel BOSS spectrographs on the Sloan 2.5-meter telescope \citep{2006AJ....131.2332G,2013AJ....146...32S} to obtain IFS data with a field of view ranging from $12^{\prime\prime}$ to $32^{\prime\prime}$, effective spatial resolution of $\text{FWHM}\sim2.5^{\prime\prime}$, and spectral resolution of $R\sim 2000$ in the wavelength range from 3622\AA\ to 10354\AA\ \citep{2015AJ....150...19L}. The observational data reach an $r$-band continuum signal-to-noise ratio (S/N) of $4-8$ per \AA\ per fiber at $1-2$ effective radii ($R{e}$) of galaxies. MaNGA raw data are reduced with the Data Reduction Pipeline \citep[DRP;][]{2016AJ....152...83L,2021AJ....161...52L} to produce a data cube for each galaxy, with a spaxel size of $0.5^{\prime\prime}\times0.5^{\prime\prime}$ and absolute flux calibration better than 5\% for more than 80\% of the wavelength range \citep{2016AJ....152..197Y, 2016AJ....151....8Y}. Additionally, the Data Analysis Pipeline performs full spectral fitting to the DRP datacubes, providing measurements of stellar kinematics, emission lines, and spectral indices \citep{2019AJ....158..231W, 2019AJ....158..160B}. The DRP and DAP data products of all the MaNGA galaxies are released as part of the final data release of SDSS-IV \citep[DR17;][]{2022ApJS..259...35A}.

All 559 galaxies in SwiM\_v4.2 have UVOT observations in {\tt uvw1} and {\tt uvw2}, while 490 also have observations in {\tt uvm2}. In addition to a catalog that provides integrated photometry from Swift, SwiM\_v4.2 also includes two-dimensional maps of numerous emission lines and spectral indices as measured by the MaNGA DAP, as well as Swift/UVOT images in {\tt uvw1} and {\tt uvw2} (also {\tt uvm2} when available) and SDSS images in {\tt ugriz}. All maps and images are convolved and resampled to match the spatial resolution and sampling of the {\tt uvw2} band, which has a point spread function (PSF) with a full width at half maximum (FWHM) of $2.^{\prime\prime}92$ and a pixel size of $1^{\prime\prime}$. \autoref{fig:global} displays all the galaxies in SwiM\_v4.2 as grey dots in the color-mass diagram ($g-r$ versus $\log_{10}M_\ast$; left panel) and the diagram of star formation rate (SFR) and mass ($\log_{10}\text{SFR}$ versus $\log_{10}M_\ast$; right). Here, $g-r$ and $M_\ast$ are taken from the NSA, and SFRs are from the GSWLC-X2 catalog \citep{2018ApJ...859...11S}. In the figure, the MaNGA DR17 sample is plotted as background contours for comparison. As can be seen, the SwiM\_v4.2 catalog spans similarly wide ranges in these parameters. Plotted in red circles are the galaxies in the earlier SwiM\_v3.1 catalog, which was described in detail in \citet{2020ApJS..251...11M} and has been used in several works to study spatially resolved dust attenuation in nearby galaxies \citep{2020MNRAS.494.4751M, 2023MNRAS.526..904D, 2023ApJ...957...75Z}. 

For each of the 559 SwiM galaxies we retrieved its $K_S$-band (2.16 $\mu\text{m}$) atlas image from the 2MASS data archive \citep{2006AJ....131.1163S}. Compared to the SwiM images, the 2MASS images have the same sampling of $1^{\prime\prime}$ per pixel and  a similar spatial resolution of FWHM$\sim 2.5^{\prime\prime}-3.5^{\prime\prime}$. Following Paper I, we simply resample the 2MASS images to match the sampling of the SwiM images, without performing additional resolution matching,  which would lead to flux differences of $<1\%$ due to the similar spatial resolutions. 

\subsection{Deriving and Characterizing Dust Attenuation Curves} \label{subsec:deriving}

For each spaxel of each galaxy in SwiM\_v4.2, we apply the same methodology as described in Paper I to derive the attenuation curve across the full wavelength range from NUV to NIR. Readers are referred to Paper I for a detailed description of our method and tests. Here, we provide a brief overview of the process, which consists of three steps. First, we use the technique of \citet{2020ApJ...896...38L} to derive a relative attenuation curve from the MaNGA optical spectrum and apply it to obtain an attenuation-corrected spectrum with an arbitrary flux unit. Next, we employ the Bayesian spectral fitting code \citep[BIGS;][]{2019MNRAS.485.5256Z} to fit the attenuation-corrected spectrum, yielding the best-fit model spectrum that spans the full wavelength range from NUV to NIR. Finally, assuming that dust attenuation in the NIR band is negligible, we use the $K_S$-band image to calibrate the absolute flux of the dust-free best-fit model spectrum. We then derive the absolute attenuation curve in the optical and determine the attenuation in the three NUV bands by comparing the observed spectrum and images with the model spectrum.

To characterize the dust attenuation curves, we measure the following parameters from each curve: (1) $A_V$, the optical opacity, defined as the attenuation in the $V$-band; (2) $A_B/A_V$, the optical slope of the attenuation curve, given by the attenuation ratio between the $B$ ($\lambda=4400$\AA) and $V$ ($\lambda=5500$\AA) bands; and (3) $A_{\tt w2}/A_{\tt w1}$, the NUV slope, defined as the attenuation ratio between the {\tt uvw2} and {\tt uvw1} bands. In the literature, another commonly used parameter for the optical slope is the total-to-selective attenuation ratio in the $V$-band, defined as $R_V\equiv A_V/E(B-V)$, where $E(B-V)\equiv A_B-A_V$ is the color excess. By this definition, $A_B/A_V$ is related to $R_V$ as $A_B/A_V=1+R_V^{-1}$. For example, the standard Calzetti curve with $R_V=4.05$ gives $A_B/A_V=1.25$ \citep{1994ApJ...429..582C,2000ApJ...533..682C}, while the Milky Way-like curve with $R_V=3.1$ yields $A_B/A_V=1.32$ \citep{1989ApJ...345..245C}. The NUV slope is $A_{\tt w2}/A_{\tt w1}=1.19$ for the Calzetti curve and 1.24 for the Milky Way curve.

Tests on mock spectra, as conducted in Paper I, demonstrated that our method can reliably recover the average dust attenuation properties in individual spaxels of SwiM galaxies, provided that the spectral signal-to-noise ratio (S/N) is sufficiently high. We adopt the same selection criteria as in Paper I to identify spaxels for this study: $S/N>20$ and $A_V>0.25$. Here, $S/N$ refers to the signal-to-noise ratio in the continuum, measured around 5500\AA. Applying these criteria, we obtain a total of 2487 spaxels across 91 galaxies. These galaxies are represented as blue dots in \autoref{fig:global}, where the size of each dot is proportional to the number of selected spaxels contributed by the galaxy. Notably, the selected galaxies are predominantly blue and star-forming, and they tend to be more massive than those in the parent sample.

\subsection{Measuring Stellar Populations and Emission Lines} \label{subsec:stellar_character}

Using the attenuation curves derived above, we correct the observed MaNGA spectrum in each spaxel for dust attenuation. We then perform full spectral fitting on the dust-corrected spectrum, extracting both stellar population parameters from the best-fit stellar spectrum and emission line parameters from the starlight-subtracted spectrum. The methodology is described in detail in \citet{2021ApJ...917...72L}. In brief, we fit the dust-corrected spectra using a set of 150 simple stellar populations (SSPs) selected from the SSP library of \citet[][BC03]{2003MNRAS.344.1000B}. This library provides model spectra for 1326 SSPs at a spectral resolution of $3\AA$, covering 221 ages from 0 to 20 Gyr and six metallicities ranging from 0.005$Z_\odot$ to 2.5$Z_\odot$. The models are computed using the initial mass function (IMF) of \citet{2003PASP..115..763C} and the Padova evolutionary tracks \citep{1994A&AS..106..275B}. Each spectrum is fitted with a linear combination of the SSPs, with the effect of stellar velocity dispersion taken into account by convolving the spectra of the SSPs with a Gaussian. During the fitting, we have carefully masked out all the detected emission lines following the scheme described in \citet{2005AJ....129..669L}. 

Based on the best-fit stellar spectrum and the corresponding fractional contributions of the SSPs, we then measure the following stellar population parameters: 
\begin{enumerate}
  \item $\log_{10}\Sigma_{\ast}$---logarithm of the stellar mass surface density in units of $\text{M}_{\odot}~\text{kpc}^{-2}$.
  \item $\log_{10}t_{L}$ and $\log_{10}t_{M}$---logarithm of the luminosity-weighted and mass-weighted stellar age in units of yr.
  \item $Z_{L}$ and $Z_{M}$---logarithm of the luminosity-weighted and mass-weighted stellar metallicity in units of solar metallicity $Z_{\odot}$, where $Z_{\odot}=0.02$.
  \item $D_{n}4000$---the narrow-band version of the 4000\AA\ break \citep{1999ApJ...527...54B}. 
  \item EW(H$\delta_\text{A}$)---the equivalent width of the $H\delta$ absorption line in units of \AA.
\end{enumerate}

\begin{figure*}[ht!]
  \centering
   \includegraphics[width=0.45\textwidth]{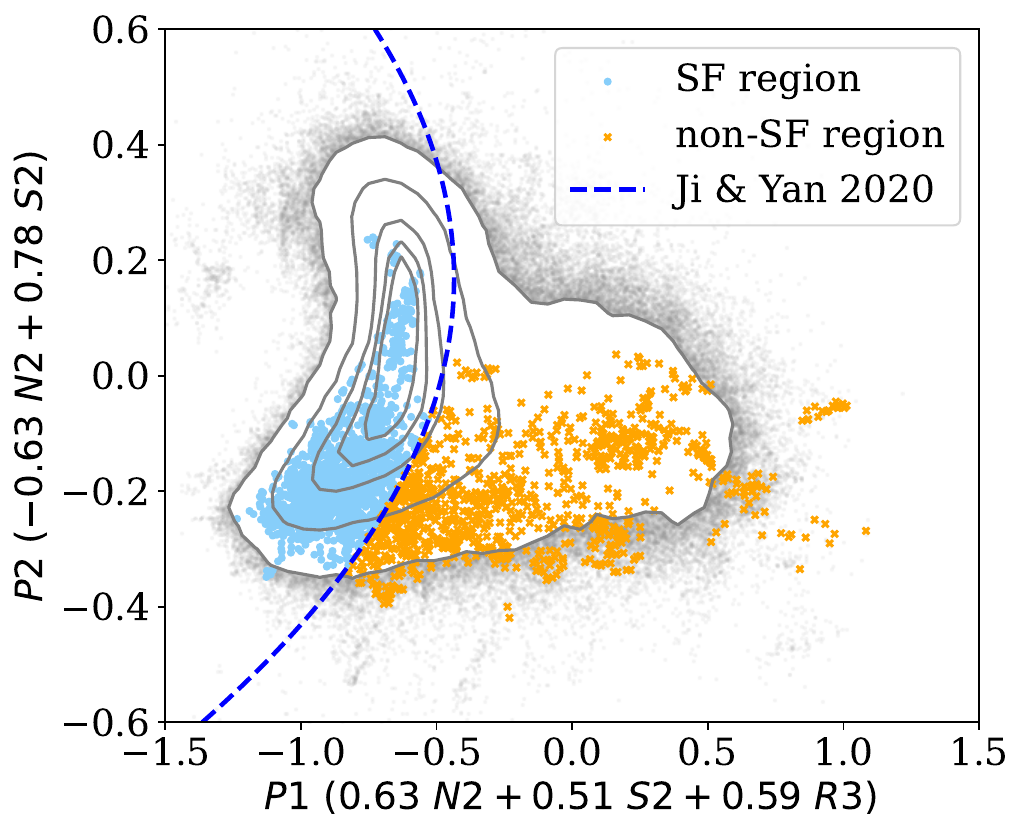}
   \includegraphics[width=0.45\textwidth]{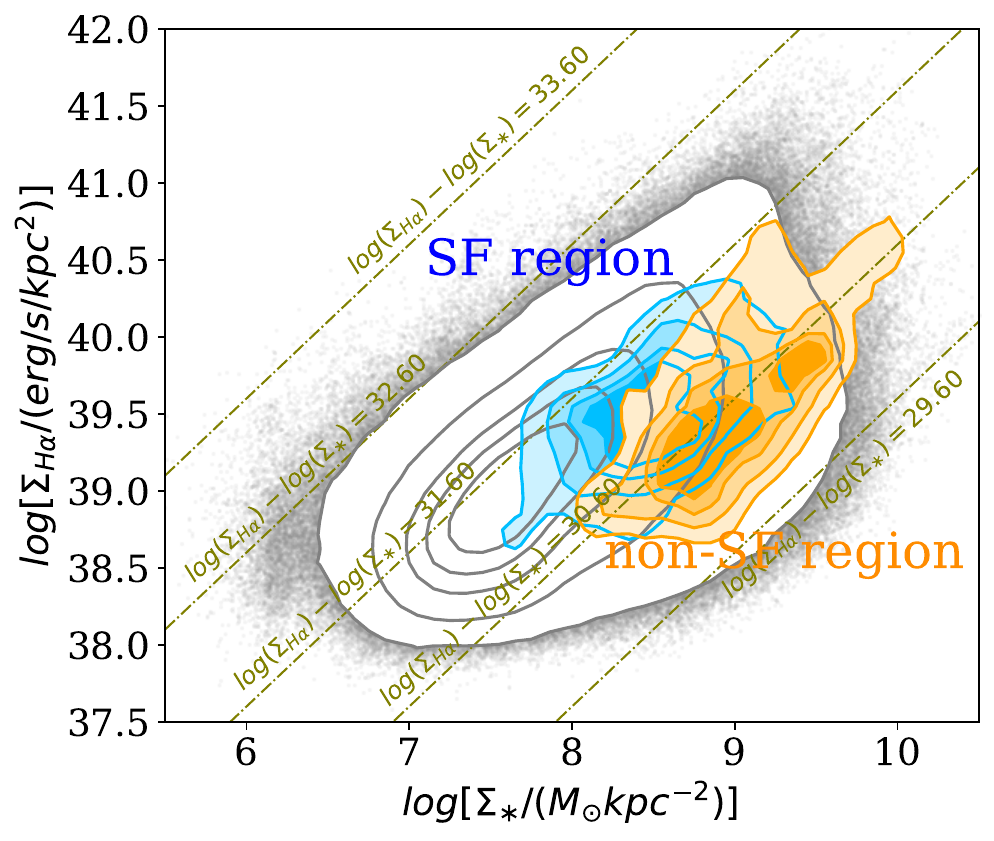}
  \caption{\emph{Left}: The reprojected BPT diagram from \citet{2020MNRAS.499.5749J}, with the blue dashed line indicating the division between star-forming (SF) and non-SF regions. \emph{Right}: The diagram of H$\alpha$ surface brightness ($\Sigma_{\text{H}\alpha}$) versus surface stellar mass density ($\Sigma_\ast$), where the parallel dashed lines denote constant values of specific H$\alpha$ surface brightness, i.e., $\Sigma_{\text{H}\alpha}/\Sigma_\ast$. In both panels, blue dots/contours and yellow crosses/contours represent the SF and non-SF regions used in this study, respectively. The gray contours show the distribution of all spaxels with $S/N>3$ from the full MaNGA sample, with the outermost contour enclosing 95\% of the sample, while gray dots represent individual spaxels beyond that contour level.}
  \label{fig:P1_P2}
\end{figure*}

We subtract the best-fit model spectrum from the observed spectrum to isolate the emission-line spectrum. Each emission line is then fitted with either a single Gaussian or a double Gaussian function, providing measurements of the flux, surface brightness, central wavelength, line width, and equivalent width (EW). To correct for the effects of gas attenuation, we calculate the Balmer decrement using the observed H$\alpha$/H$\beta$ flux ratio and assume case-B recombination. From H$\alpha$/H$\beta$, we estimate the gas attenuation, quantified by the color excess $E(B-V)_\text{gas}$, and use it to correct the flux of all emission lines. Based on these emission line measurements, we derive the following parameters to characterize the gas-related properties of each spaxel:
\begin{enumerate}
  \item $\log_{10}\Sigma_{\text{H}\alpha}$---logarithm of the surface brightness of the H$\alpha$ emission line in units of $\text{erg~s}^{-1}~\text{kpc}^{-2}$. 
  \item $\log_{10}\text{sSFR}$---logarithm of the specific star formation rate (sSFR) defined as the ratio of SFR to stellar mass in a given spaxel. We estimate the SFR for each spaxel from the H$\alpha$ luminosity using the estimator from \citet{1998ARA&A..36..189K}. It is worth noting that the SFR estimated in this way is reliable only for star-forming (SF) regions where H$\alpha$ emission is dominantly contributed by young massive stars. 
  \item $\log_{10}(\Sigma_{\text{H}\alpha}/\Sigma_{\ast})$---logarithm of the specific surface brightness of H$\alpha$. For SF regions, this parameter is equivalent to $\log_{10}\text{sSFR}$ defined above. 
  \item $\log_{10}\text{EW}(\text{H}\alpha)$---logarithm of the  equivalent width of the H$\alpha$ emission line in units of \AA. 
  \item $12+\log_{10}(\text{O/H})_\text{O3N2}$---the gas-phase metallicity estimated using the O3N2 indicator:    $\text{O3N2}\equiv([\text{OIII}]5007/\text{H}\beta)/([\text{NII}]6583/\text{H}\alpha)$ \citep[][]{2013A&A...559A.114M}.
  \item $12+\log_{10}(\text{O/H})_\text{R23}$---the gas-phase metallicity estimated using the R23 indicator:  $\text{R23}\equiv([\text{OIII}]5007,4959+[\text{OII}]3727)/\text{H}\beta$ \citep{2022ApJS..262....3N}.
  \item N2S2---logarithm of the flux ratio between emission lines [NII]$\lambda6583$ and [SII]$\lambda6717$, a parameter that is sensitive to both gas-phase metallicity and ionization parameter. 
\end{enumerate}

\subsection{Classification of SF and Non-SF Regions} \label{subsec:sample_select}

We classify all spaxels in our sample into two categories: star-forming (SF) and non-SF regions, based on the diagnostic diagram developed by \citet{2020MNRAS.499.5749J}. This diagram is derived by projecting the traditional BPT diagrams \citep{1981PASP...93....5B} into a new parameter space, allowing for a clearer separation of different ionization models. Derived this way, the new parameters, $P_1$ and $P_2$, are linear combinations of $\text{N2} \equiv \log_{10}([\text{NII}]6583/\text{H}\alpha)$, $\text{S2} \equiv \log_{10}(([\text{SII}]6717+[\text{SII}]6731)/\text{H}\alpha)$, and $\text{R3} \equiv \log_{10}([\text{OIII}]5007/\text{H}\beta)$.
The left panel of \autoref{fig:P1_P2} shows the diagram of $P_1$ versus $P_2$ for  the SF and non-SF spaxels in our sample, separated by the blue dashed line from \citet{2020MNRAS.499.5749J} and plotted as blue and yellow dots, respectively. For comparison, the gray contours/dots represent the distribution of all spaxels with S/N$>3$ from the full sample of MaNGA galaxies. The right panel presents the SF and non-SF regions on the $\log_{10} \Sigma_{\text{H}\alpha}$ versus $\log_{10} \Sigma_\ast$ plane. The dashed lines represent different constant values of $\log_{10} (\Sigma_{\text{H}\alpha}/\Sigma_\ast)$, which, for SF regions, corresponds to $\log_{10} \text{sSFR}$. As expected, while the two types of regions exhibit some degree of separation in this diagram, they also show significant overlap. 
Compared to the parent MaNGA sample, our sample is dominated by relatively dense regions with $\Sigma_\ast\gtrsim10^{7.5}~\text{M}_\odot~\text{kpc}^{-2}$.

\begin{figure*}[ht!]
  \centering
   \includegraphics[width=0.32\textwidth]{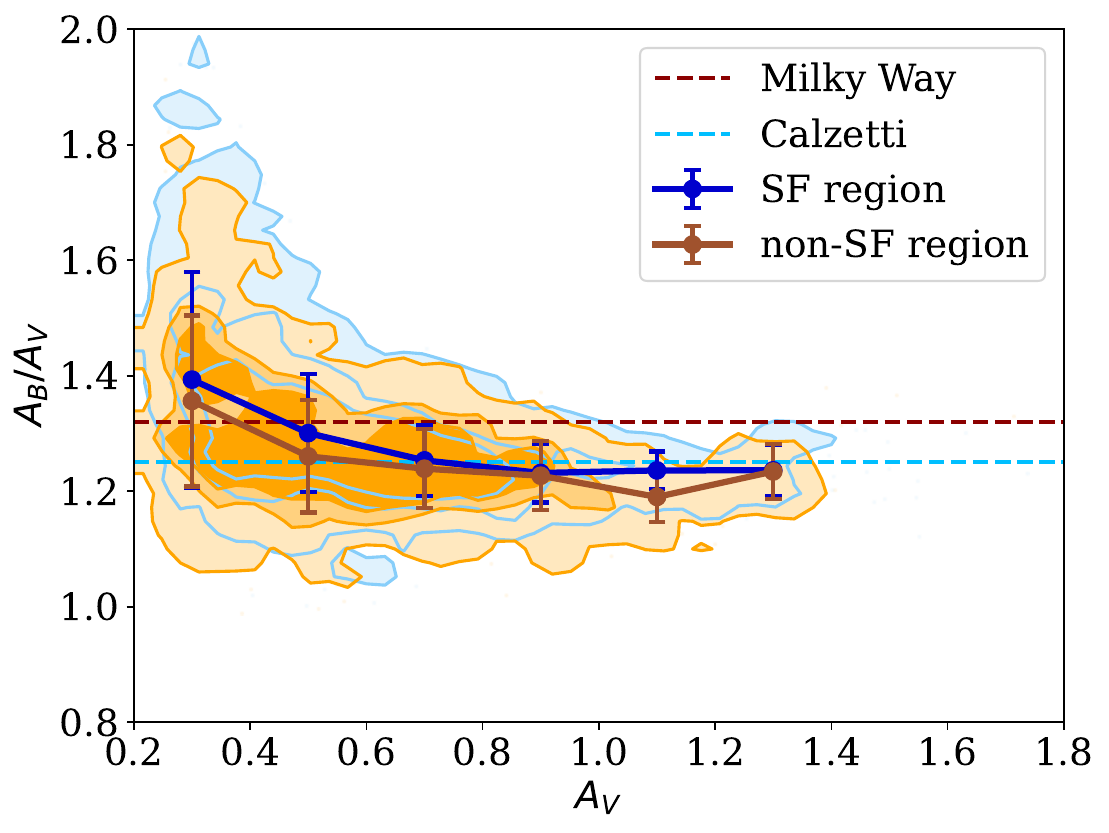}
   \includegraphics[width=0.32\textwidth]{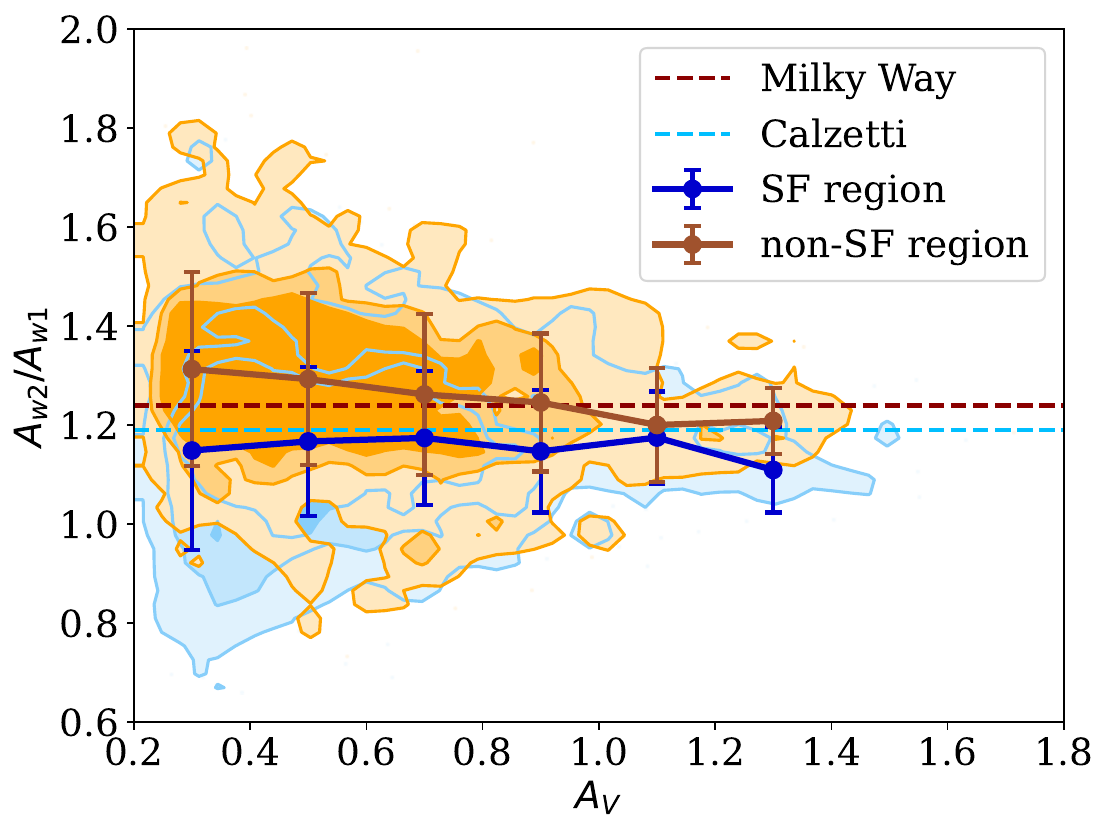}
   \includegraphics[width=0.32\textwidth]{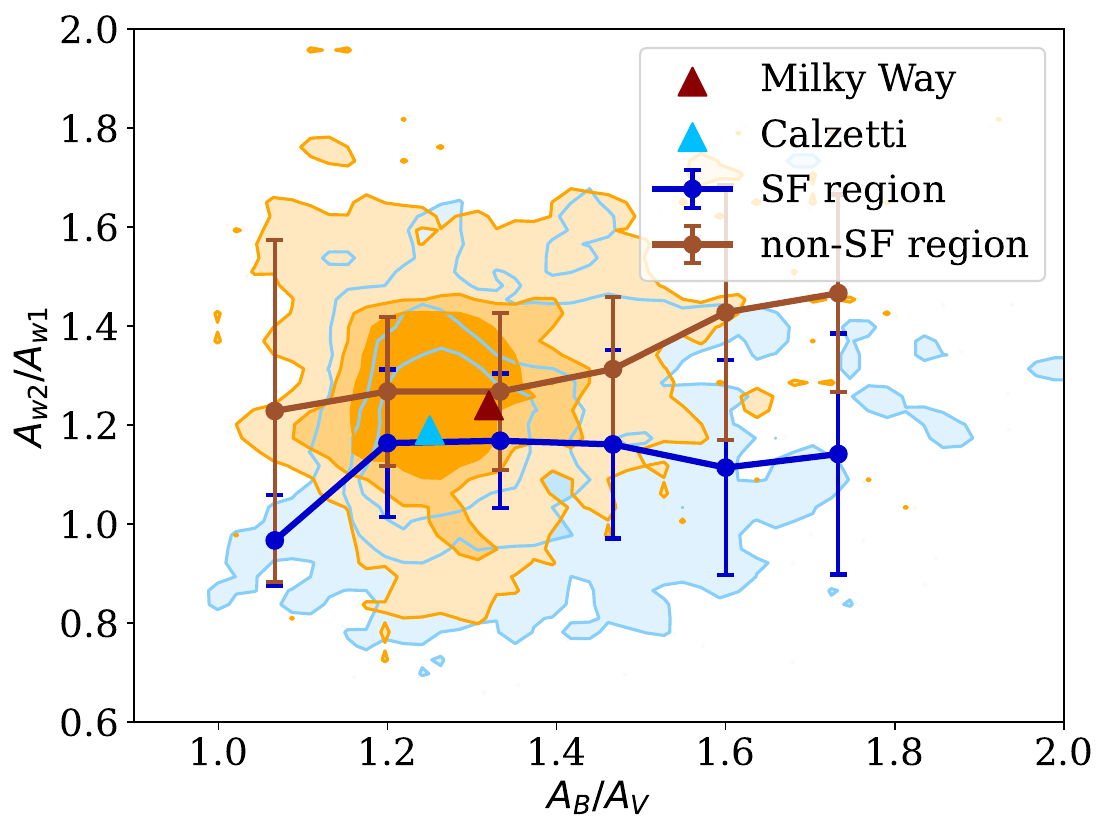}
   \caption{Mutual correlations among $A_V$, $A_B/A_V$, and $A_{\tt w2}/A_{\tt w1}$ for SF regions (light blue contours) and non-SF regions (orange contours) in our sample. The blue and brown symbols with error bars represent the median values and the associated $1\sigma$ scatter of these correlations. For comparison, the horizontal dashed lines in the left and middle panels and the triangles in the right panel indicate the corresponding values from the Calzetti curve and Milky Way-type curves.}
  \label{fig:each_other}
\end{figure*}

\begin{figure*}[h!]
  \centering
   \includegraphics[width=0.32\textwidth]{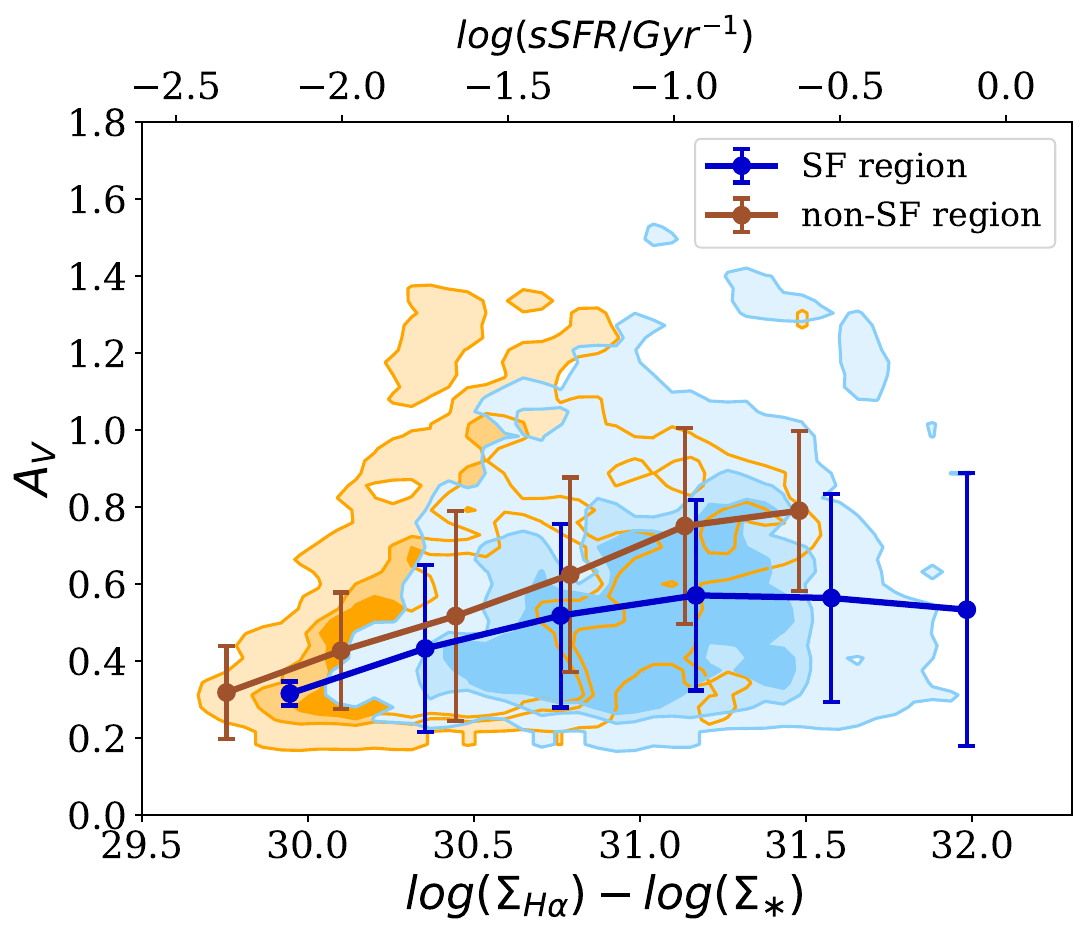}
   \includegraphics[width=0.32\textwidth]{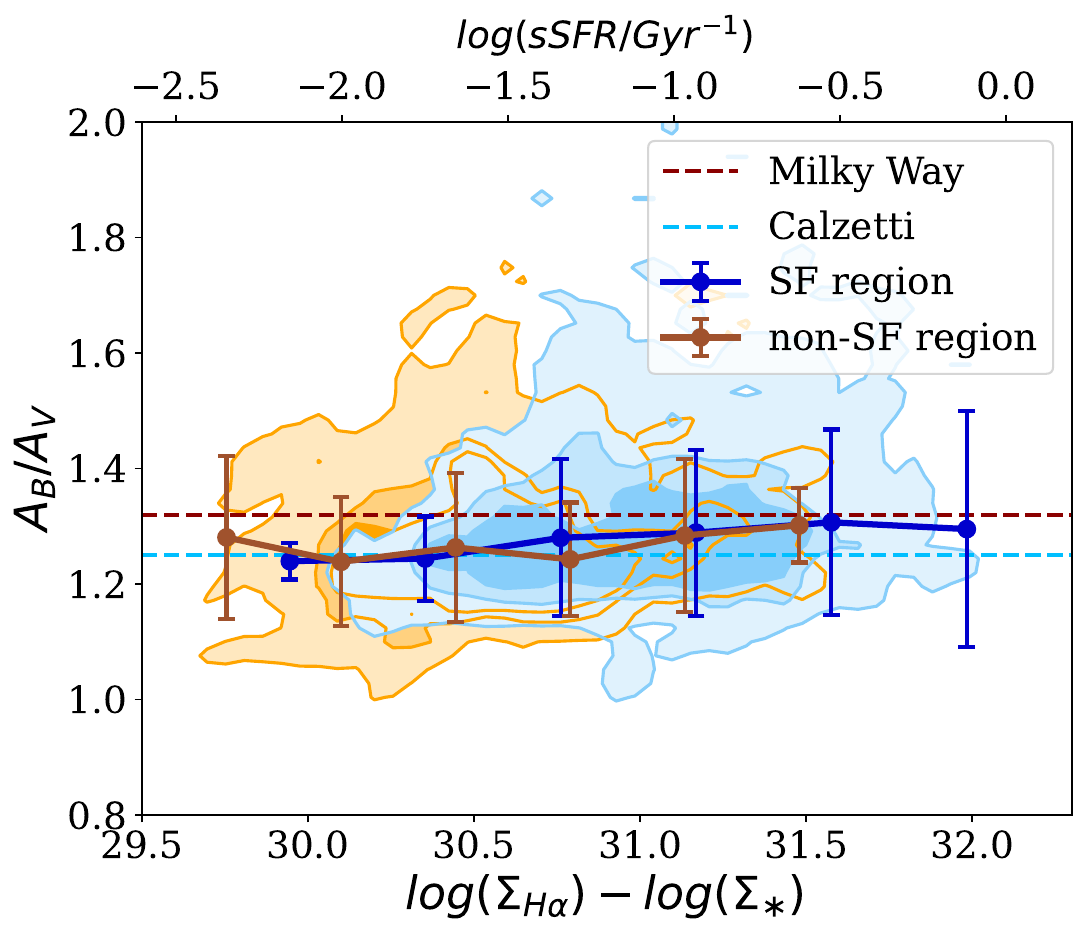}
   \includegraphics[width=0.32\textwidth]{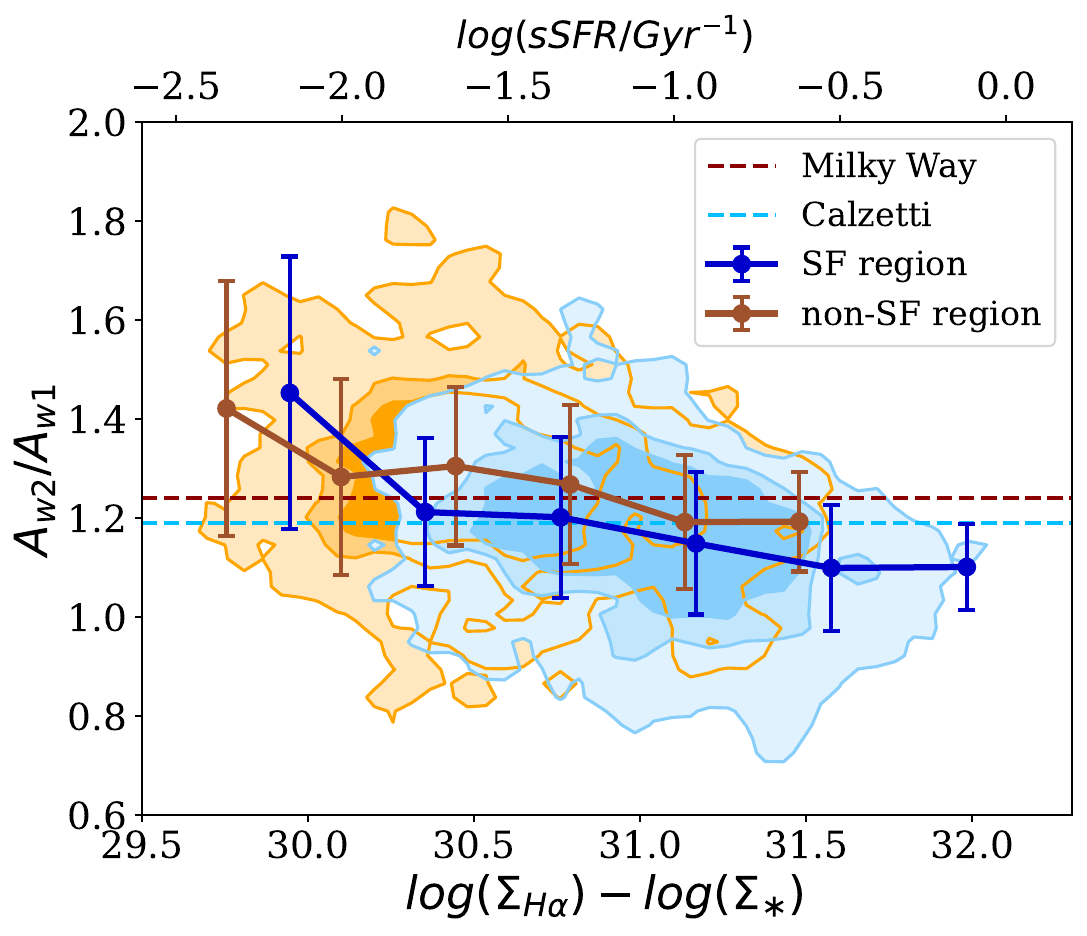}
   \includegraphics[width=0.32\textwidth]{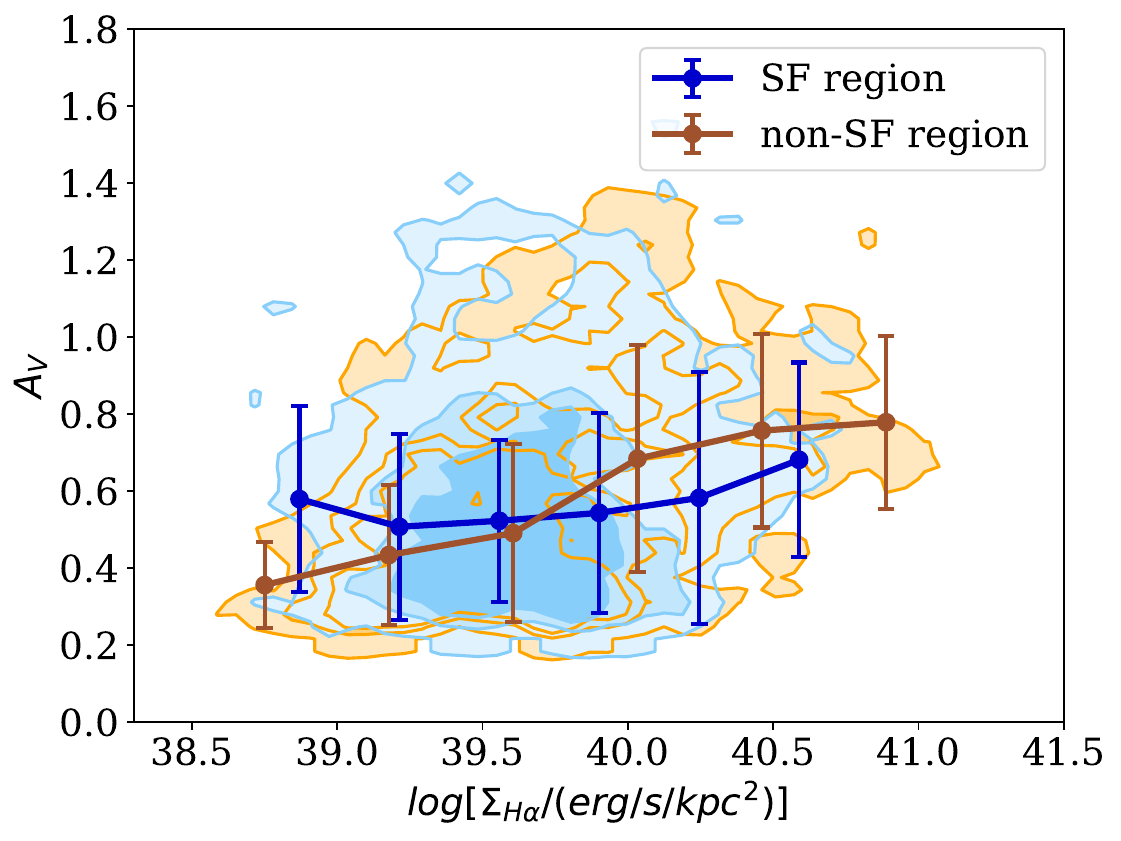}
   \includegraphics[width=0.32\textwidth]{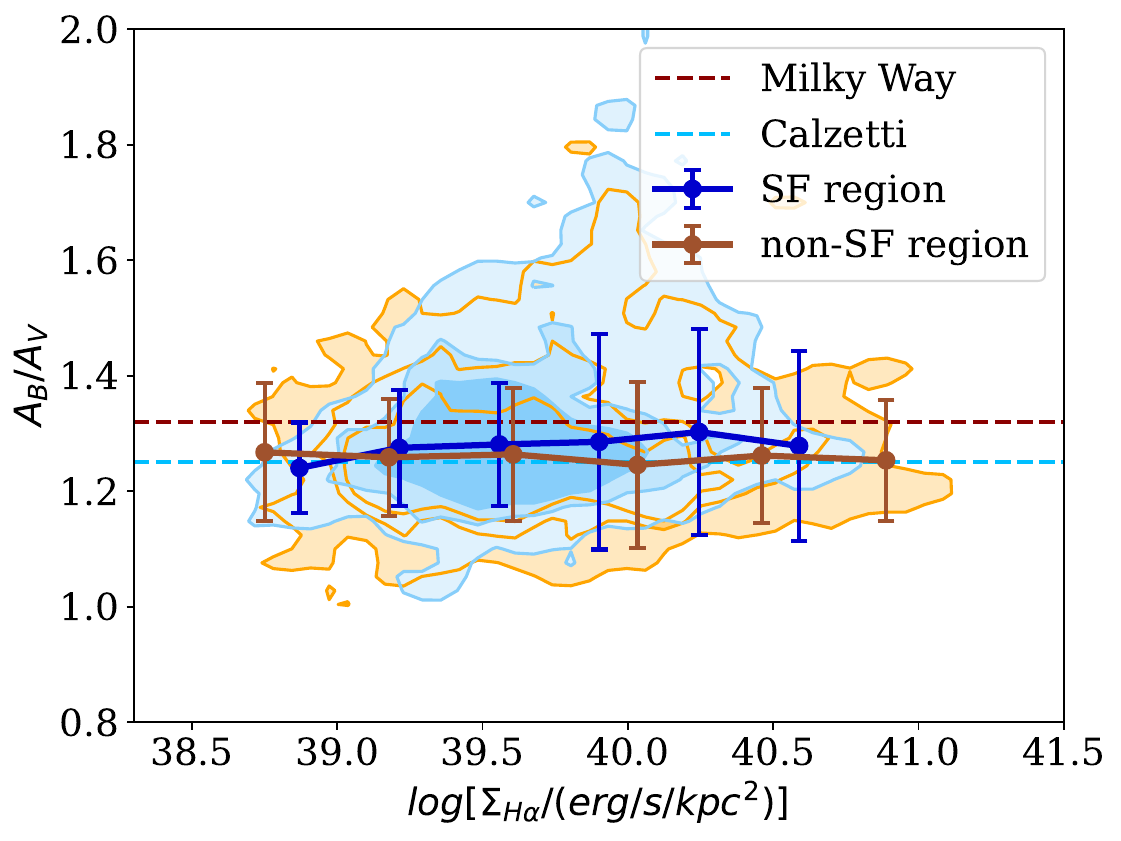}
   \includegraphics[width=0.32\textwidth]{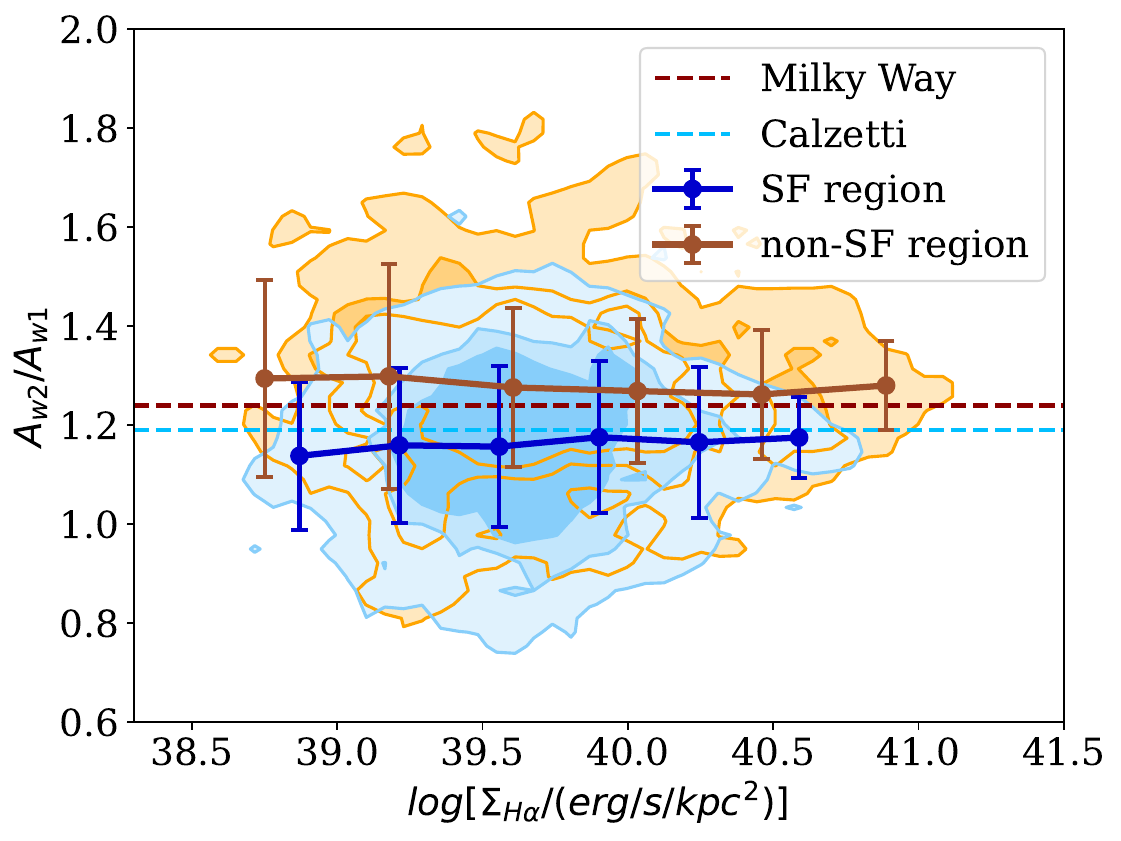}
   \includegraphics[width=0.32\textwidth]{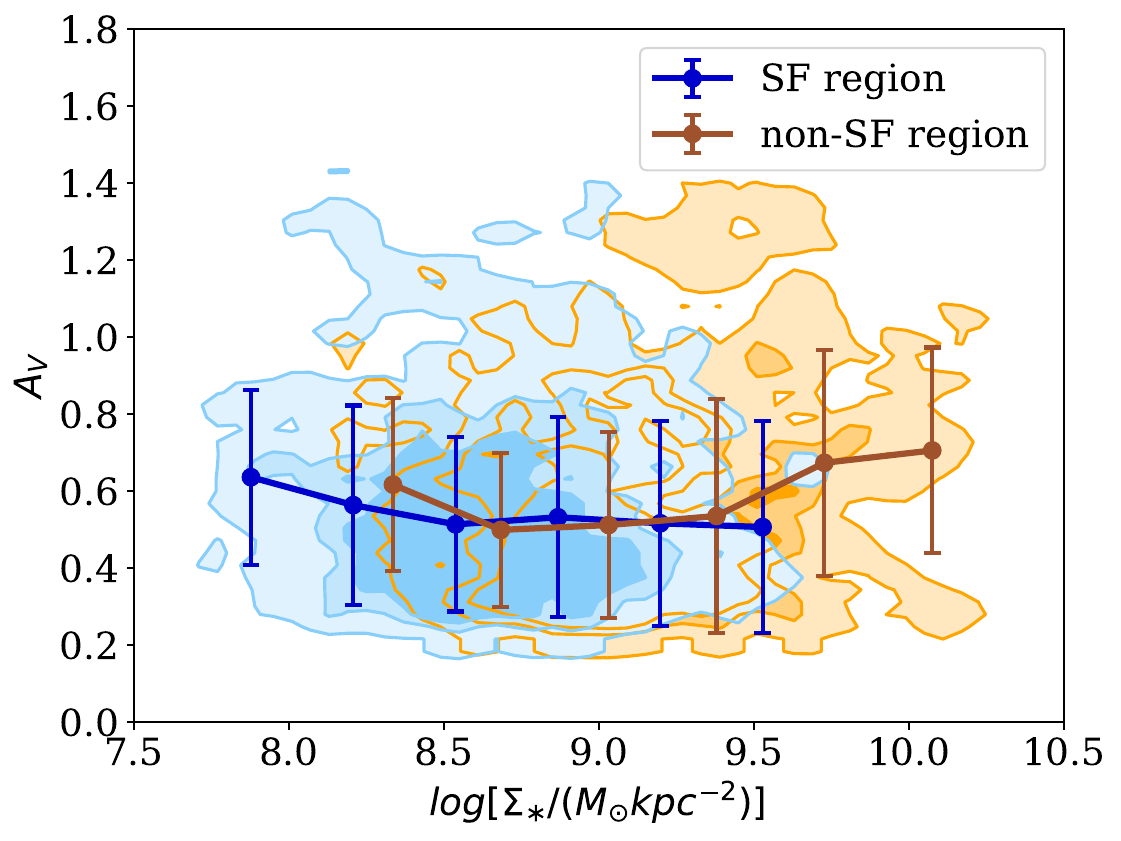}
   \includegraphics[width=0.32\textwidth]{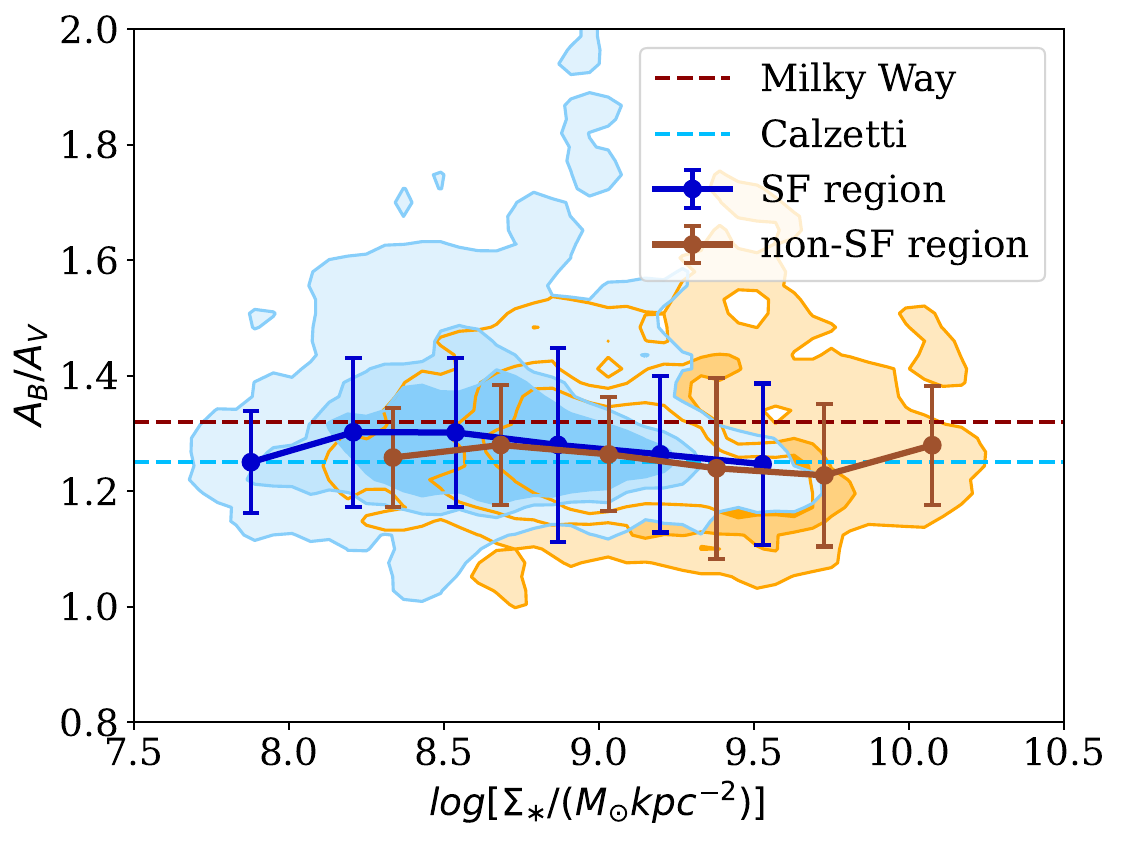}
   \includegraphics[width=0.32\textwidth]{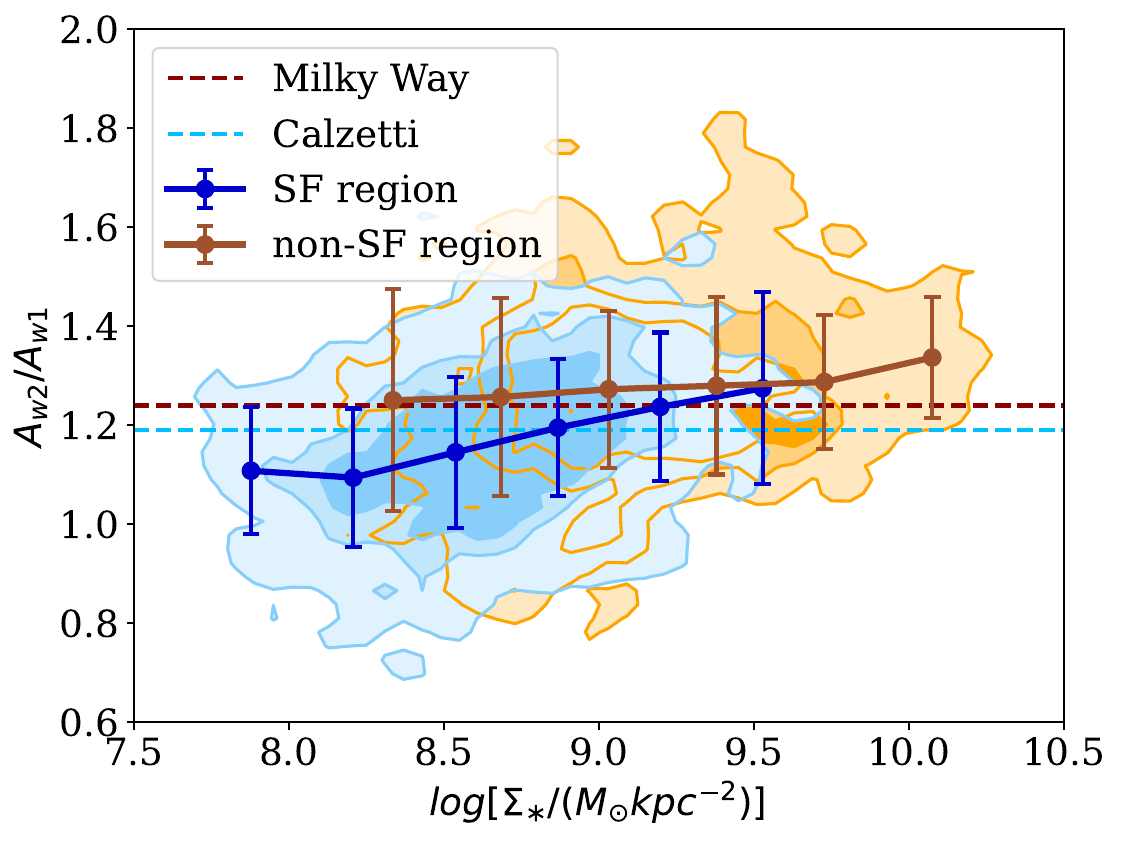}   
  \caption{The three attenuation properties—$A_V$ (left panels), $A_B/A_V$ (middle panels), and $A_{\tt w2}/A_{\tt w1}$ (right panels)—are shown as functions of the logarithm of specific H$\alpha$ surface brightness ($\Sigma_{\text{H}\alpha}/\Sigma_\ast$, top row), H$\alpha$ surface brightness ($\Sigma_{\text{H}\alpha}$, middle row), and surface stellar mass density ($\Sigma_\ast$, bottom row). The results are presented separately for SF regions and non-SF regions, with the same symbols/lines/colors as in \autoref{fig:each_other}.}
  \label{fig:dep_ssfr}
\end{figure*}

\begin{figure*}[h!]
  \centering
   \includegraphics[width=0.32\textwidth]{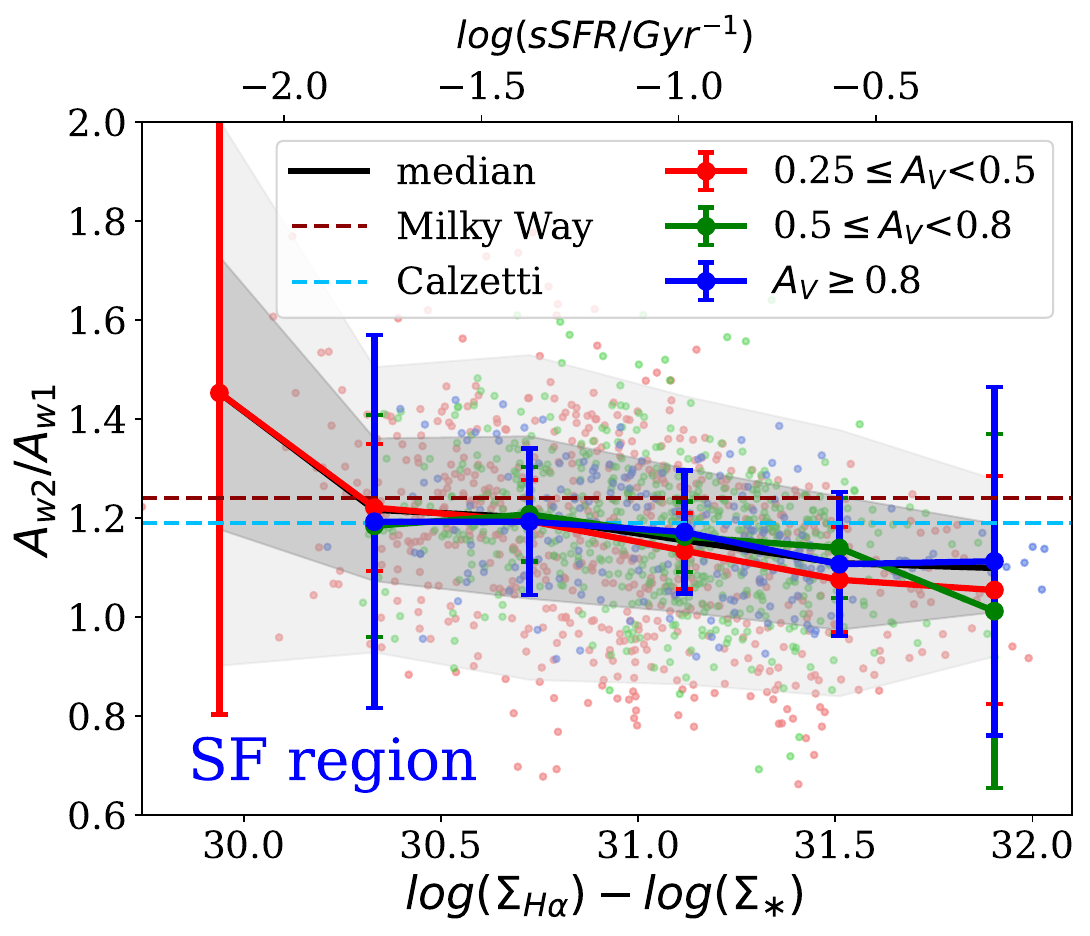}
   \includegraphics[width=0.32\textwidth]{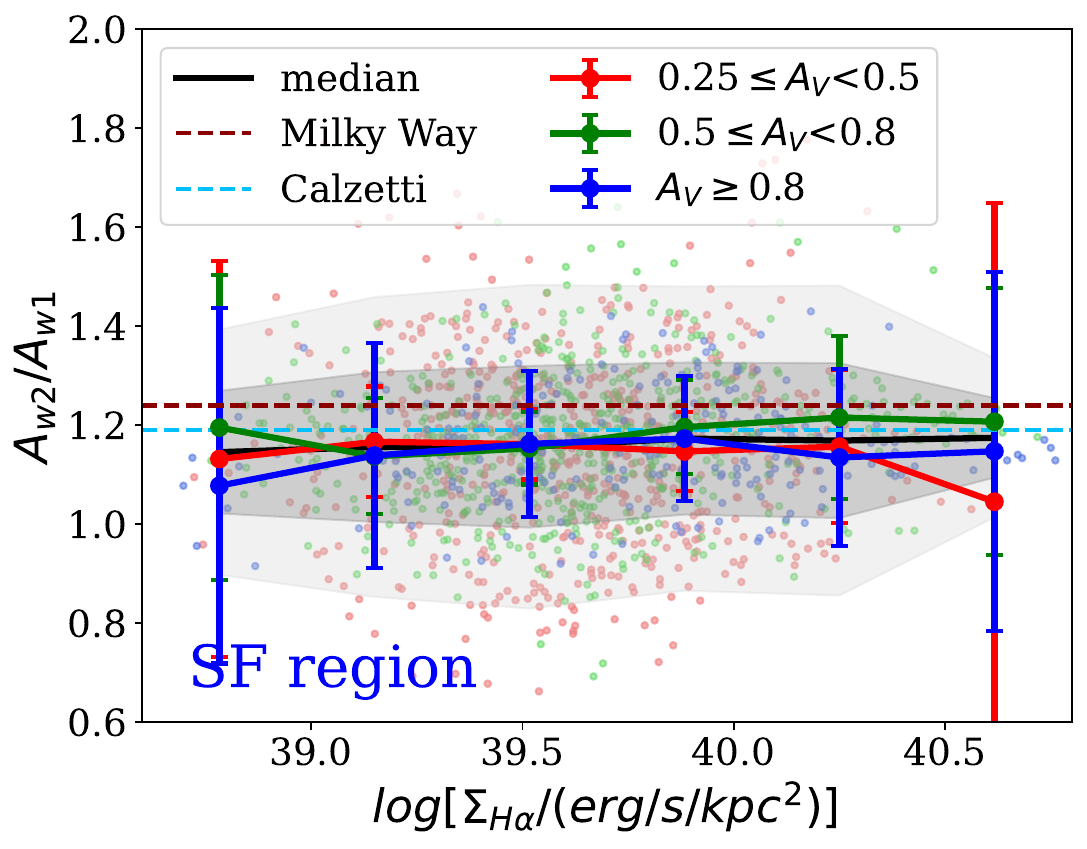}
   \includegraphics[width=0.32\textwidth]{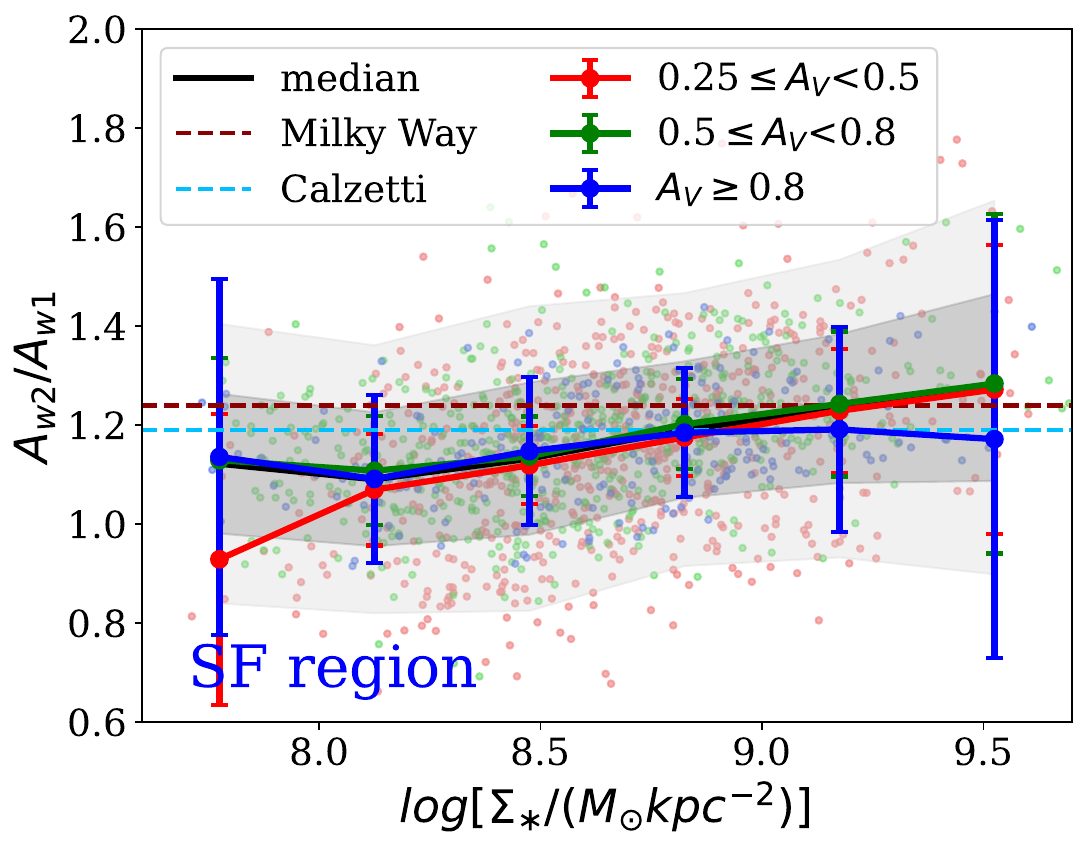}   
   \includegraphics[width=0.32\textwidth]{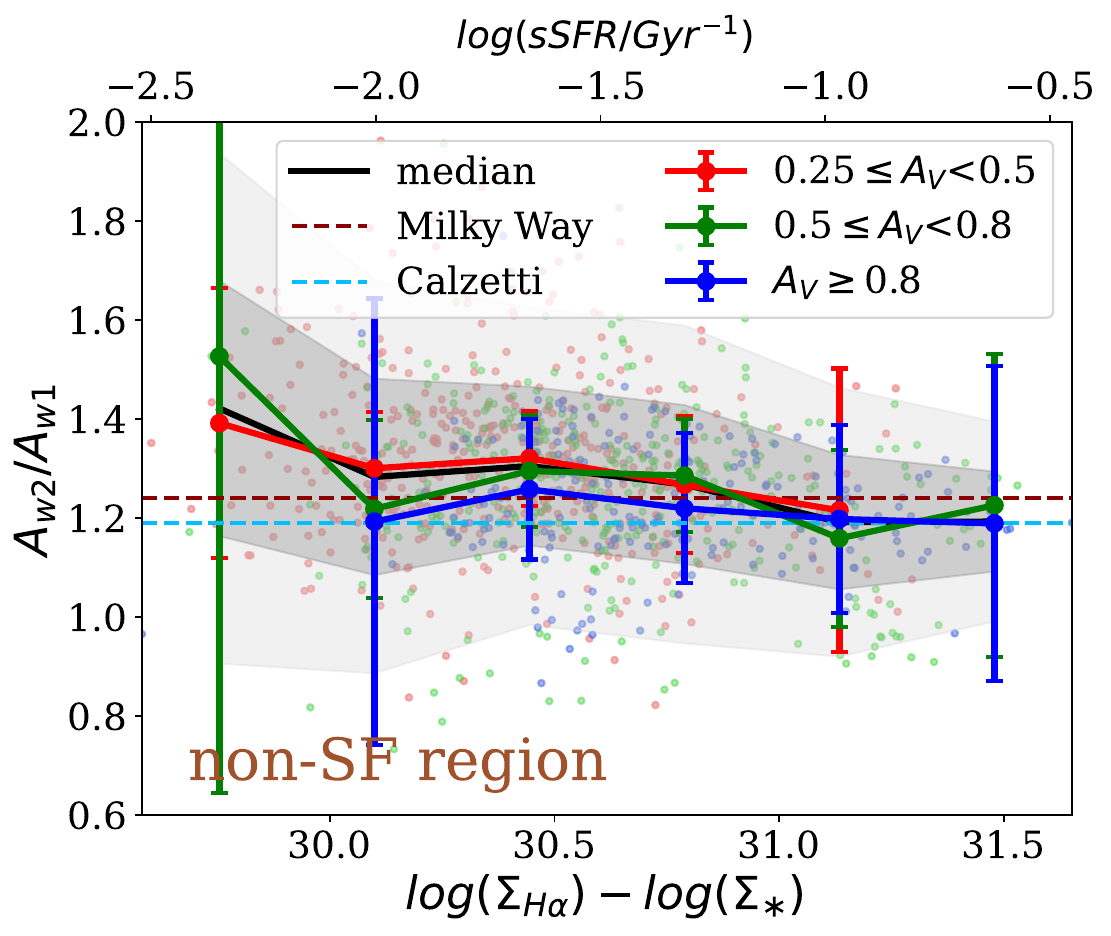}
   \includegraphics[width=0.32\textwidth]{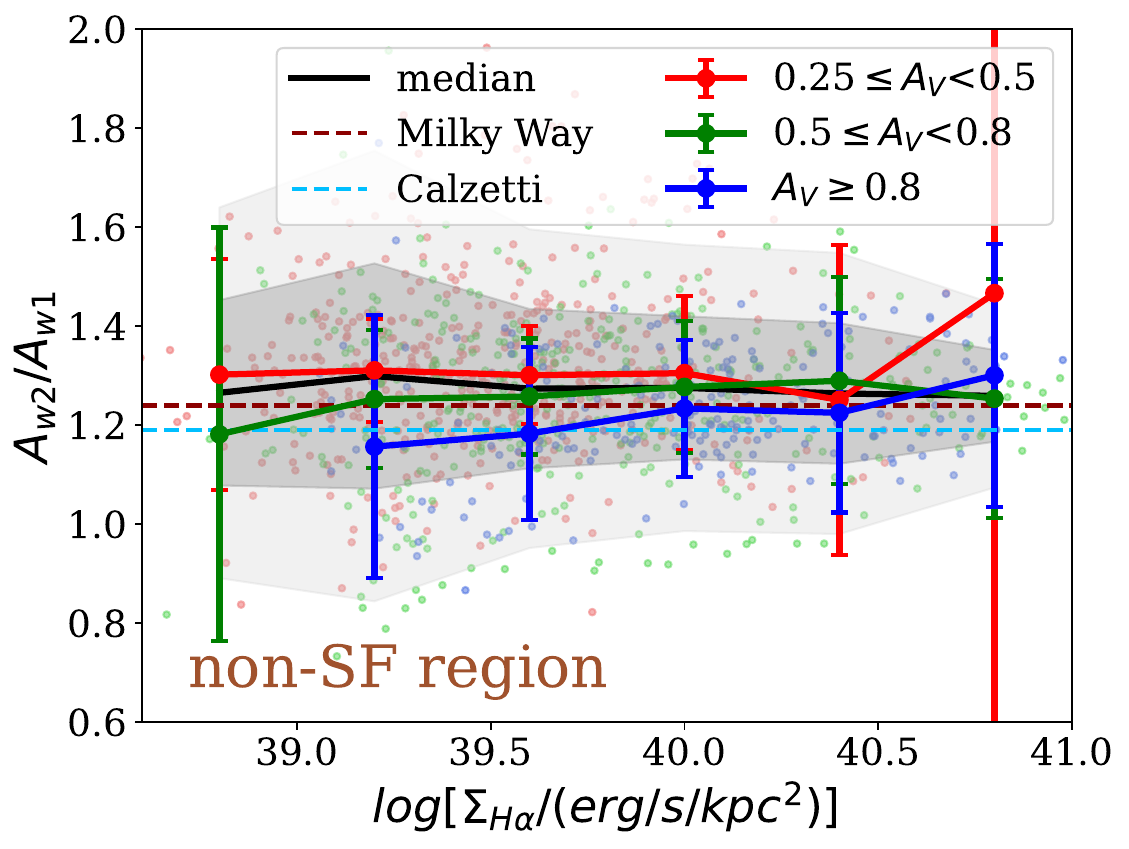}
   \includegraphics[width=0.32\textwidth]{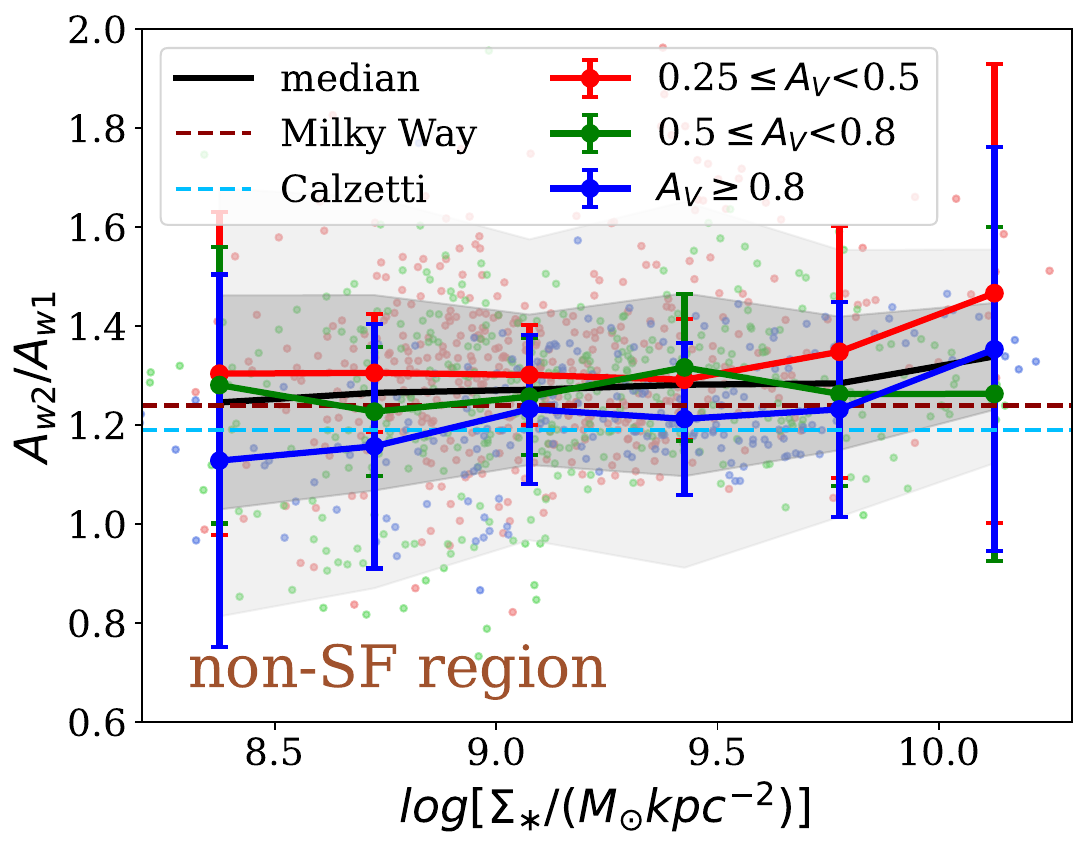}
  \caption{The NUV slope ($A_{\tt w2}/A_{\tt w1}$) of attenuation curves is shown as a function of specific H$\alpha$ surface brightness (left panels), H$\alpha$ surface brightness (middle panels), and surface stellar mass density (right panels) for SF regions (upper panels) and non-SF regions (lower panels). In each panel, the dark and light gray regions represent the $1\sigma$ and $2\sigma$ scatter around the median relation of the full sample, respectively. The red, green, and blue dots correspond to individual regions within different ranges of optical opacity ($A_V$), as indicated. The colored symbols and lines denote the median relation of the subsamples, with error bars representing Poisson errors. The horizontal dashed lines indicate the slopes of the Calzetti and Milky Way-type attenuation curves.}
  \label{fig:sSFR_slope_with_Av}
\end{figure*}

\section{Results} \label{sec:results}

\subsection{Mutual relaition between $A_V$, $A_B/A_V$ and $A_{\tt w2}/A_{\tt w1}$}

We begin by examining the interrelationships between optical opacity, characterized by $A_V$, and the slopes of the attenuation curves in the optical ($A_B/A_V$) and NUV ($A_{\tt w2}/A_{\tt w1}$), as shown in \autoref{fig:each_other}. The results are presented separately for star-forming (SF) and non-SF regions. In both cases, the optical slope $A_B/A_V$ exhibits an inverse correlation with $A_V$. The NUV slope $A_{\tt w2}/A_{\tt w1}$ shows a similar but weaker anti-correlation with $A_V$ in non-SF regions, while this correlation is nearly absent in SF regions. Additionally, the optical and NUV slopes display a slight positive correlation in non-SF regions but appear uncorrelated in SF regions. Overall, these findings are consistent with those reported in Paper I (see their Fig. 7), where similar trends were observed across all regions collectively. Beyond the results of Paper I, we also find that non-SF regions generally exhibit steeper NUV slopes than SF regions, even when $A_V$ and $A_B/A_V$ are restricted to narrow ranges.

\begin{figure*}[ht!]
  \centering
   \includegraphics[width=0.3\textwidth]{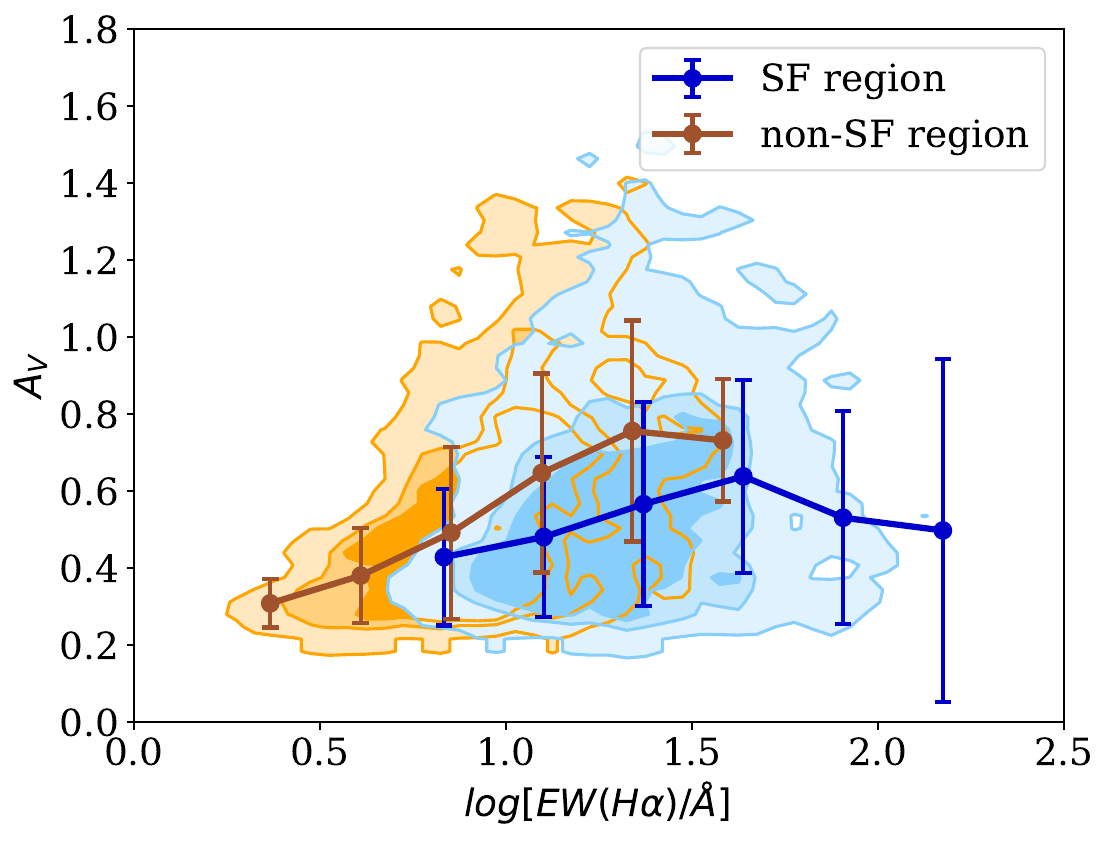}
   \includegraphics[width=0.3\textwidth]{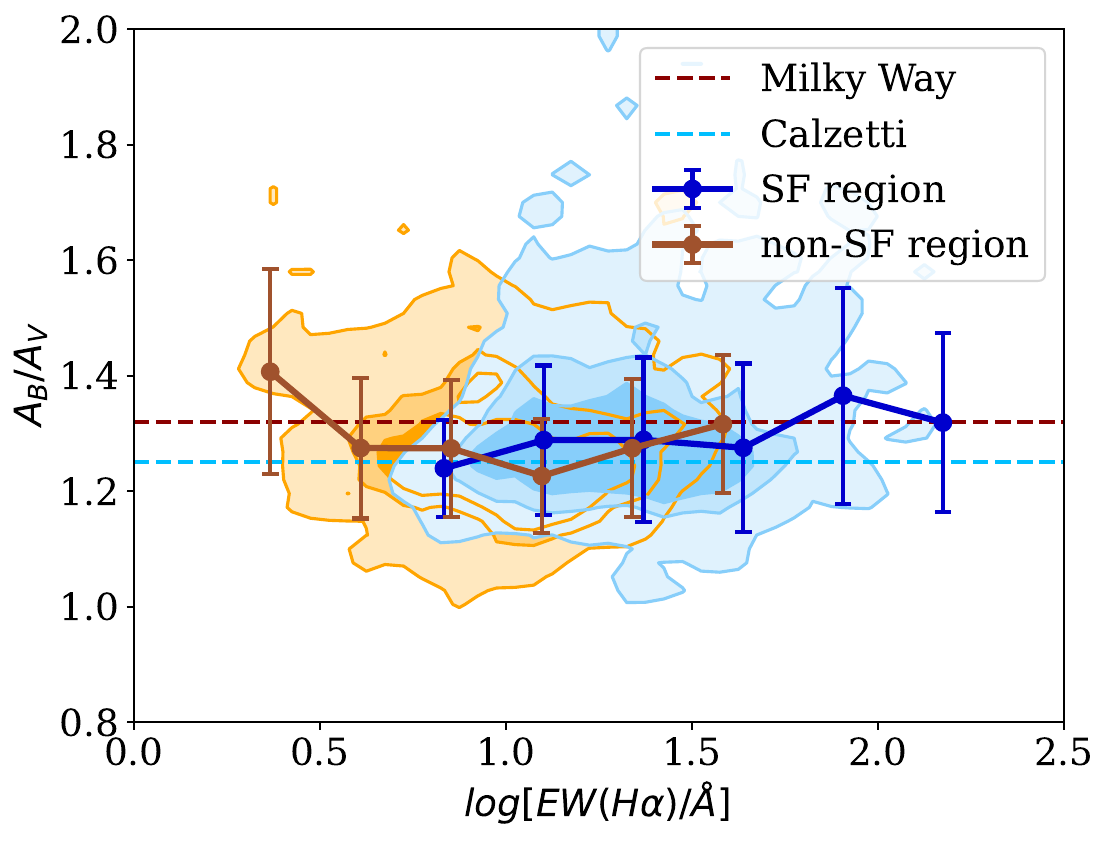}
   \includegraphics[width=0.3\textwidth]{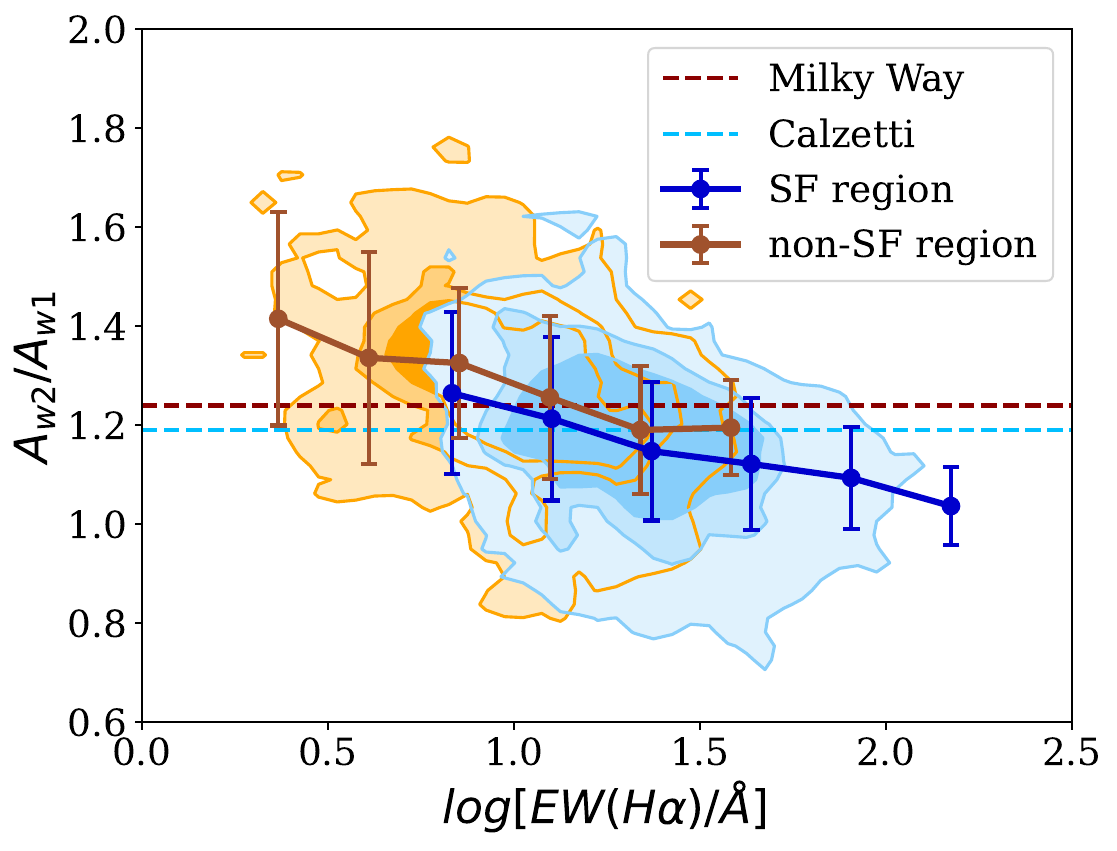}
   \includegraphics[width=0.3\textwidth]{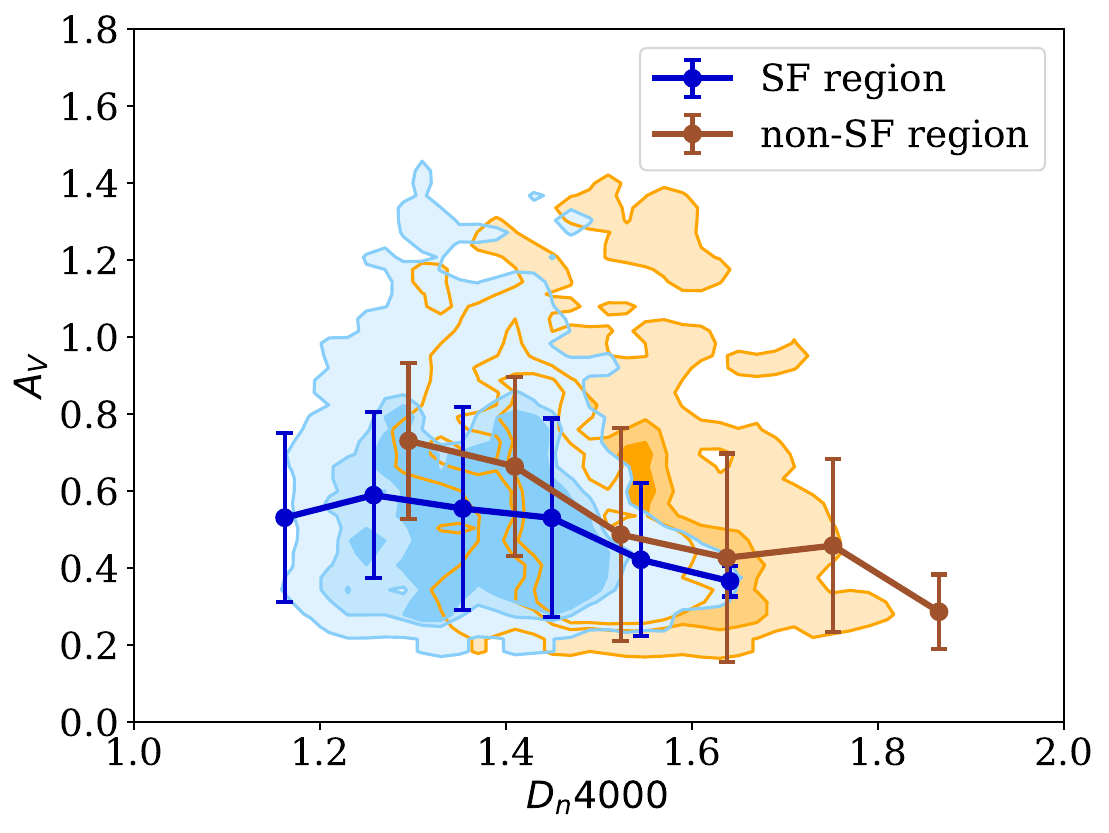}
   \includegraphics[width=0.3\textwidth]{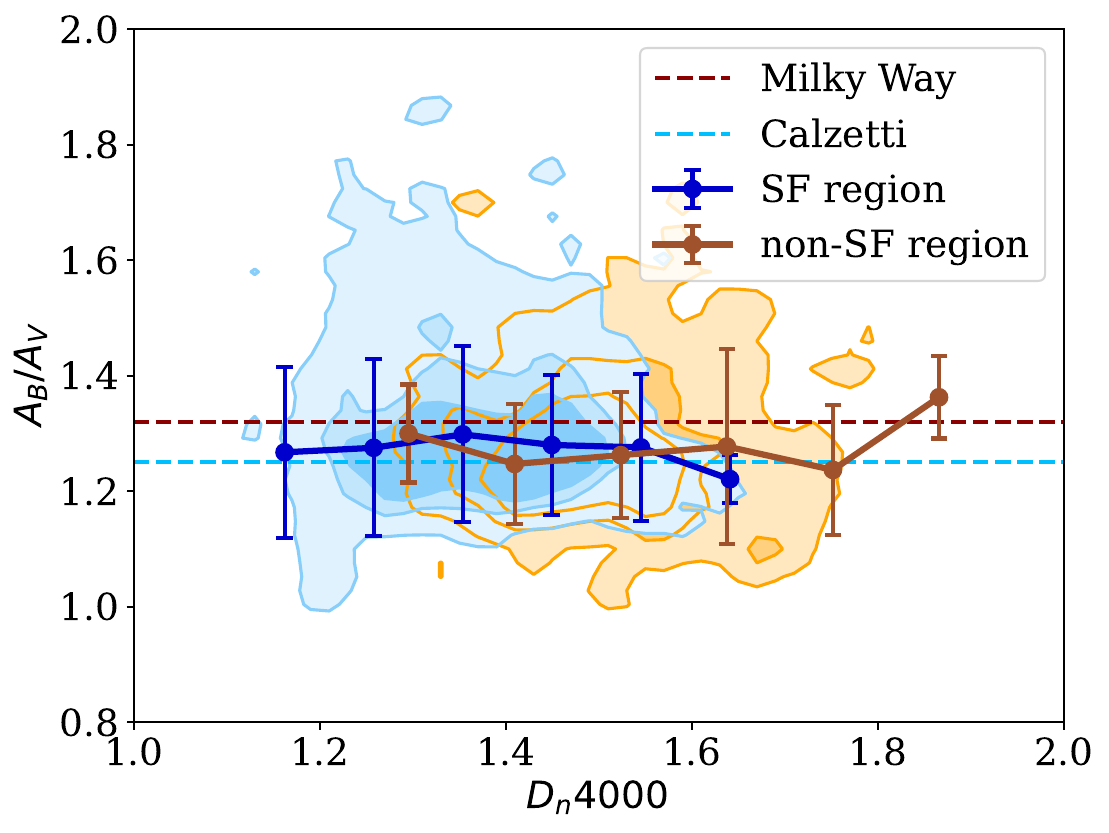}
   \includegraphics[width=0.3\textwidth]{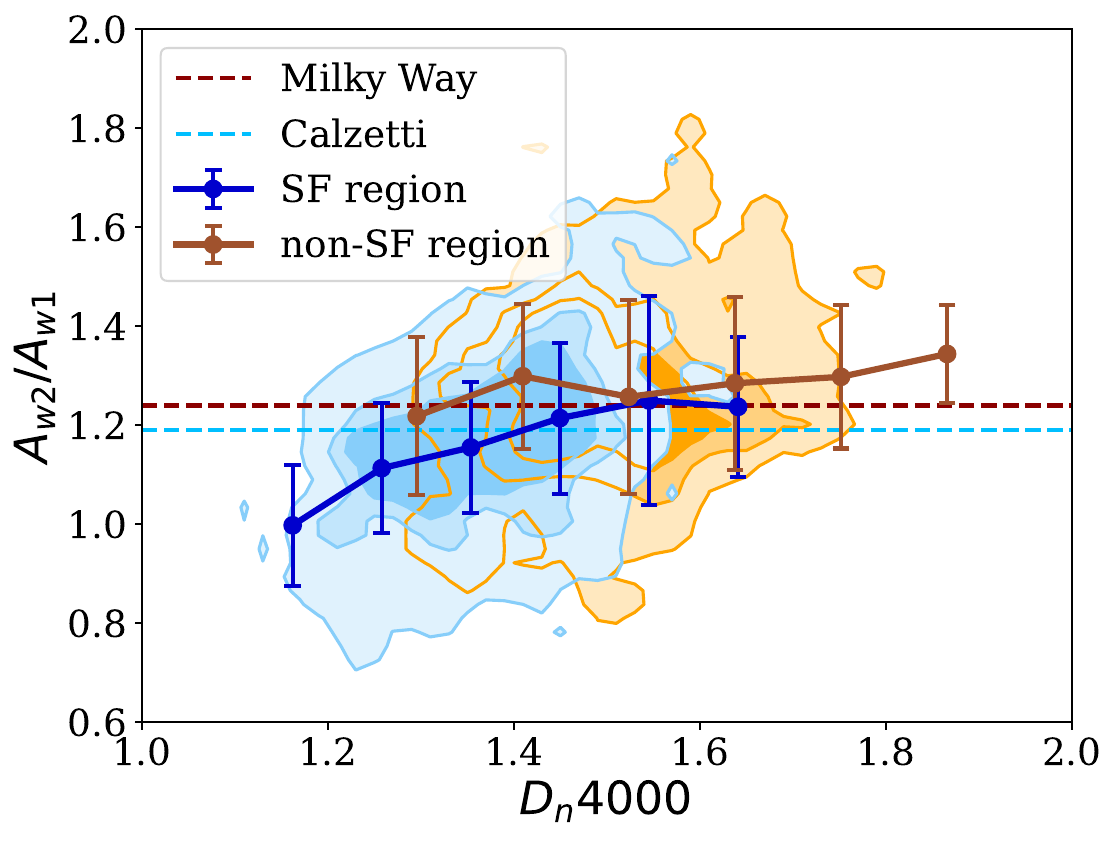}
   \includegraphics[width=0.3\textwidth]{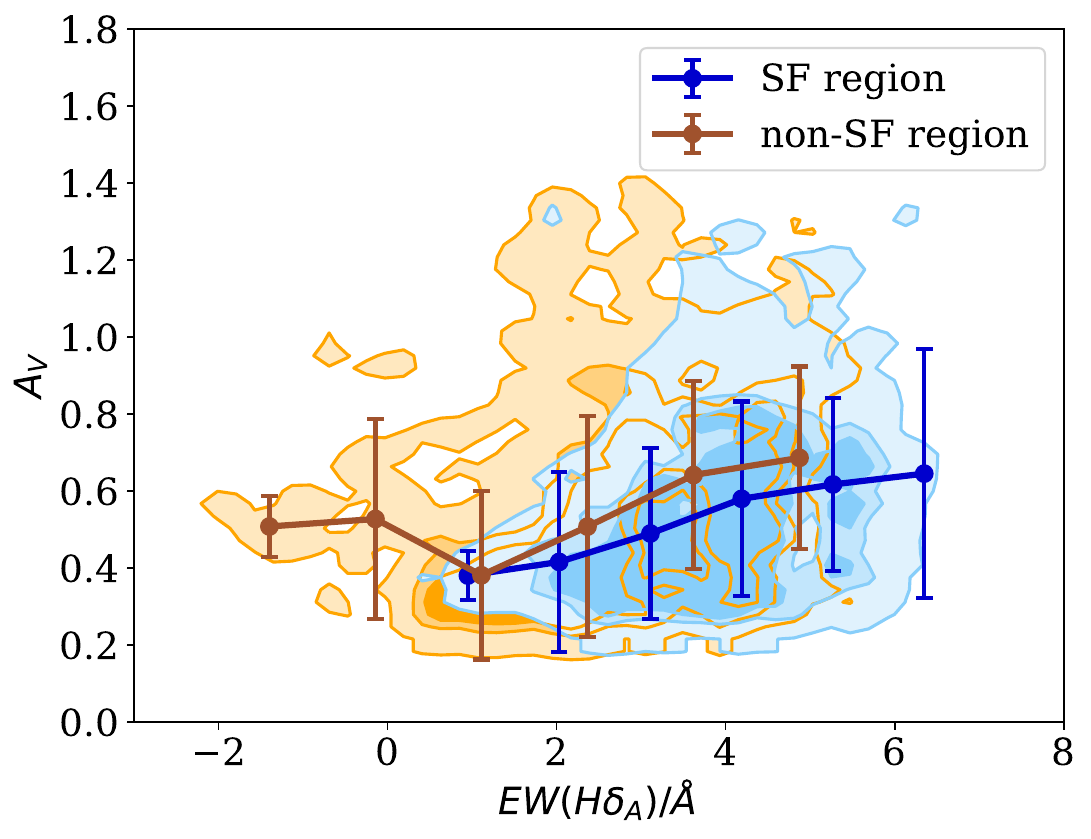}
   \includegraphics[width=0.3\textwidth]{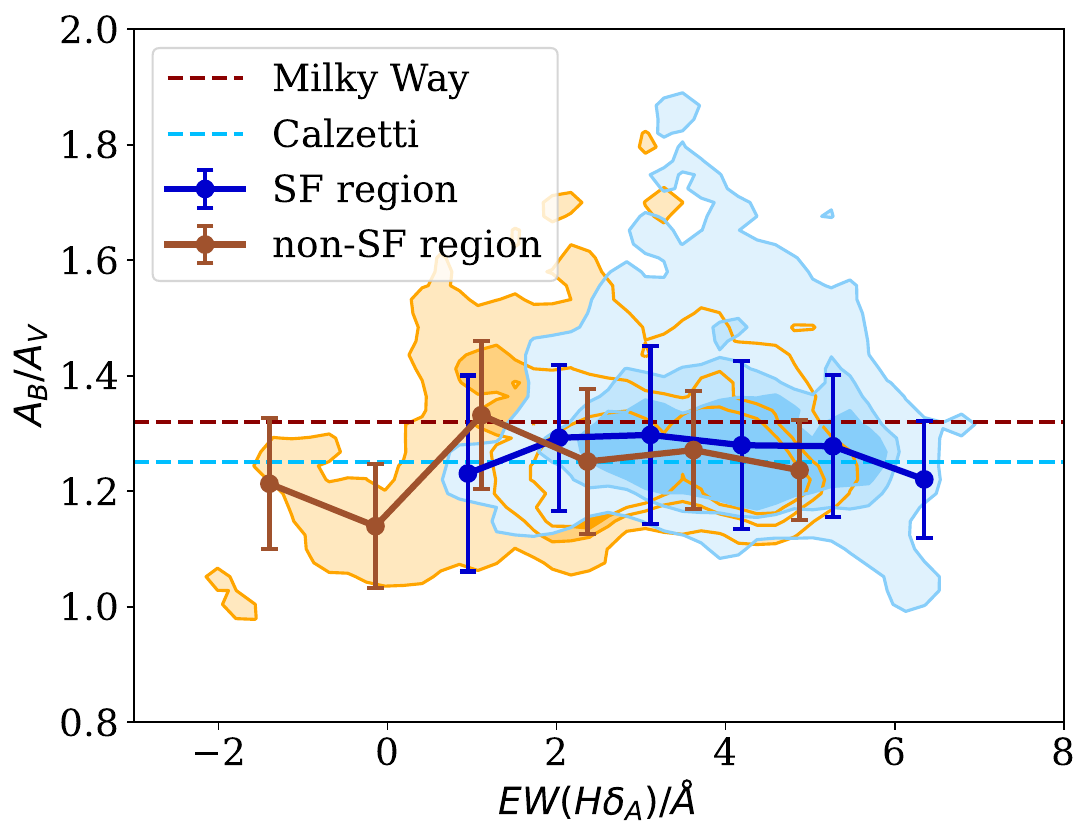}
   \includegraphics[width=0.3\textwidth]{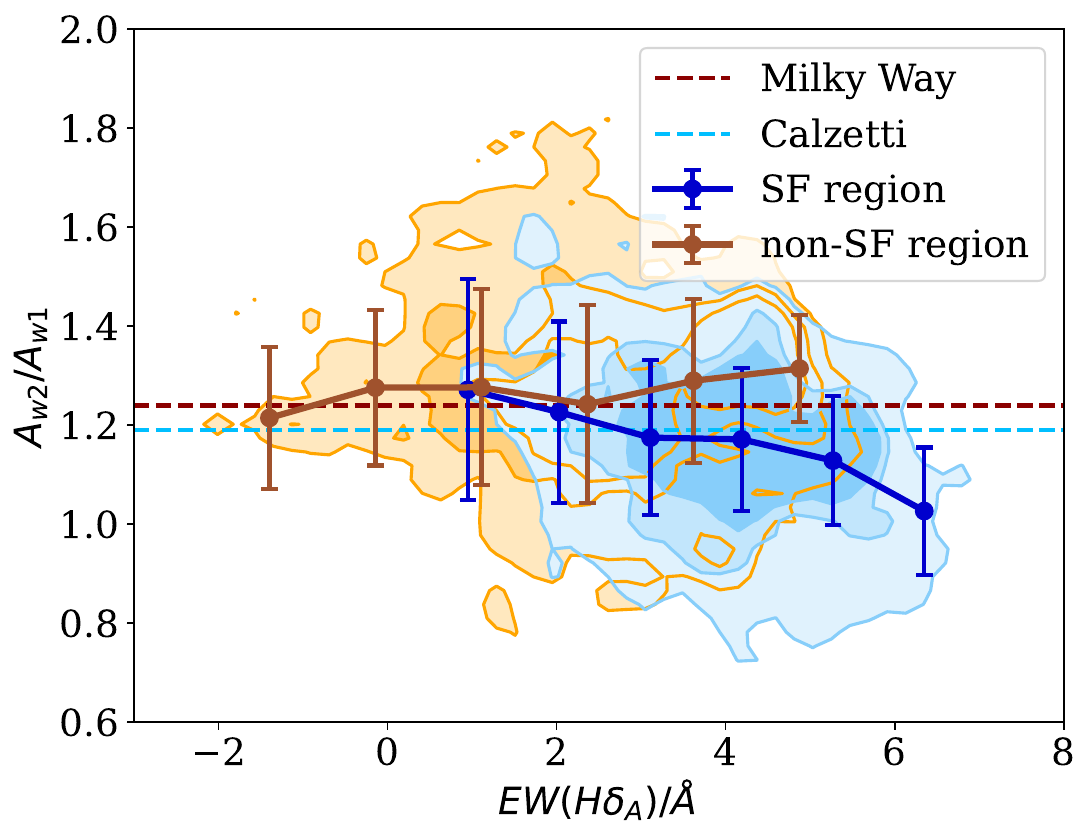}
  \caption{The three attenuation properties—$A_V$ (left panels), $A_B/A_V$ (middle panels), and $A_{\tt w2}/A_{\tt w1}$ (right panels)—are shown as functions of the logarithm of EW(H$\alpha$) (top row), $D_n4000$ (middle row), and EW(H$\delta_{\text{A}}$) (bottom row). The results are presented separately for SF regions and non-SF regions, with the same symbols/lines/colors as in \autoref{fig:each_other}.}
  \label{fig:SFH}
\end{figure*}

\begin{figure*}[ht!]
  \centering
   \includegraphics[width=0.3\textwidth]{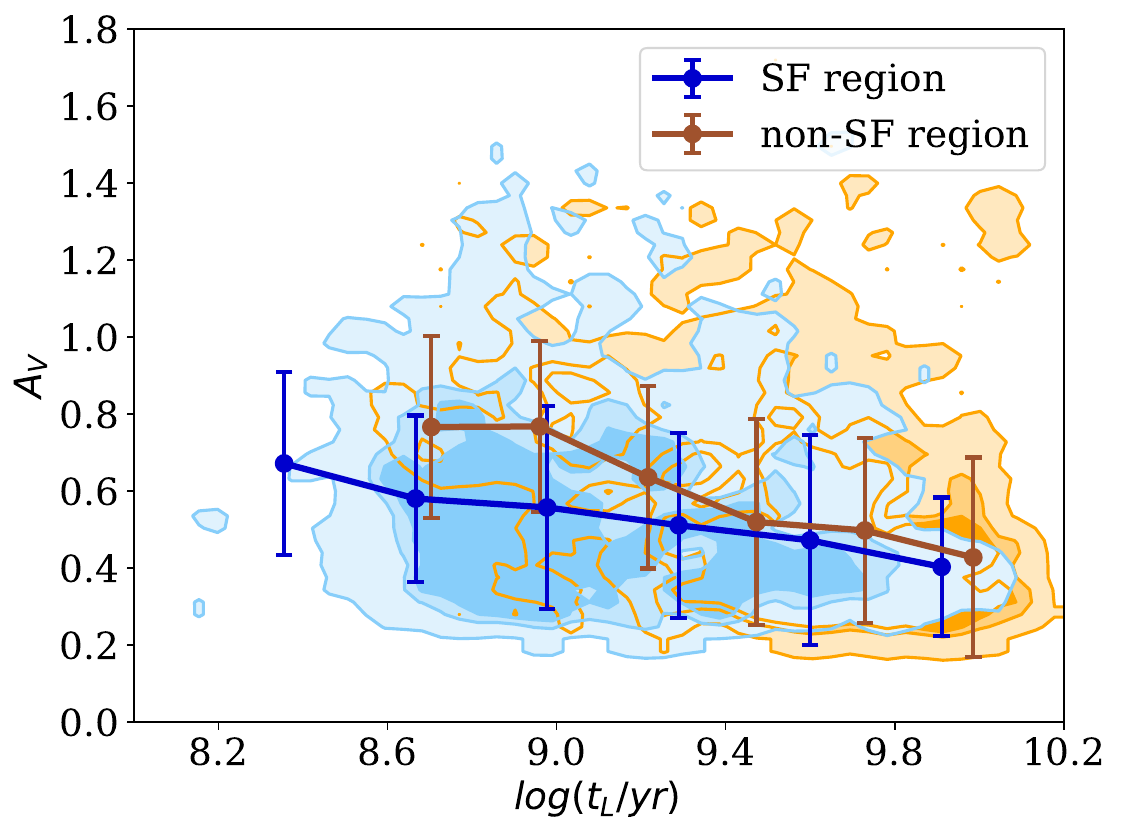}
   \includegraphics[width=0.3\textwidth]{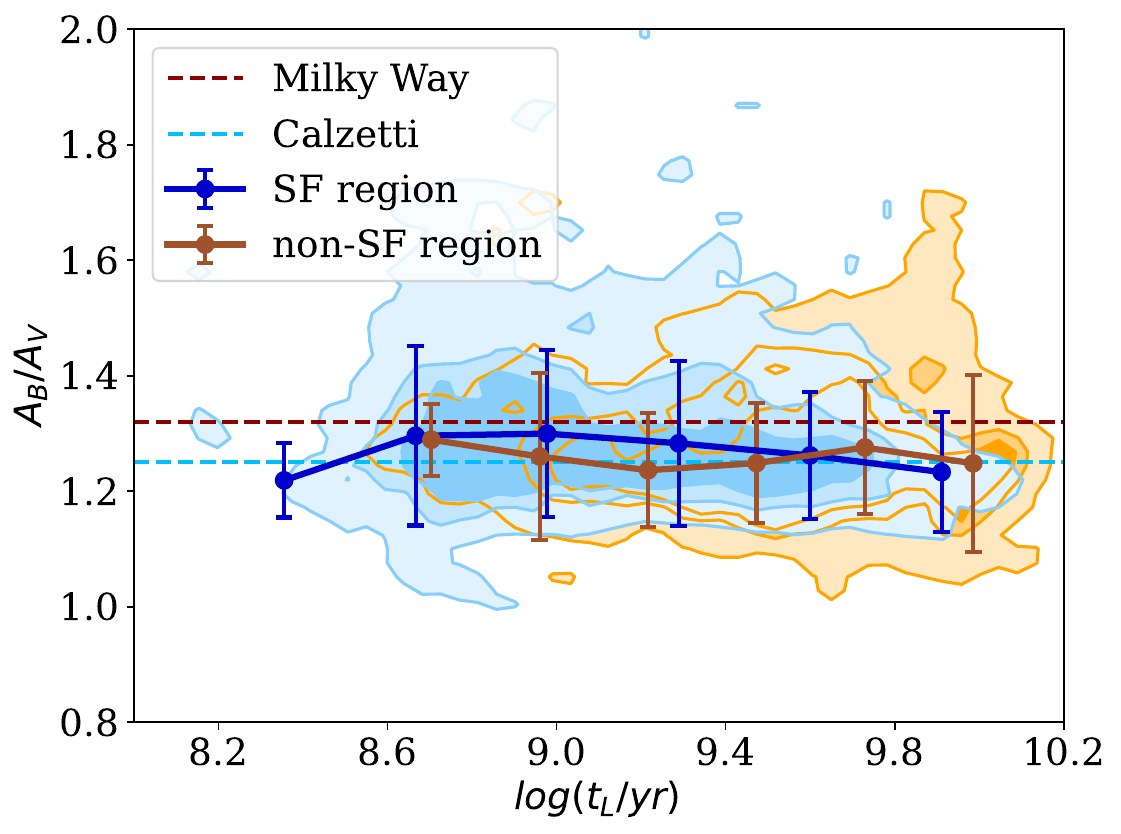}
   \includegraphics[width=0.3\textwidth]{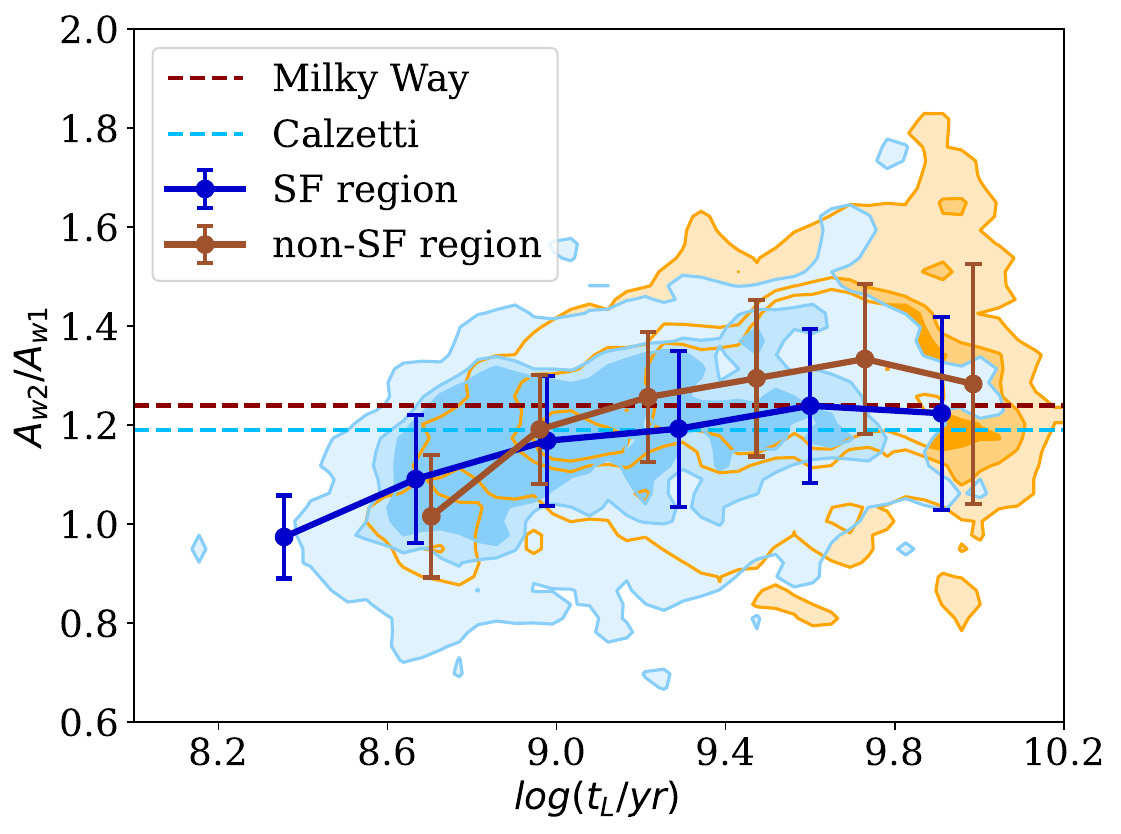}
   \includegraphics[width=0.3\textwidth]{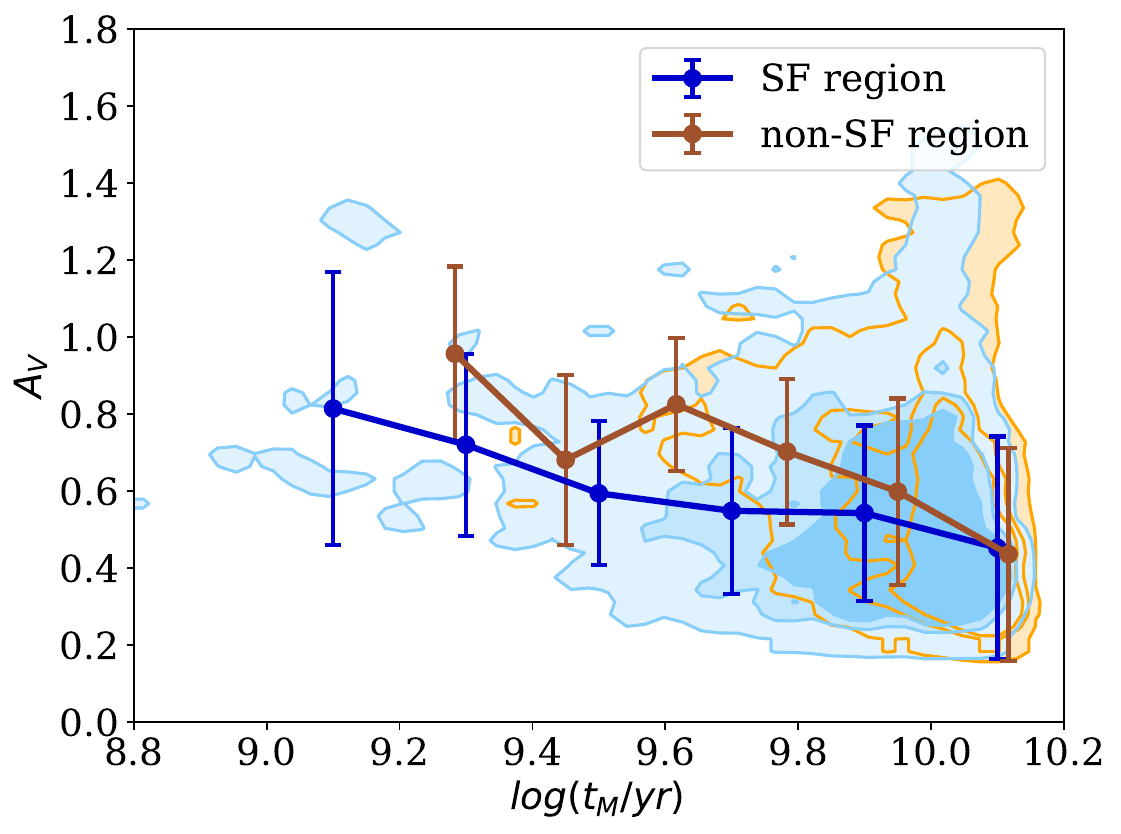}
   \includegraphics[width=0.3\textwidth]{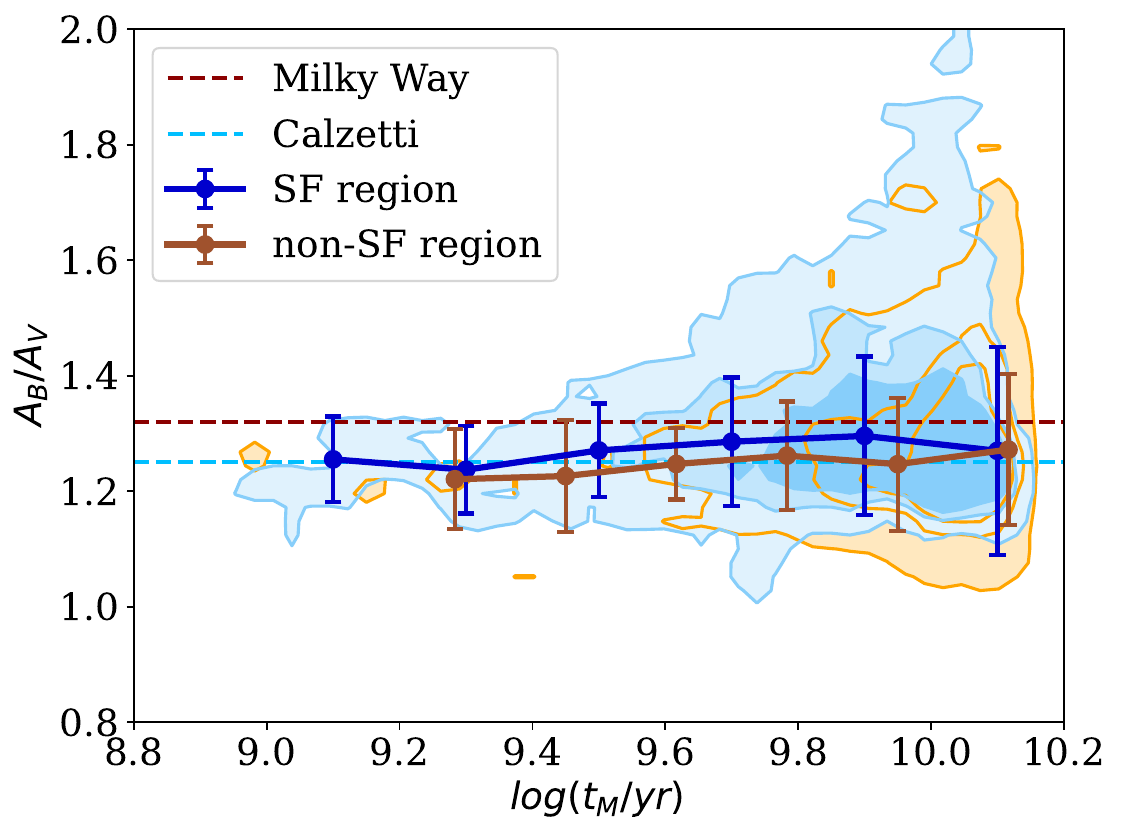}
   \includegraphics[width=0.3\textwidth]{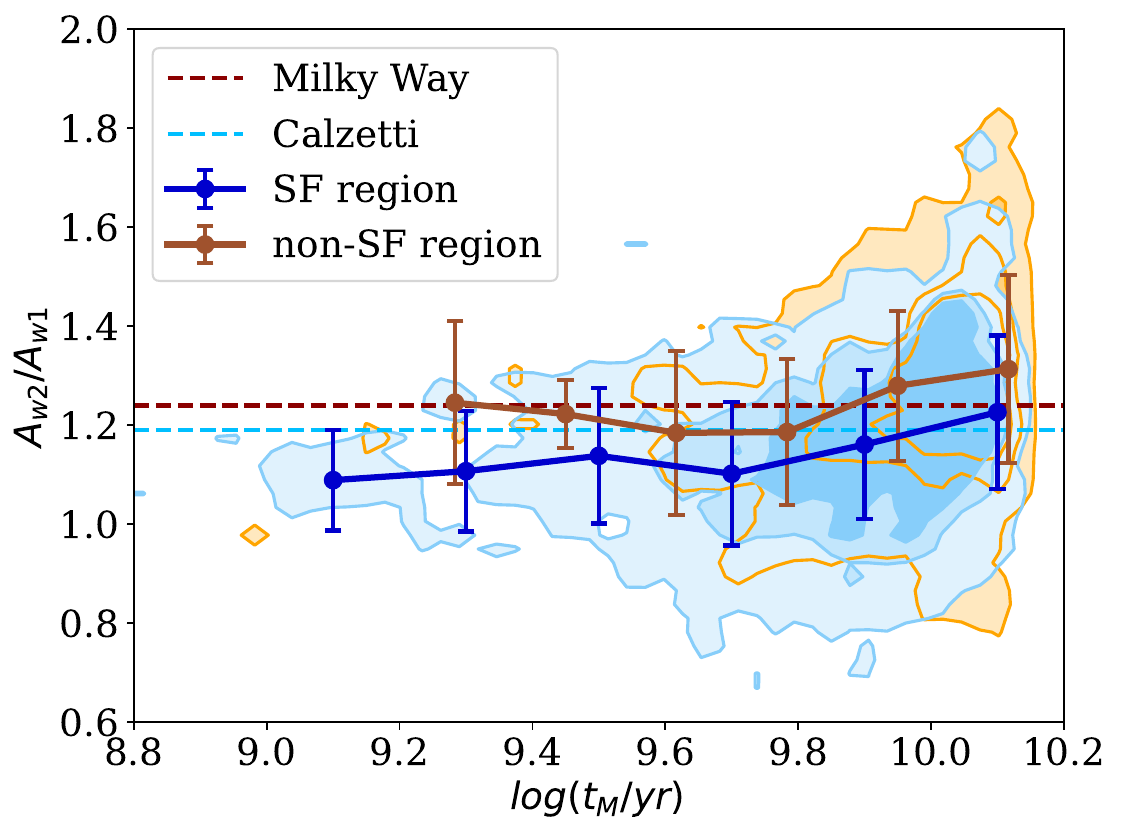}
  \caption{The three attenuation properties—$A_V$ (left panels), $A_B/A_V$ (middle panels), and $A_{\tt w2}/A_{\tt w1}$ (right panels)—are shown as functions of the logarithm of luminosity-weighted stellar age (top panels) and mass-weighted stellar age (bottom panels). The results are presented separately for SF regions and non-SF regions, with the same symbols/lines/colors as in \autoref{fig:each_other}.}
  \label{fig:stellar_age}
\end{figure*}

\subsection{Dependence on sSFR and recent star formation history}

In the top panels of \autoref{fig:dep_ssfr}, we examine the correlations between attenuation properties—characterized by optical opacity ($A_V$) and the slopes of the attenuation curves in the optical ($A_B/A_V$) and NUV ($A_{\tt w2}/A_{\tt w1}$)—and the specific surface brightness of the H$\alpha$ emission line, $\Sigma_{\text{H}\alpha}/\Sigma_\ast$. In SF regions, this parameter is essentially equivalent to the sSFR, as noted earlier. SF and non-SF regions are plotted in different colors. For clarity, we label the bottom x-axis with $\Sigma_{\text{H}\alpha}/\Sigma_\ast$ and the top x-axis with sSFR to facilitate direct comparison between these parameters. Consistent with Paper I, we find that in both SF and non-SF regions, $A_V$ increases with $\Sigma_{\text{H}\alpha}/\Sigma_\ast$, while the optical slope $A_B/A_V$ shows no clear correlation with this parameter. However, unlike Paper I, which found the NUV slope $A_{\tt w2}/A_{\tt w1}$ to be nearly independent of sSFR, we identify a significant anti-correlation between these two properties in both SF and non-SF regions. This discrepancy is likely due to several factors in this study, including the larger (and potentially different) sample, the SF/non-SF classification, and the slightly wider sSFR range.

In the middle and bottom panels of \autoref{fig:dep_ssfr}, we present the three attenuation curve parameters as functions of the H$\alpha$ surface brightness ($\Sigma_{\text{H}\alpha}$) and the surface density of stellar mass  ($\Sigma_\ast$). For the optical slope $A_B/A_V$, we find no clear dependence on either $\Sigma_{\text{H}\alpha}$ or $\Sigma_\ast$. In contrast, the optical opacity $A_V$ shows a positive correlation with $\Sigma_{\text{H}\alpha}$ but exhibits little to no correlation with $\Sigma_\ast$. This suggests that the positive correlation between $A_V$ and $\Sigma_{\text{H}\alpha}/\Sigma_\ast$ seen in the top-left panel is primarily driven by the relationship between $A_V$ and $\Sigma_{\text{H}\alpha}$. For the NUV slope $A_{\tt w2}/A_{\tt w1}$, we find no or weak dependence on $\Sigma_{\text{H}\alpha}$ and a positive correlation with $\Sigma_\ast$. Therefore, the anti-correlation between $A_{\tt w2}/A_{\tt w1}$ and $\Sigma_{\text{H}\alpha}/\Sigma_\ast$ as observed in the top-right panel is primarily driven by the correlation of $A_{\tt w2}/A_{\tt w1}$ with $\Sigma_\ast$, rather than with $\Sigma_{\text{H}\alpha}$. This result indicates that the steeper NUV slopes in non-SF regions compared to SF regions, as seen in \autoref{fig:each_other}, are primarily attributed to the relatively high surface stellar mass densities in non-SF regions. 

One might question whether the observed anti-correlation between $A_{\tt w2}/A_{\tt w1}$ and $\Sigma_{\text{H}\alpha}/\Sigma_\ast$ is an independent relationship or merely a consequence of the correlations between $A_{\tt w2}/A_{\tt w1}$ and $A_V$ (middle panel of \autoref{fig:each_other}) and between $A_V$ and $\Sigma_{\text{H}\alpha}/\Sigma_\ast$ (top-left panel of \autoref{fig:dep_ssfr}). To investigate this possibility, we divide both SF and non-SF regions into subsamples based on $A_V$ and examine how $A_{\tt w2}/A_{\tt w1}$ correlates with $\Sigma_{\text{H}\alpha}/\Sigma_\ast$, $\Sigma_{\text{H}\alpha}$, and $\Sigma_\ast$ within these subsamples. The results, shown in \autoref{fig:sSFR_slope_with_Av}, reveal that all previously identified correlations in the full samples of SF and non-SF regions remain unchanged even when $A_V$ is restricted to narrow ranges. This finding confirms that the anti-correlation between $A_{\tt w2}/A_{\tt w1}$ and $\Sigma_{\text{H}\alpha}/\Sigma_\ast$ is intrinsic and not driven by the other two correlations.

In \autoref{fig:SFH}, we further investigate the dependence of attenuation curve properties on recent star formation history (SFH), using three observational diagnostics: EW(H$\alpha$), $D_n4000$, and EW(H$\delta_{\text{A}}$). The first diagnostic, EW(H$\alpha$), represents the equivalent width of the H$\alpha$ emission line. This parameter is widely used as a proxy for the specific star formation rate (sSFR), effectively measuring the strength of ongoing star formation. As expected, EW(H$\alpha$) exhibits correlations with dust attenuation properties similar to those observed in the top panels of \autoref{fig:dep_ssfr}. The second parameter, $D_n4000$, quantifies the strength of the continuum break around 4000\AA. It is particularly sensitive to stellar populations formed within the past 1–2 Gyr. The third parameter, EW(H$\delta_{\text{A}}$), measures the equivalent width of the H$\delta$ absorption line, serving as an indicator of massive stars formed over the past few Myr. Together, these three diagnostic parameters provide a broad picture of the recent SFH in a given region, covering timescales from the present back to approximately 1–2 Gyr ago. As shown in the figure, both SF and non-SF regions exhibit similar correlations between $A_V$ and $A_B/A_V$ and the recent SFH diagnostics. However, the NUV slope, $A_{\tt w2}/A_{\tt w1}$, behaves differently between the two types of regions. In non-SF regions, it shows little to no dependence on $D_n4000$ and EW(H$\delta_{\text{A}}$), whereas in SF regions, a significant correlation is observed. Specifically, the NUV slope of attenuation curves becomes flatter as $D_n4000$ decreases or EW(H$\delta_{\text{A}}$) increases, both of which indicate the presence of recently formed young stellar populations.

\begin{figure*}[ht!]
  \centering
   \includegraphics[width=0.32\textwidth]{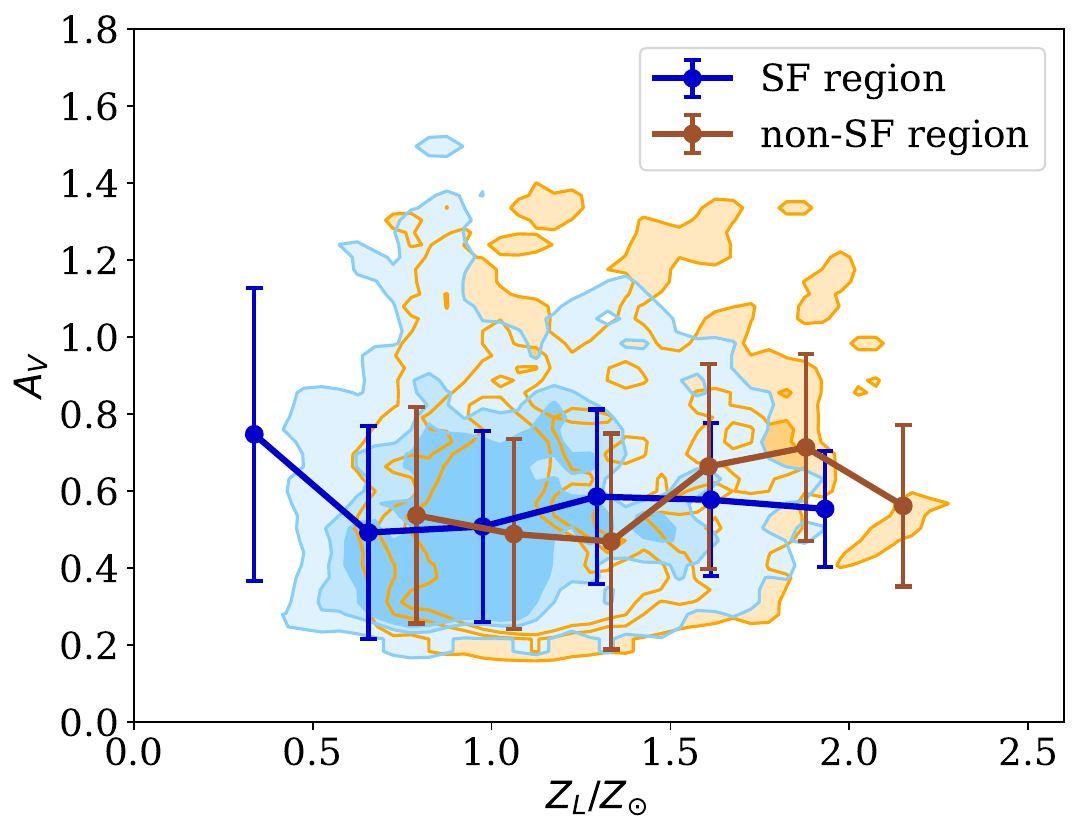}
   \includegraphics[width=0.32\textwidth]{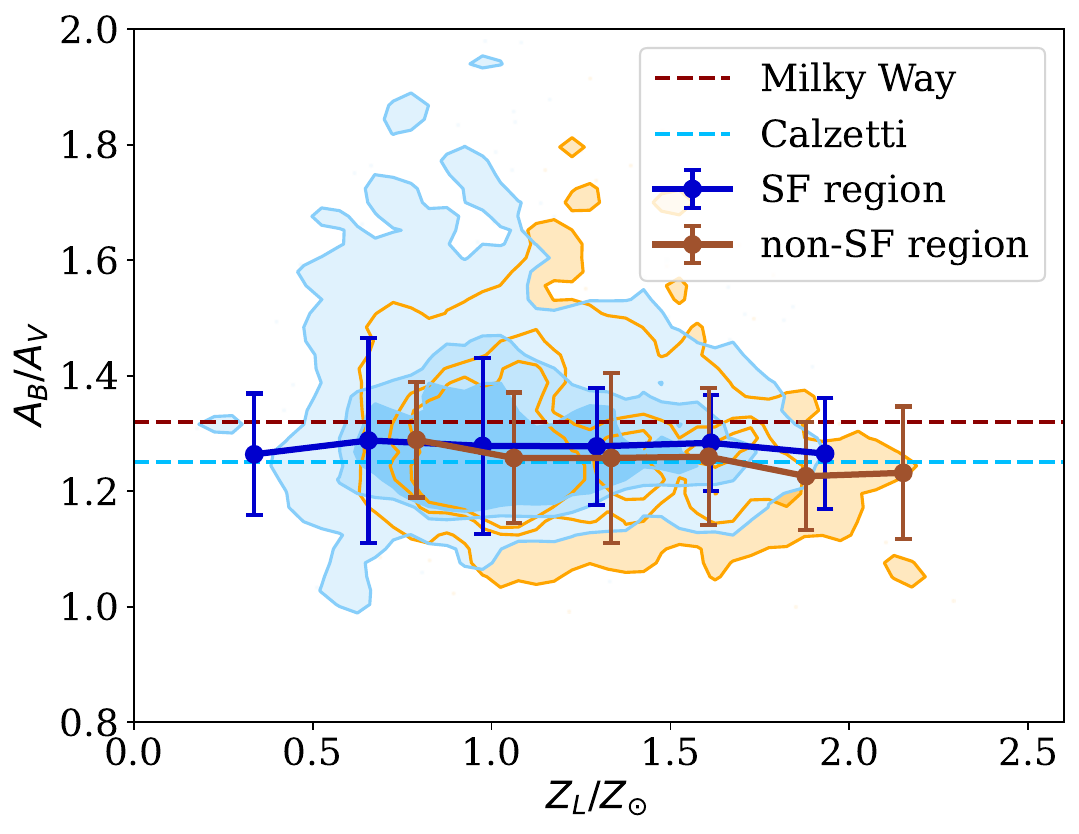}
   \includegraphics[width=0.32\textwidth]{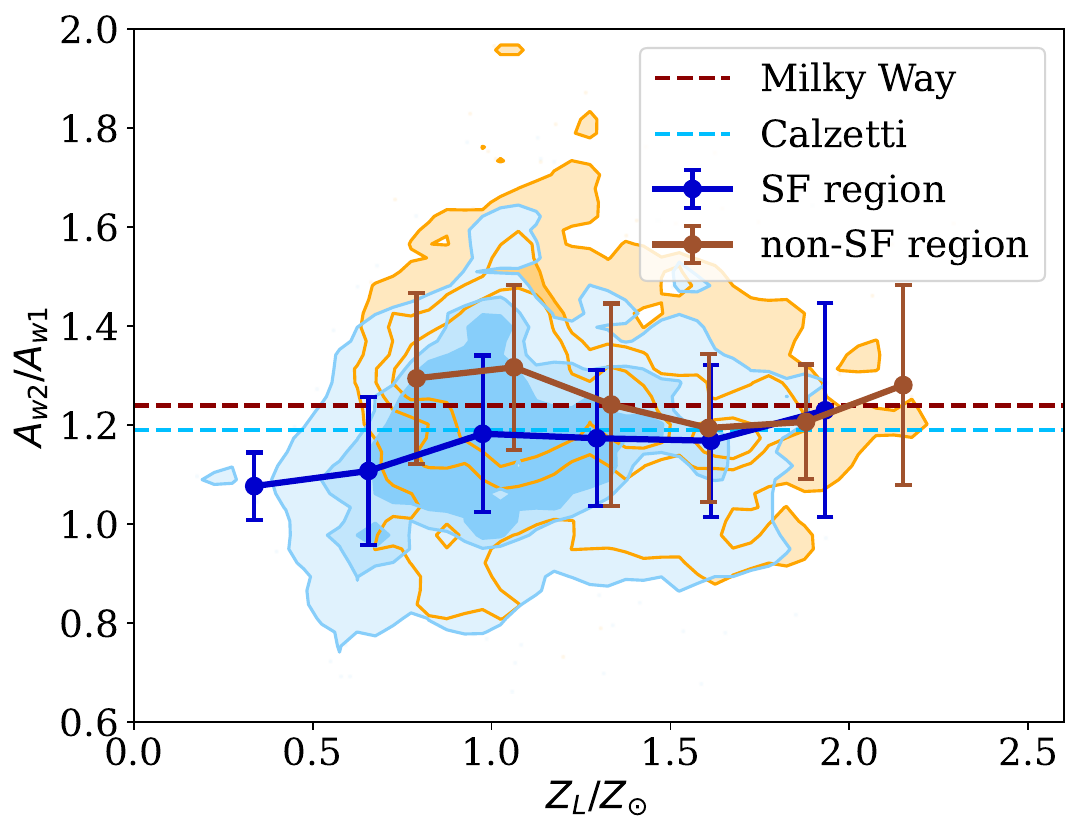}
   \includegraphics[width=0.32\textwidth]{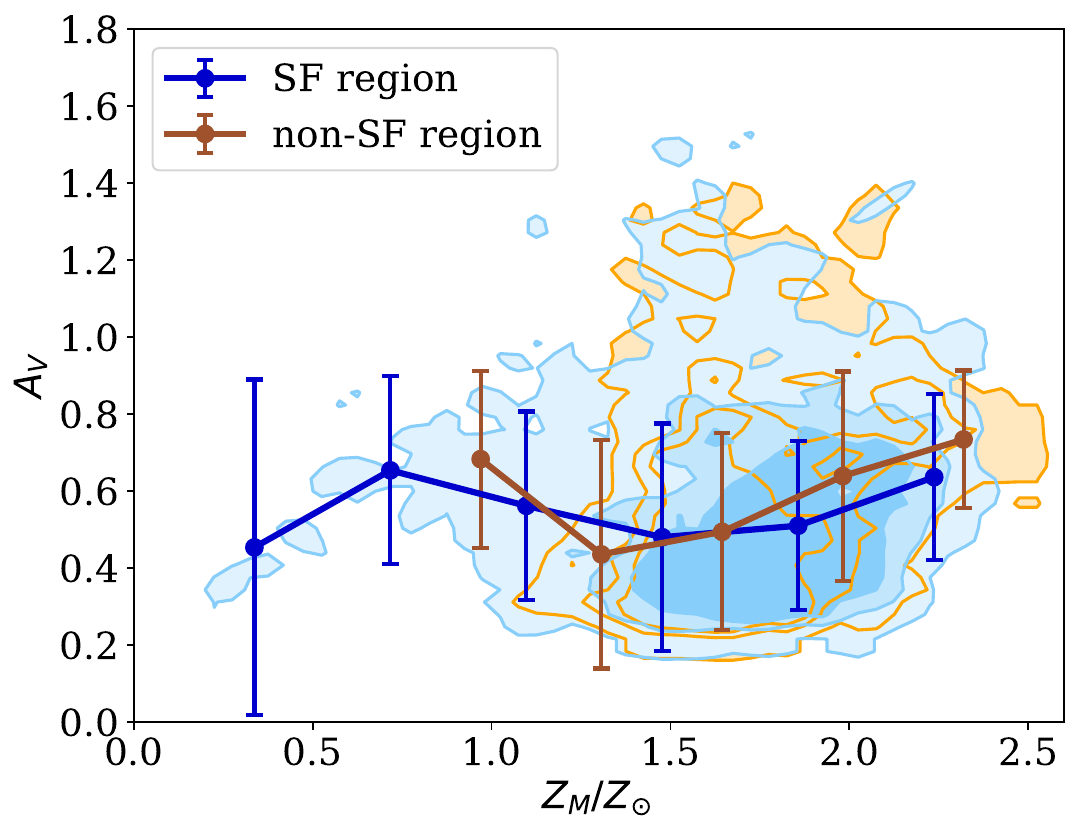}
   \includegraphics[width=0.32\textwidth]{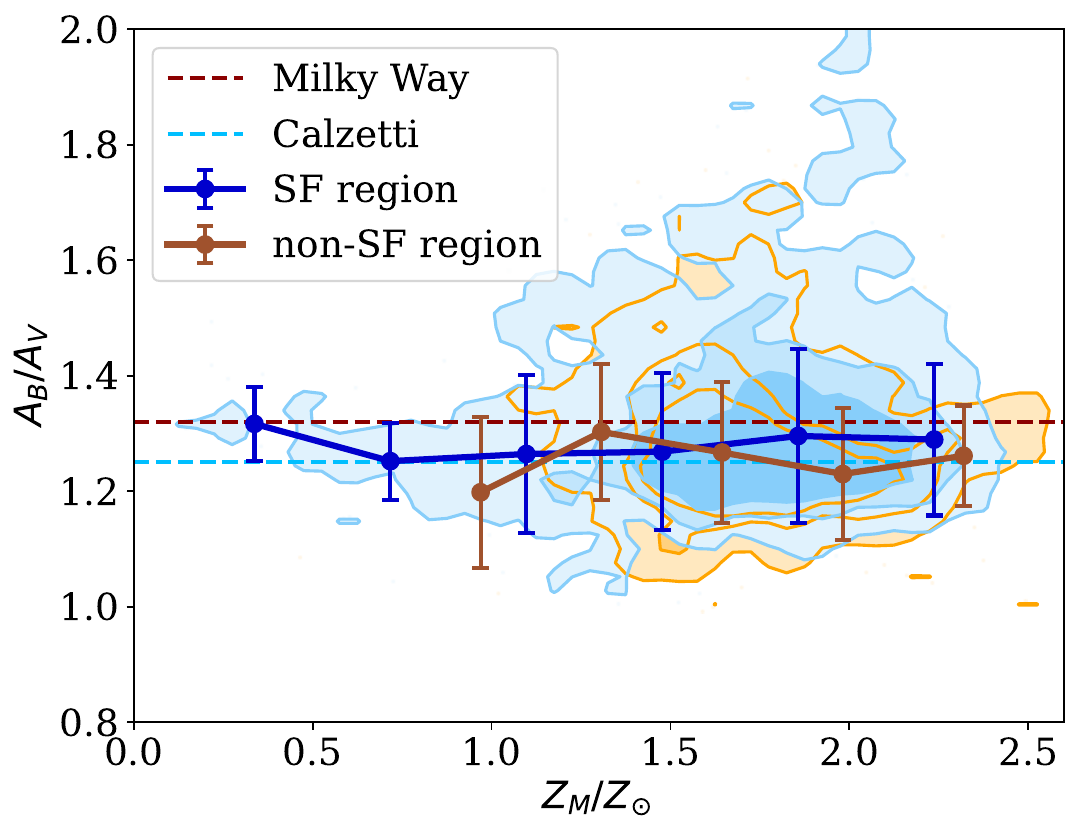}
   \includegraphics[width=0.32\textwidth]{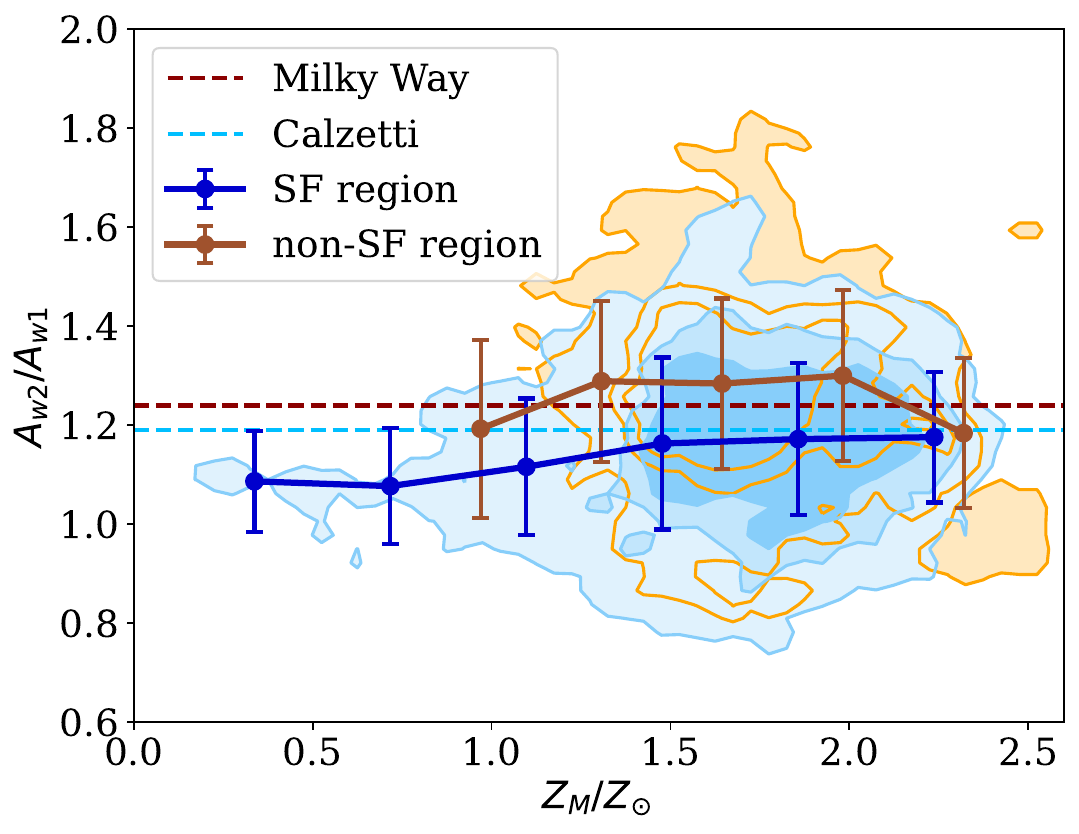}
   \includegraphics[width=0.32\textwidth]{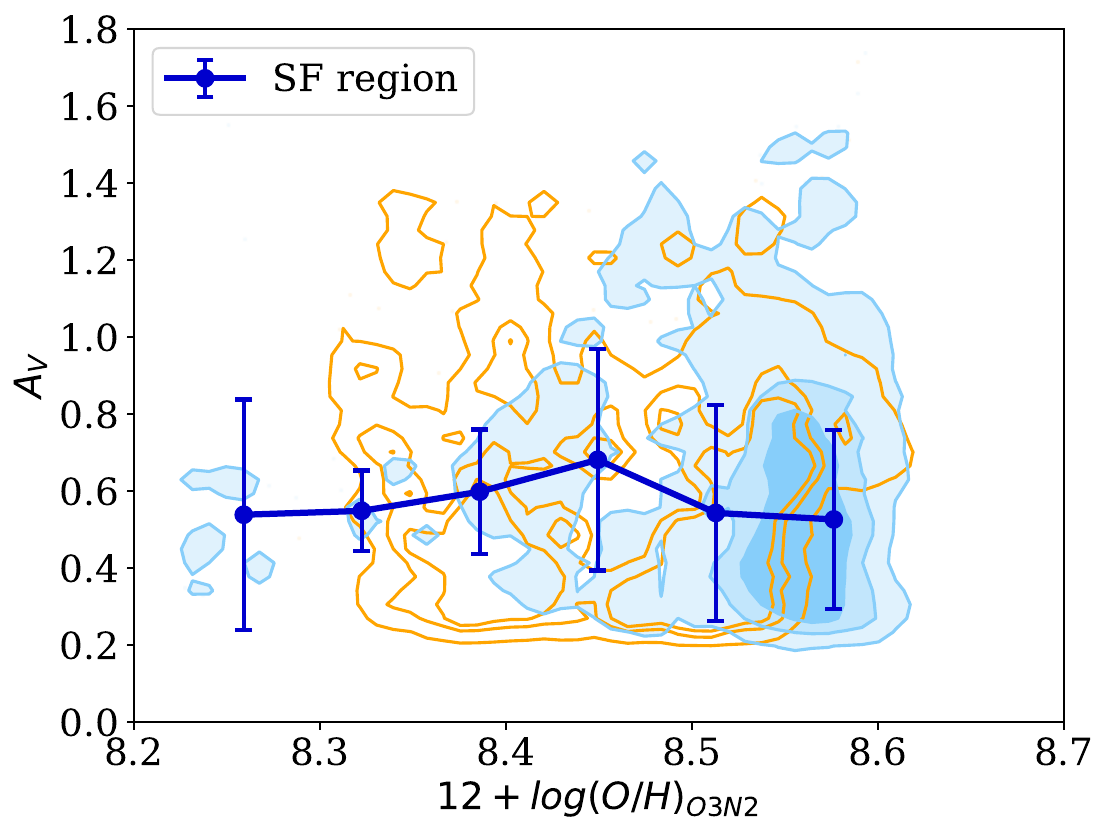}
   \includegraphics[width=0.32\textwidth]{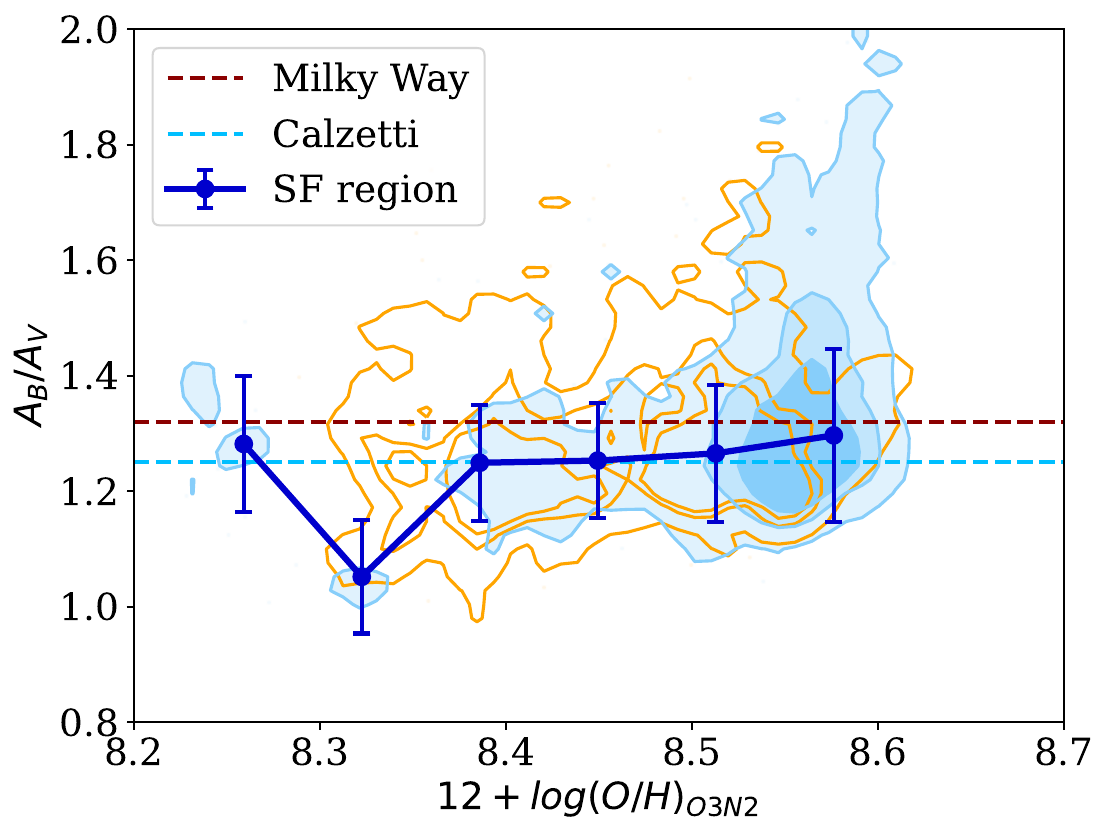}
   \includegraphics[width=0.32\textwidth]{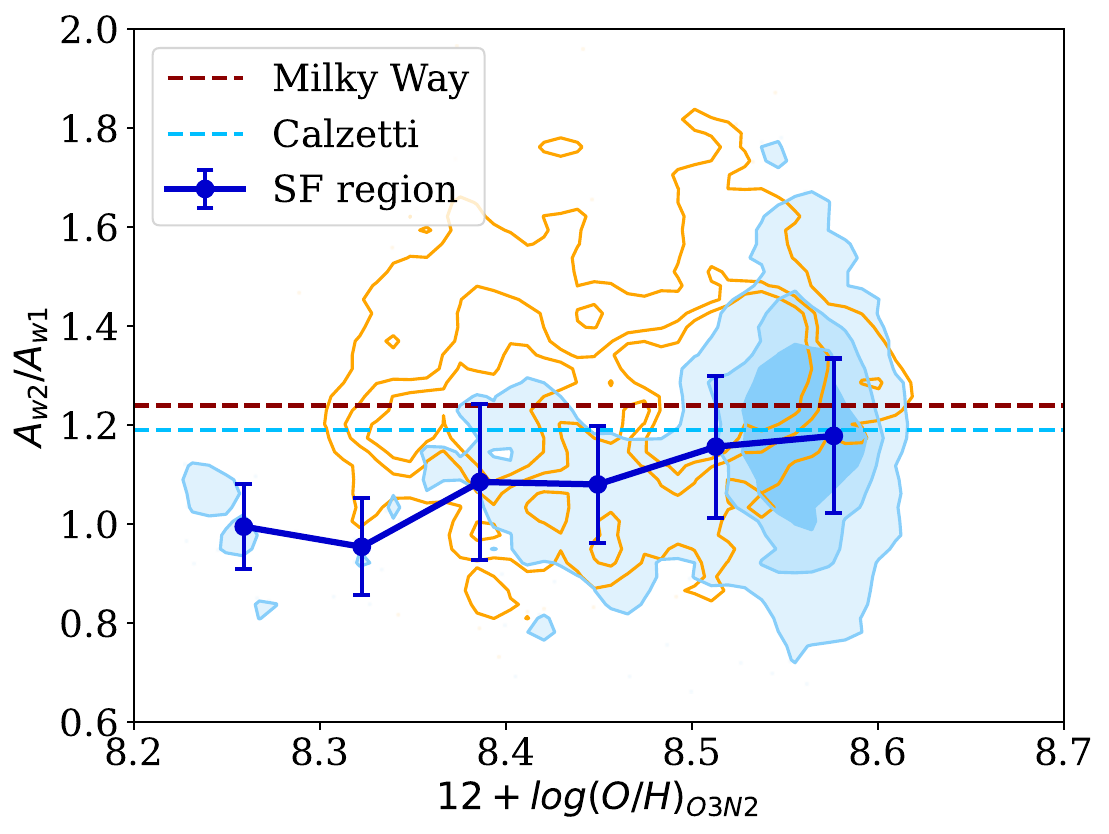}
   \includegraphics[width=0.32\textwidth]{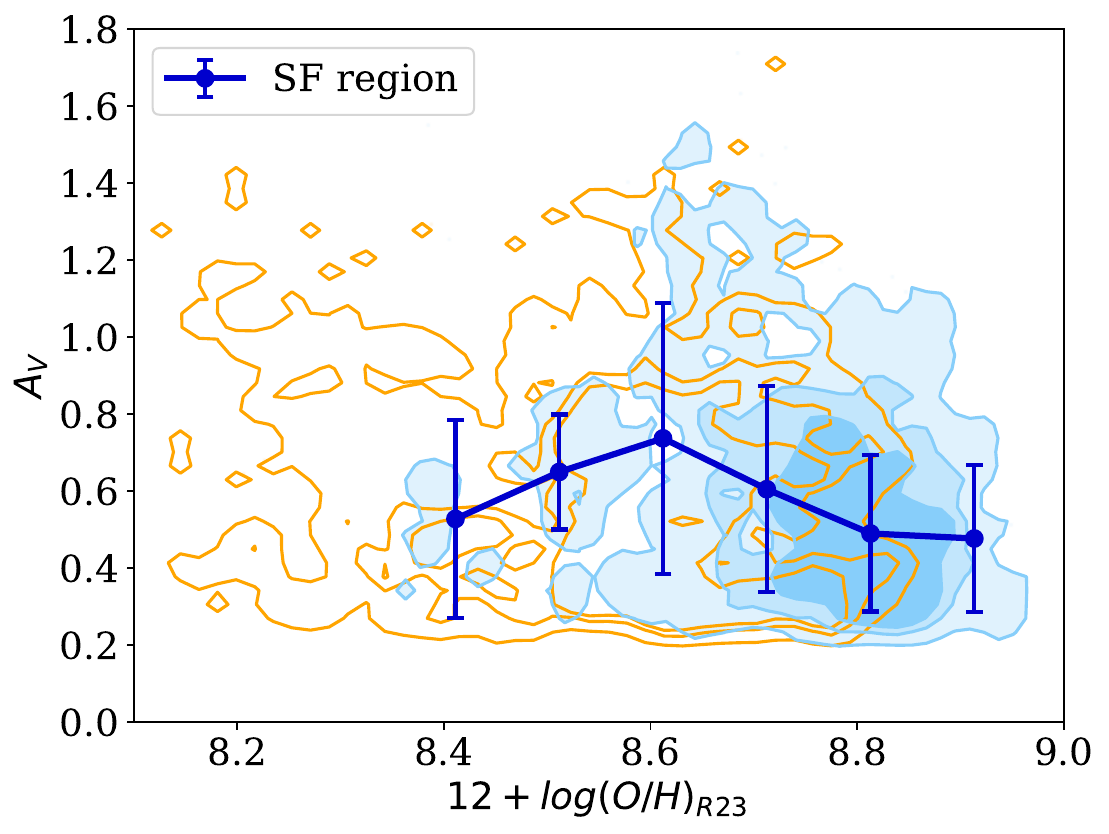}
   \includegraphics[width=0.32\textwidth]{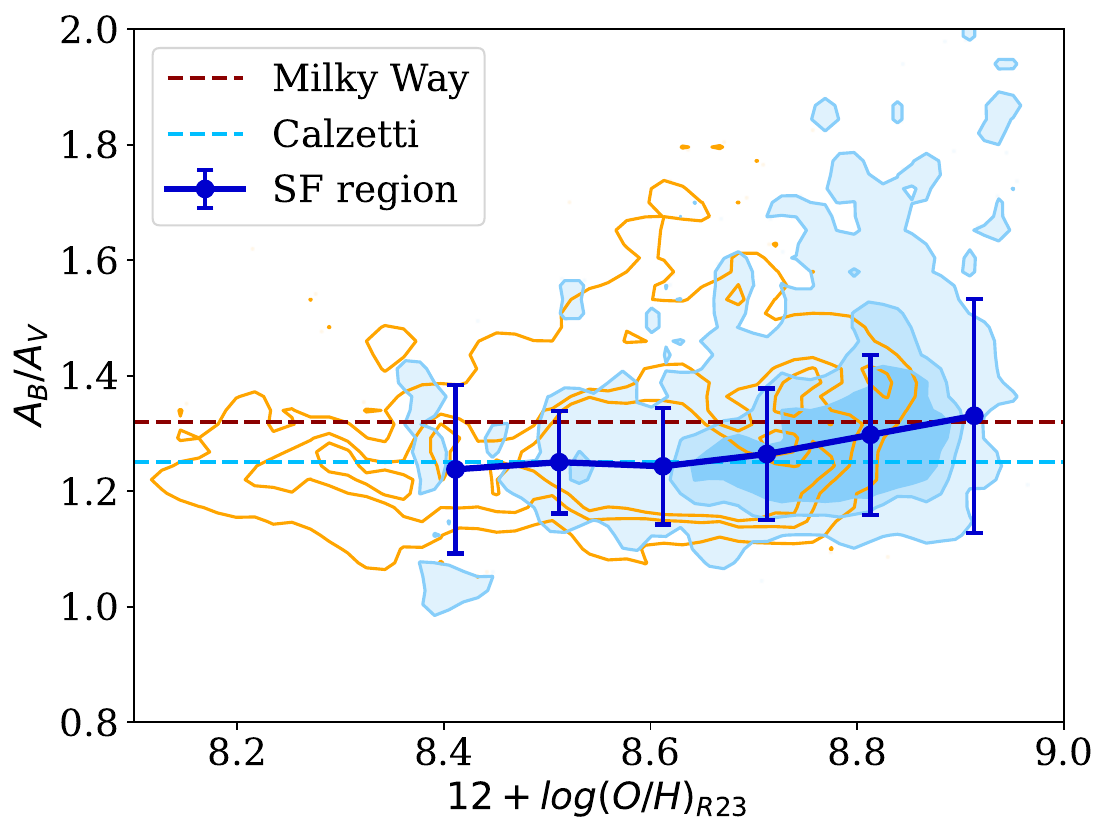}
   \includegraphics[width=0.32\textwidth]{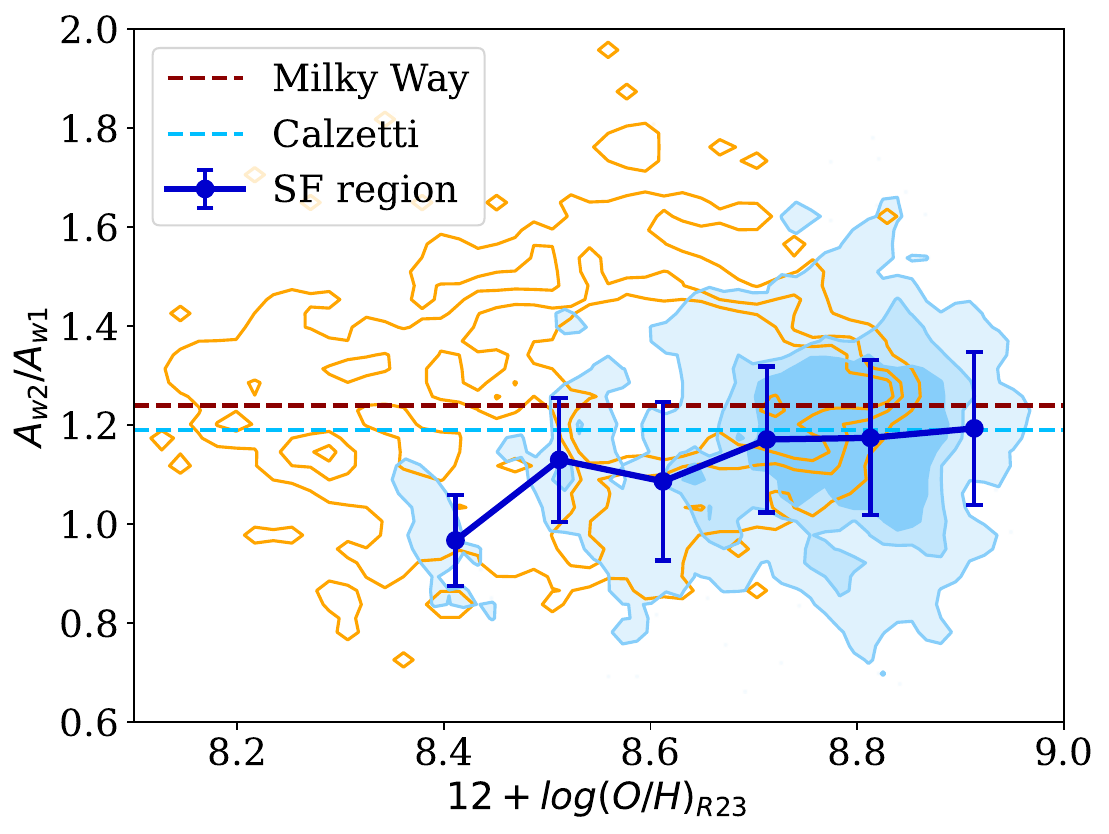}
   \includegraphics[width=0.32\textwidth]{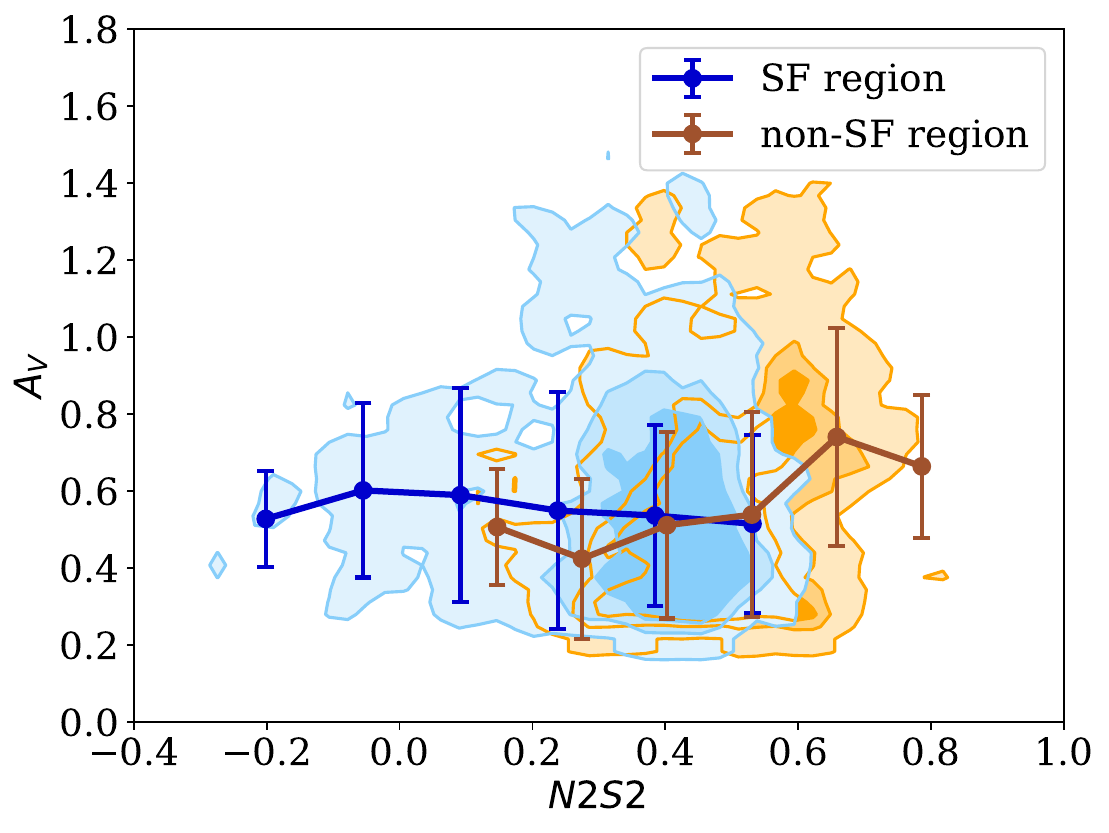}
   \includegraphics[width=0.32\textwidth]{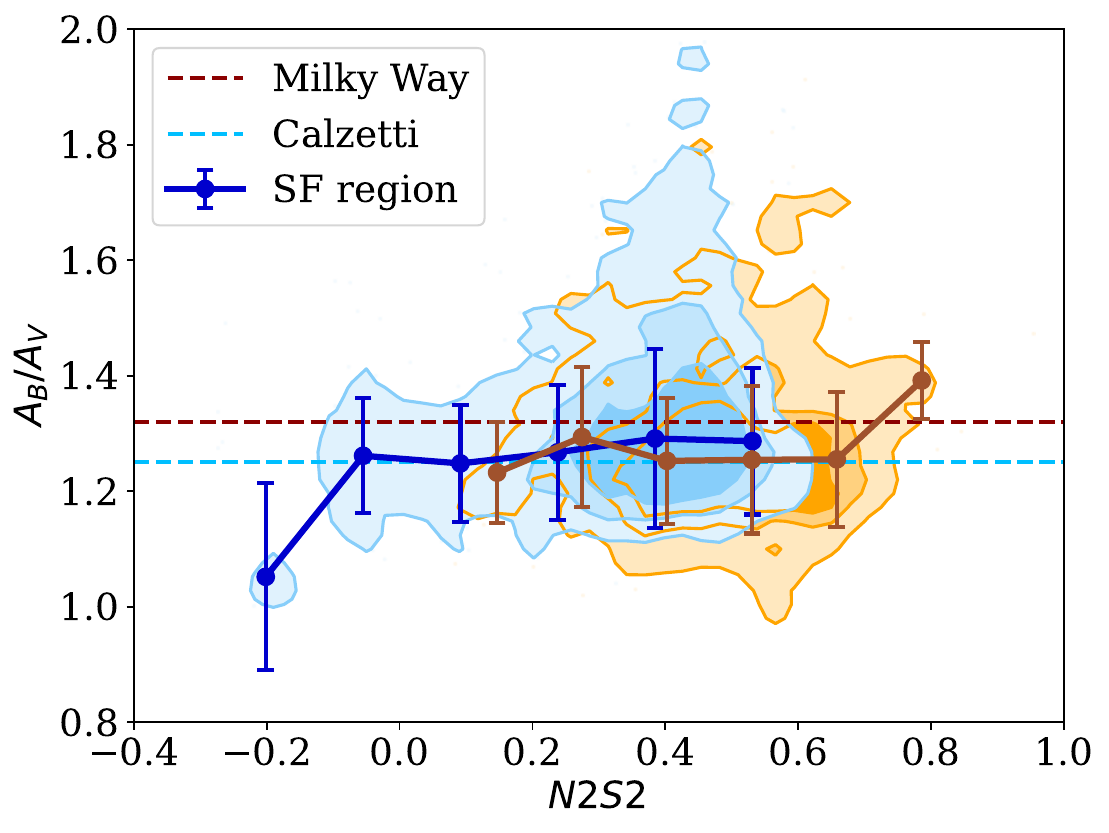}
   \includegraphics[width=0.32\textwidth]{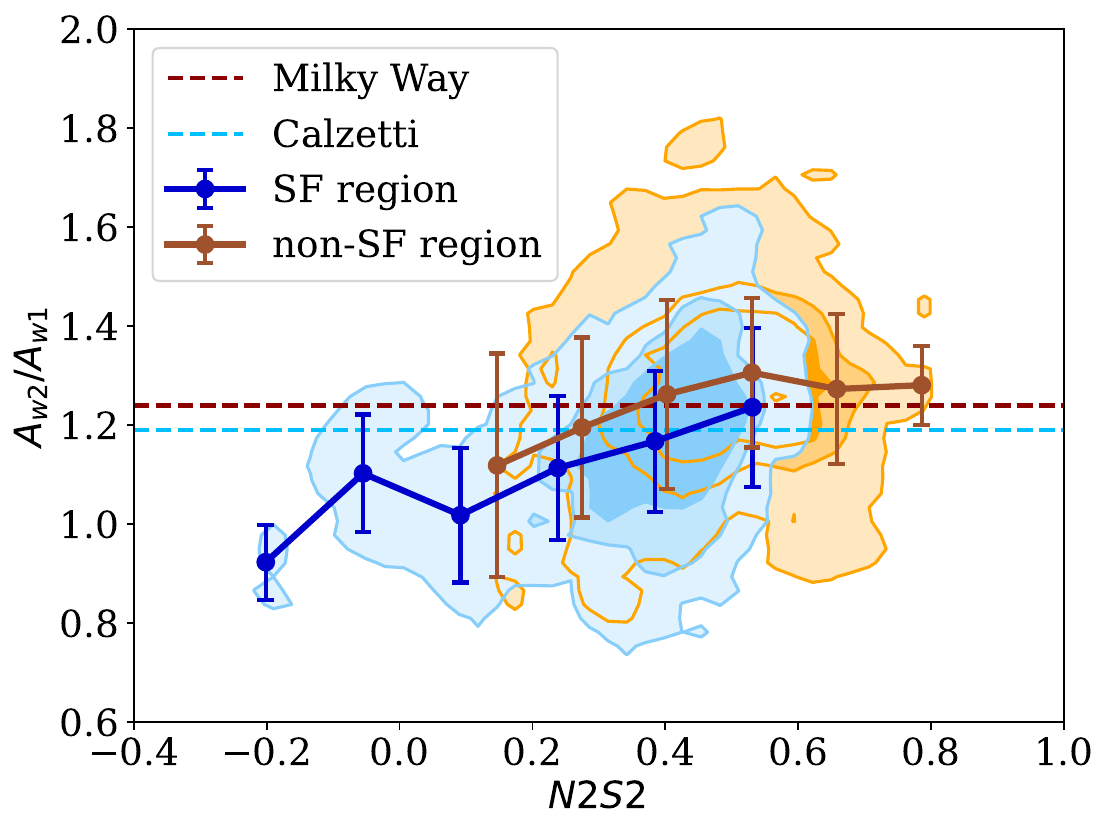}
  \caption{The three attenuation properties—$A_V$ (left panels), $A_B/A_V$ (middle panels), and $A_{\tt w2}/A_{\tt w1}$ (right panels)—are shown as functions of (panels from top to bottom) luminosity-weighted and mass-weighted stellar metallicity, gas-phase metallicity estimated using the O3N2 and R23 indicators, and the N2S2 line ratio. The results are presented separately for SF regions and non-SF regions, with the same symbols/lines/colors as in \autoref{fig:each_other}.}
  \label{fig:metallicity}
\end{figure*}

\subsection{Dependence on age and metallicity}

In \autoref{fig:stellar_age}, we explore how dust attenuation properties depend on the average stellar age, weighted either by luminosity ($t_L$) or mass ($t_M$). Results are shown separately for SF and non-SF regions. The optical slope $A_B/A_V$ exhibits little to no dependence on either $t_L$ or $t_M$, across both SF and non-SF regions. In contrast, we observe a negative correlation between $A_V$ and stellar age, as well as a positive correlation between $A_{\tt w2}/A_{\tt w1}$ and stellar age. This aligns with the previously noted dependence of $A_V$ and $A_{\tt w2}/A_{\tt w1}$ on $D_n4000$, which also serves as an indicator of the mean stellar age of the population.

In \autoref{fig:metallicity}, we present the attenuation parameters as functions of metallicity, both stellar and gas-phase. The first two rows of panels show the results for luminosity-weighted ($Z_L$) and mass-weighted ($Z_M$) stellar metallicity. Overall, the dependence on stellar metallicity is weak for all three dust attenuation parameters across both SF and non-SF regions, except for a slight increase in $A_{\tt w2}/A_{\tt w1}$ with increasing metallicity in SF regions. 

In the next two rows of panels in \autoref{fig:metallicity}, we further explore the relationship between dust attenuation parameters and gas-phase metallicity, using the oxygen abundance $12+\log_{10}\text{(O/H)}$ as estimated by O3N2 and R23 indicators. While we display spaxels from both SF and non-SF regions, oxygen abundance estimates are meaningful only for SF regions. Therefore, we present the median and scatter only for SF regions in each panel. As shown, the dust attenuation parameters exhibit a rather weak dependence on oxygen abundance, consistent with the trends observed for stellar metallicity. However, a positive correlation between $A_{\tt w2}/A_{\tt w1}$ and gas-phase metallicity is also noticeable in SF regions.

In the bottom panels of \autoref{fig:metallicity}, we examine the three attenuation parameters as functions of N2S2, defined as $\log_{10}([\text{NII}]\lambda6583/[\text{SII}]\lambda6717)$. This parameter is sensitive to both gas-phase metallicity and the ionization parameter. At a fixed ionization level, metal-poorer regions tend to have smaller N2S2 values \citep{2020ApJ...888...88L}. Similar to the results for stellar metallicity and oxygen abundance, both the optical opacity and optical slope show little to no dependence on N2S2. However, the NUV slope exhibits a positive correlation with N2S2.

Together, these results consistently suggest that while the optical properties of attenuation curves remain largely unaffected by metallicity, the NUV slope shows a weak but noticeable dependence on both stellar and gas-phase metallicities.

\begin{figure*}[ht!]
  \centering
   \includegraphics[width=0.32\textwidth]{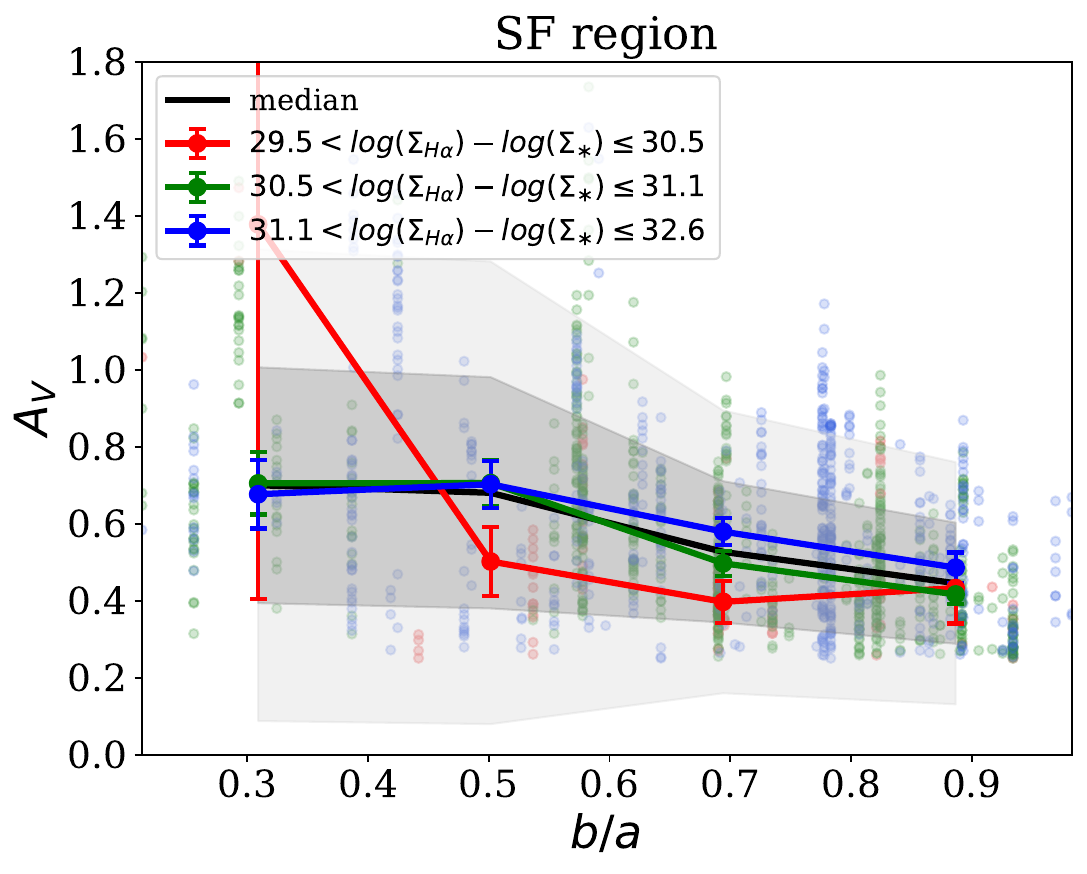}
   \includegraphics[width=0.32\textwidth]{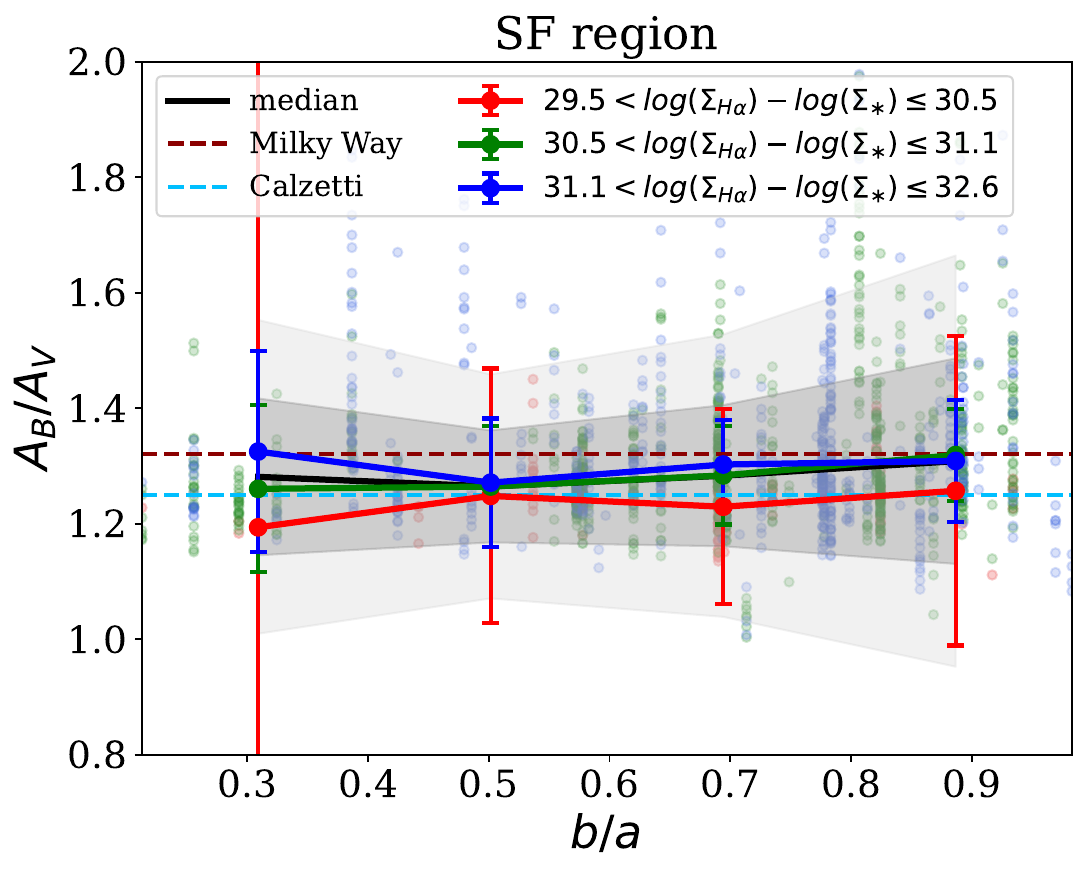}
   \includegraphics[width=0.32\textwidth]{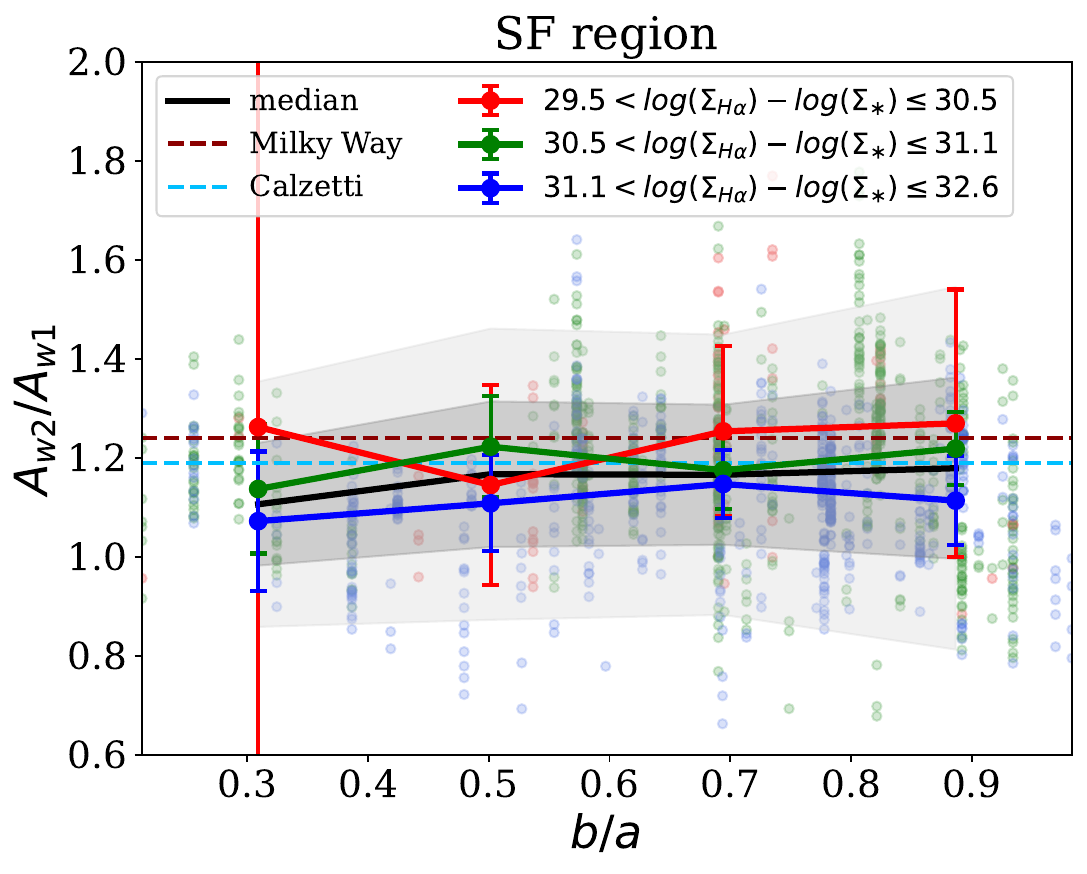}   
   \includegraphics[width=0.32\textwidth]{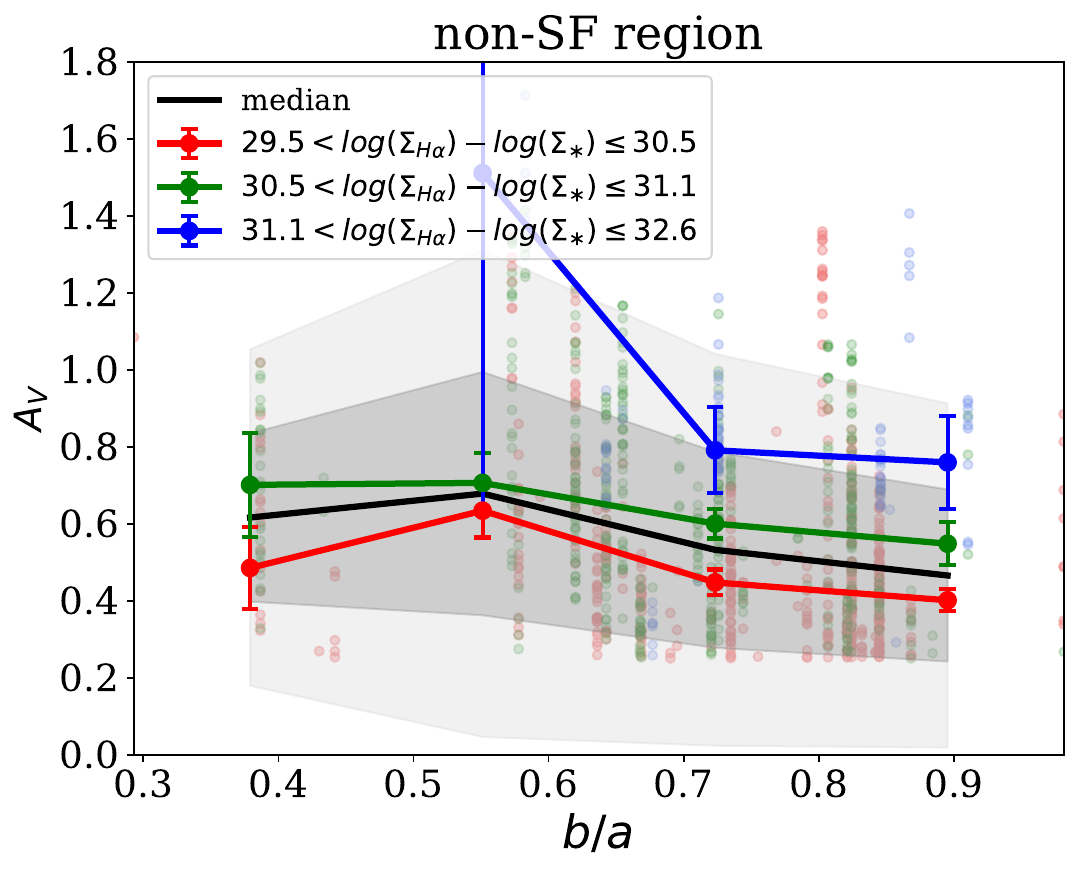}
   \includegraphics[width=0.32\textwidth]{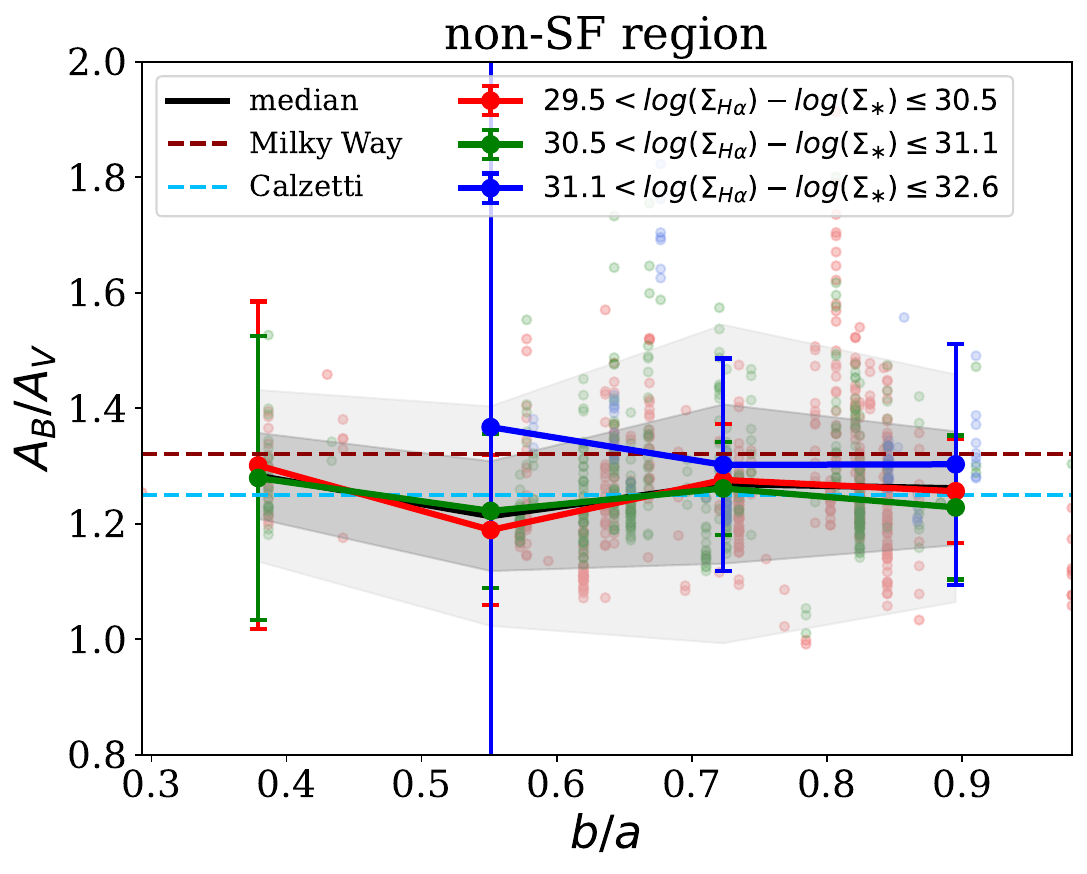}
   \includegraphics[width=0.32\textwidth]{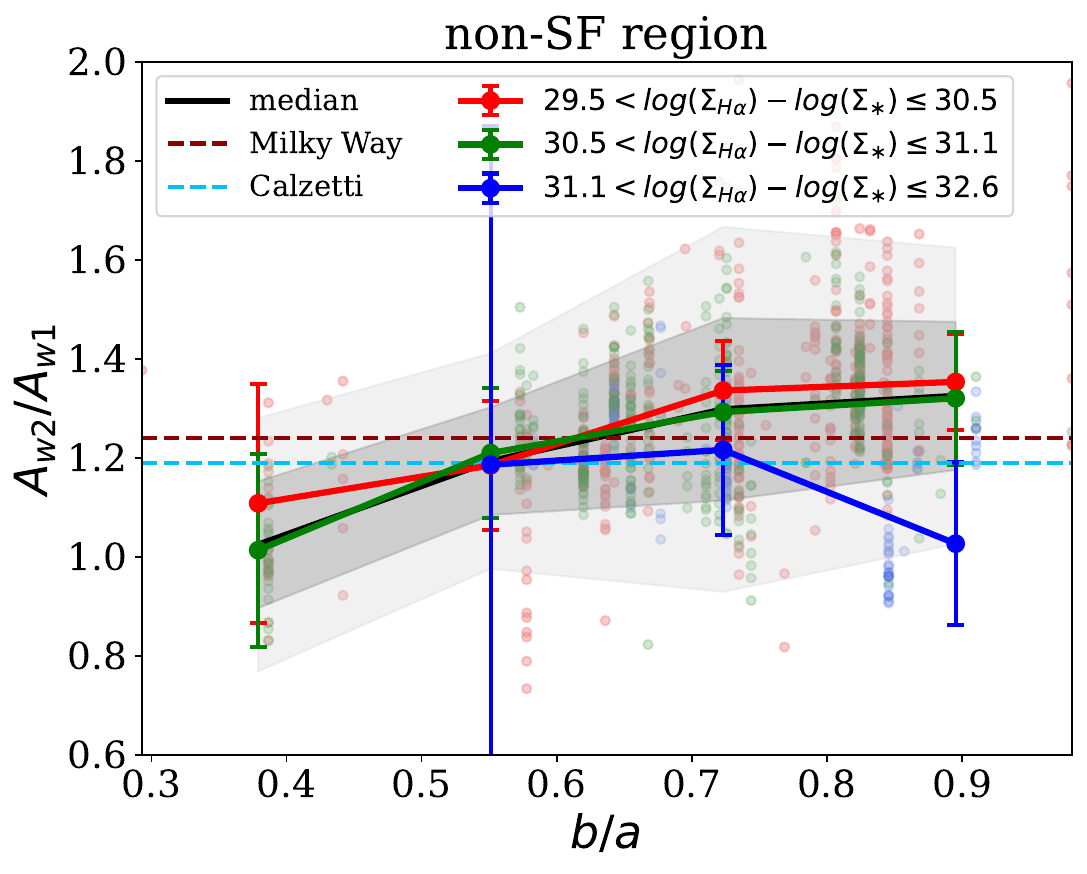}
  \caption{The three attenuation properties—$A_V$ (left panels), $A_B/A_V$ (middle panels), and $A_{\tt w2}/A_{\tt w1}$ (right panels)—are shown as functions of minor-to-major axis ratio ($b/a$) for SF regions (upper panels) and non-SF regions (lower panels). In each panel, the dark and light gray regions represent the $1\sigma$ and $2\sigma$ scatter around the median relation of the full sample, respectively. The red, green, and blue dots correspond to individual regions within different ranges of $\log_{10}(\Sigma_{\text{H}\alpha}/\Sigma_\ast)$, as indicated. The colored symbols and lines denote the median relation of the subsamples, with error bars representing Poisson errors. The horizontal dashed lines indicate the slopes of the Calzetti and Milky Way-type curves.}
  \label{fig:dep_inclination}
\end{figure*}

\begin{figure*}[h!]
  \centering
   \includegraphics[width=0.32\textwidth]{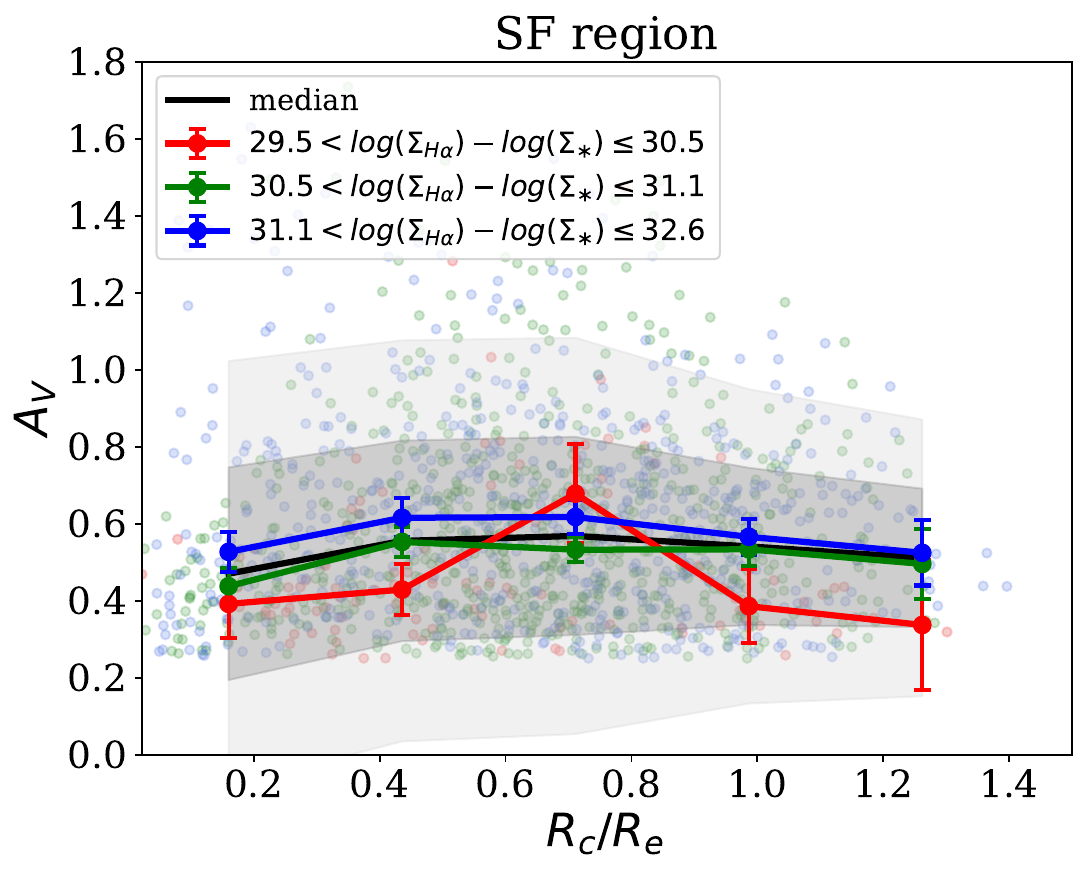}
   \includegraphics[width=0.32\textwidth]{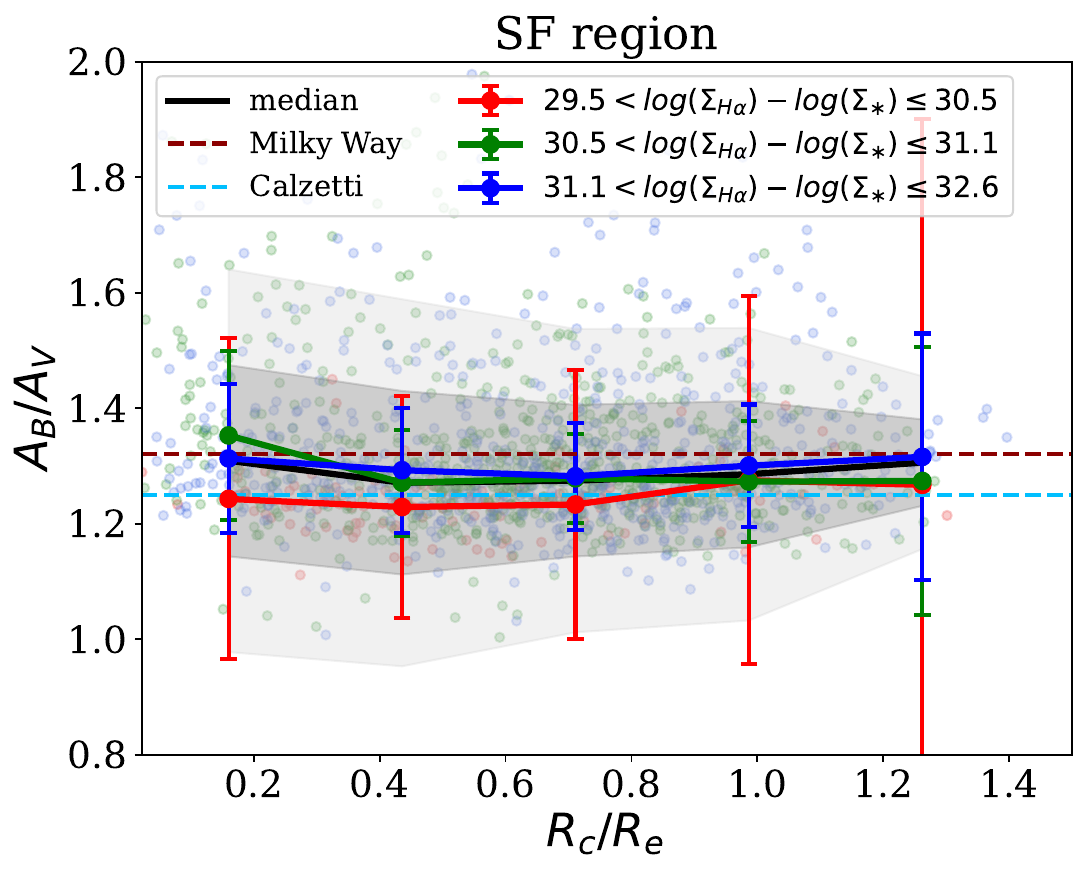}
   \includegraphics[width=0.32\textwidth]{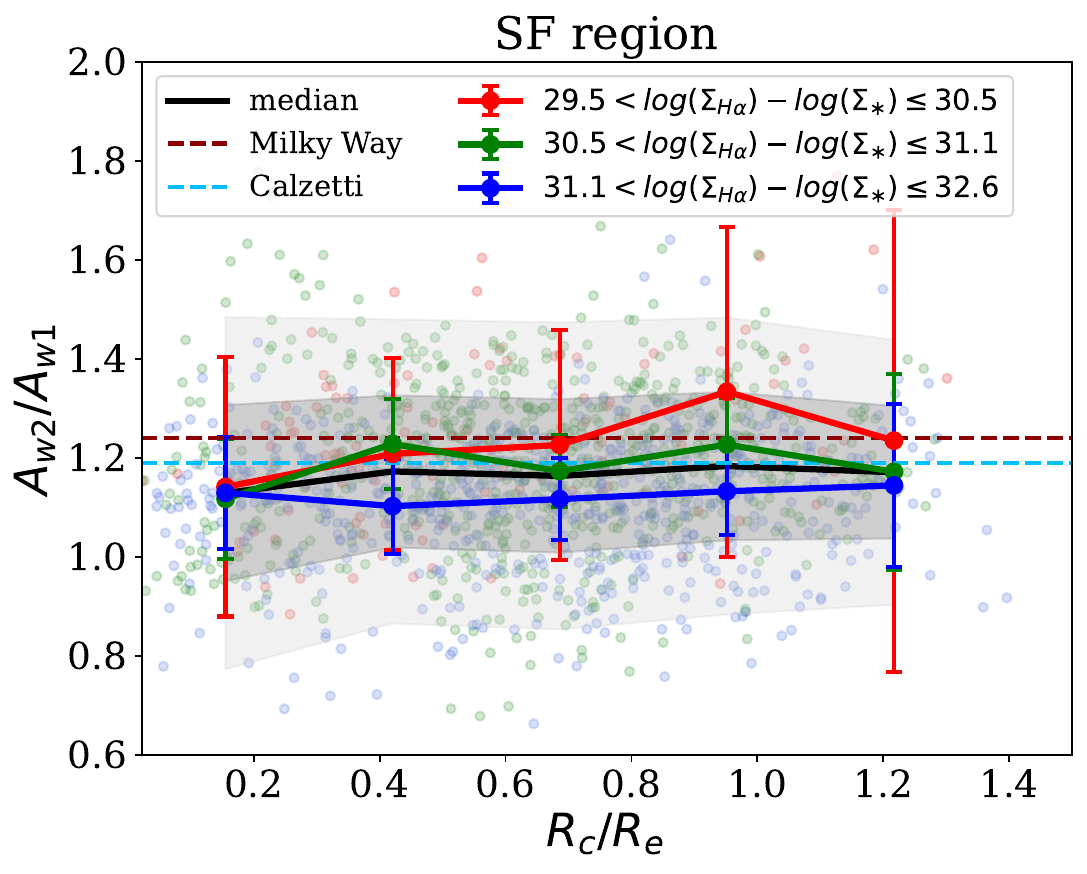}   
   \includegraphics[width=0.32\textwidth]{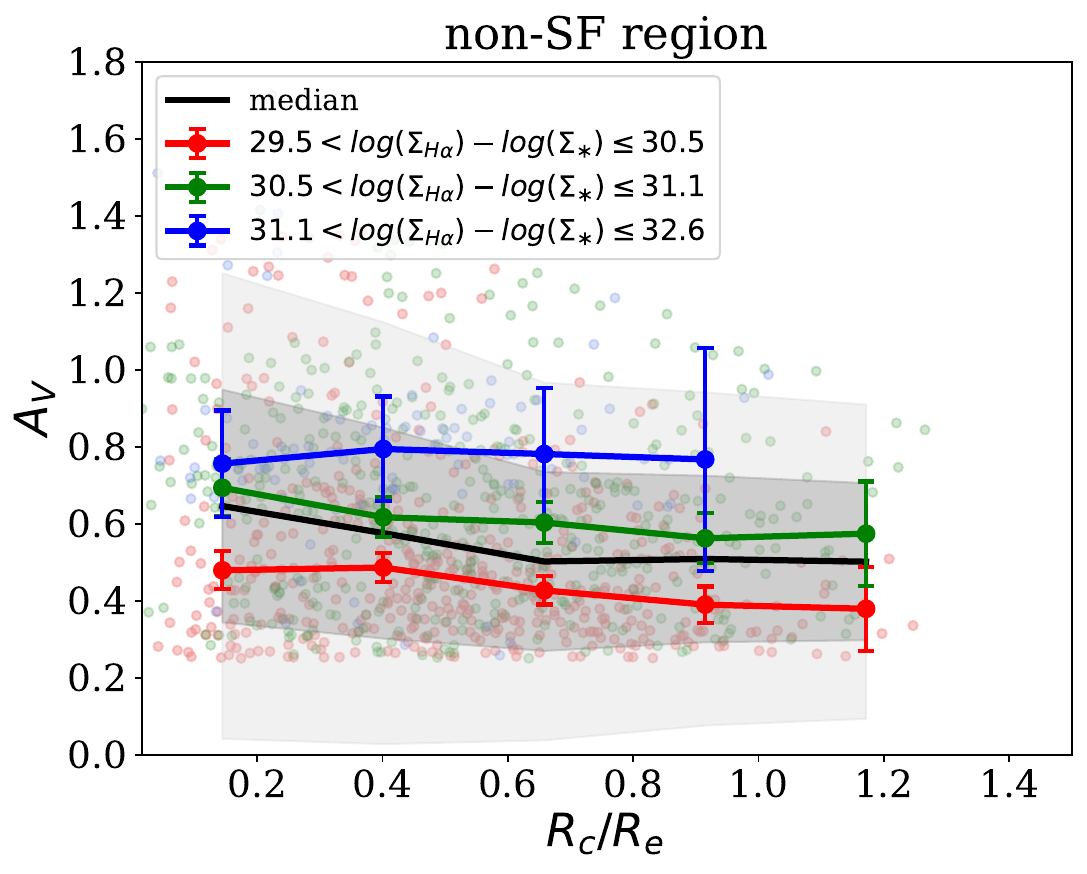}
   \includegraphics[width=0.32\textwidth]{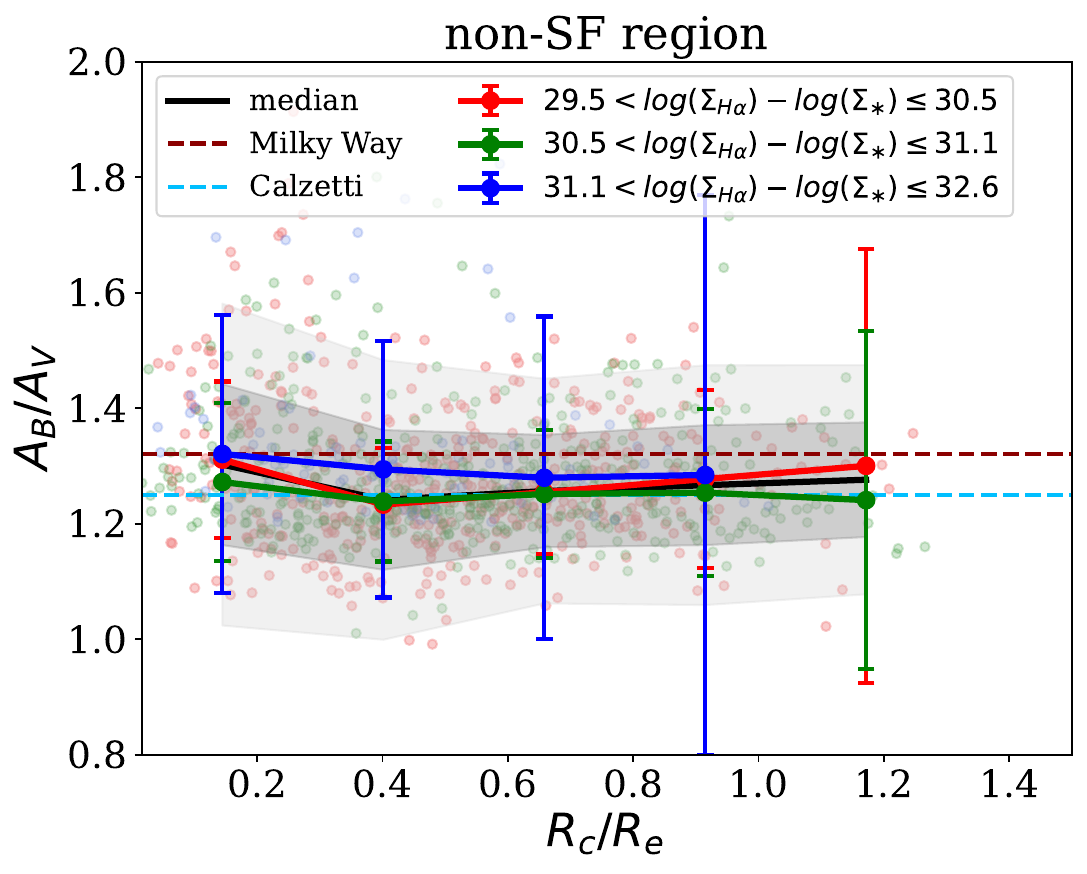}
   \includegraphics[width=0.32\textwidth]{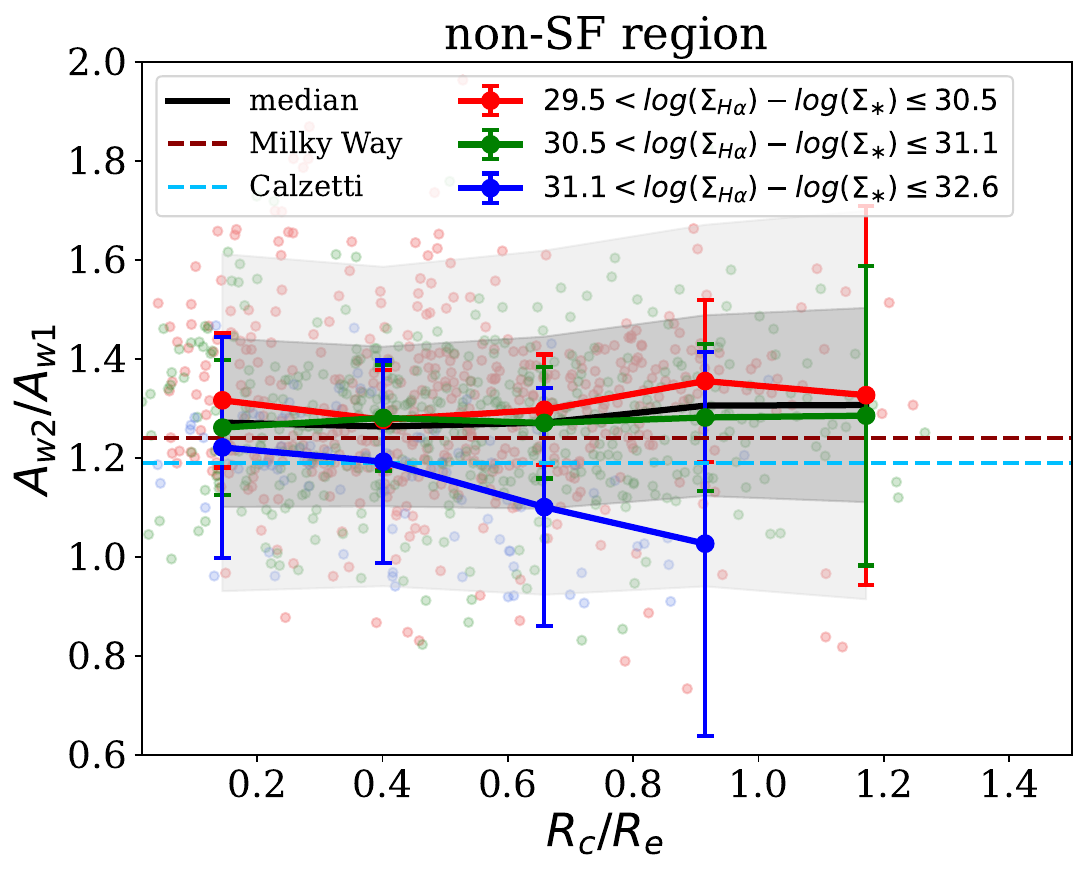}
  \caption{The three attenuation properties—$A_V$ (left panels), $A_B/A_V$ (middle panels), and $A_{\tt w2}/A_{\tt w1}$ (right panels)—are shown as functions of the galactocentric distance scaled by effective radius ($R_c/R_e$) for SF regions (upper panels) and non-SF regions (lower panels). Symbols/lines/colors are the same as in the previous figure.}
  \label{fig:dep_radius}
\end{figure*}

\subsection{Dependence on inclination and galactocentric radius}

In this subsection, we examine the potential dependence of dust attenuation parameters on the inclination of host galaxies and the galactocentric distances of the spaxels. Following Paper I, galaxy inclination is quantified by the minor-to-major axis ratio ($b/a$) measured from SDSS $r$-band images, while the galactocentric distance is given by the radius of the spaxels in the MaNGA datacube, scaled by the effective radius ($R_e$). \autoref{fig:dep_inclination} presents the three dust attenuation parameters as functions of $b/a$, shown in panels from left to right. The upper and lower  panels display results for SF and non-SF regions, respectively. In each panel, different colors represent subsamples divided by specific H$\alpha$ surface brightness ($\Sigma_{\text{H}\alpha}/\Sigma_\ast$). Overall, we find that, in both SF and non-SF regions, both the optical and NUV slopes of attenuation curves are largely independent of galaxy inclination when $\Sigma_{\text{H}\alpha}/\Sigma_\ast$ is restricted to a narrow range. However, at fixed $\Sigma_{\text{H}\alpha}/\Sigma_\ast$, $A_V$ tends to decrease as $b/a$ increases, which aligns with the expectation that more inclined galaxies experience stronger dust attenuation.

The dependence of the three attenuation parameters on the galactocentric distance, $R_c/R_e$, is presented in \autoref{fig:dep_radius}, using the same symbols and color coding as in the previous figure. Overall, we find that when $\Sigma_{\text{H}\alpha}/\Sigma_\ast$ is restricted to a narrow range, none of the dust attenuation parameters exhibit a clear correlation with $R_c/R_e$. This result is true for both SF and non-SF regions. 

Our results reinforce the conclusion from Paper I that dust attenuation is primarily governed by local processes on kpc scales or smaller, rather than by global processes at the galactic scale.

\section{Discussion}
\label{sec:discussion}

This paper extends the findings of Paper I by utilizing the updated SwiM\_v4.2 catalog, which is nearly four times larger than the earlier SwiM\_v3.1 catalog used in Paper I. While Paper I primarily explored the correlations between dust attenuation properties and sSFR across all regions collectively, this study refines the analysis by categorizing regions into SF and non-SF subsets and incorporating a broader range of stellar population and emission-line properties. Given that sSFR is meaningful only for SF regions, we replace it with the specific surface brightness of H$\alpha$ emission, $\Sigma_{\text{H}\alpha}/\Sigma_\ast$, as a proxy for the strength of H$\alpha$ emission. Importantly, all key findings from Paper I are consistently reproduced in this work. However, we uncover a distinct anti-correlation between the NUV slope $A_{\tt w2}/A_{\tt w1}$ and $\Sigma_{\text{H}\alpha}/\Sigma_\ast$, a trend that was not evident in Paper I, likely due to the smaller sample size used in that study. When dividing all regions into SF and non-SF regions, we find that both types exhibit similar trends in optical attenuation properties ($A_V$ and $A_B/A_V$), which correlate in comparable ways with all the stellar population and emission-line properties examined. However, the NUV slope $A_{\tt w2}/A_{\tt w1}$ behaves differently between the two, with SF regions generally displaying flatter slopes than non-SF regions. 

In addition to Paper I, the SwiM\_v3.1 catalog has been used to investigate the relationship between NUV stellar attenuation (characterized by the NUV power-law index, $\beta$) and optical nebular attenuation (measured via the Balmer emission line optical depth, $\tau_B^l$) by \citet{2020MNRAS.494.4751M}, as well as the infrared excess–NUV spectral index (IRX-$\beta$) relation by \citet{2023MNRAS.526..904D}. \citet{2020MNRAS.494.4751M} found that regions dominated by diffuse ionized gas (DIG) deviate from the $\beta-\tau_B^l$ relation observed in SF regions, suggesting that light from old stellar populations contributes to the scatter in this relation. Meanwhile, \citet{2023MNRAS.526..904D} reported that both $\beta$ and IRX positively correlate with $D_n4000$ in SF regions, indicating a connection between dust attenuation in the NUV and stellar populations of different ages. These findings are consistent with our result that the NUV attenuation curve slope differs between SF and non-SF regions, reinforcing the notion that dust attenuation in the NUV is influenced by both young and old stellar populations in distinct ways. The Swift/UVOT data have also been combined with far-infrared photometry from the SINGS/KINGFISH samples of nearby galaxies to investigate the impact of assumed star formation history (SFH) parameterizations on the derivation of attenuation laws from SED fitting \citep{2023ApJ...953...54B}. It was found that the NUV slope of attenuation curves varies depending on the assumed SFH form, a result that may be related to the correlation between the NUV slope and recent SFH diagnostics identified in this work.

In a more recent study, Swift/UVOT photometry was combined with mid-infrared imaging from the PHANGS-JWST survey to investigate the connection between the 2175\AA\ bump and polycyclic aromatic hydrocarbons (PAHs; \citealt{2025PASA...42...22B}). Both the 2175\AA\ bump and PAH abundance were found to be negatively correlated with sSFR. This result is fully consistent with our previous findings from Paper I, which also measured the UV bump at 2175\AA\ and explored its correlations with sSFR. As pointed out in both Paper I and \citealt{2025PASA...42...22B}, the observed anti-correlation between the 2175\AA\ bump and sSFR suggests that the carrier of the UV bump is small dust grains that are susceptible to destruction by UV photons. Theoretically, \citet{2018ApJ...869...70N} demonstrated that both the 2175\AA\ bump strength and the attenuation curve slope are primarily influenced by the fraction of unobscured young/massive stars, such that a higher fraction of these stars results in a flatter attenuation curve with a weaker 2175\AA\ bump. The positive correlations of $A_{\tt w2}/A_{\tt w1}$ with stellar age and the negative correlations of $A_{\tt w2}/A_{\tt w1}$ with specific H$\alpha$ surface brightness found in this work align well with the theoretical expectations from \citet{2018ApJ...869...70N}.

In fact, we have conducted the same analysis for the UV bump using the expanded SwiM\_v4.2 catalog. The results of this investigation will be presented in the next paper of this series (Paper III, Guo et al., in prep.). In the fourth paper of this series (Paper IV, Guo et al., in prep.), we will use the attenuation curve measurements obtained in this work to constrain a two-component dust model. This model will allow us to derive the distributions of dust grain size and the fractional contributions of dust mass from silicate and graphite compositions. As we will show, the flattening of the NUV slope in SF regions with relatively high $\Sigma_{\text{H}\alpha}/\Sigma_\ast$ and young stellar populations can be understood in terms of a lower fraction of small-sized dust grains in such regions. This aligns with the conclusion from Paper I that variations in NUV attenuation, including both the attenuation curve slope and the 2175\AA\ bump, are driven by star formation-related processes, such as the destruction of small dust grains by UV radiation in SF regions. Moreover, this work extends that conclusion to non-SF regions, indicating that the UV radiation responsible for small dust grain destruction is not solely emitted by massive stars in SF regions but can also originate from other ionizing sources in non-SF regions.

Focusing solely on the optical regime, numerous studies have investigated spatially resolved dust attenuation in nearby galaxies at kiloparsec scales using integral field spectroscopy observations \citep[e.g.][]{2013ApJ...771...62K,2016ApJ...825...34J,2017MNRAS.467..239B,2019ApJ...872...63L,2020MNRAS.495.2305G,2020ApJ...896...38L,2020ApJ...888...88L,2020ApJ...893...94T,2021ApJ...917...72L,2021MNRAS.503.4748R,2023A&A...670A.125J,2024ApJ...975..234L,2024A&A...691A.201L}. For instance, \citet{2020ApJ...893...94T} estimated the optical attenuation curves for star-forming regions in MaNGA galaxies using the empirical method of \citet{1994ApJ...429..582C}, finding that SF regions with relatively small values of $D_n4000$ ($1.1<D_n4000<1.2$) exhibit shallower average attenuation curves when they have lower surface stellar mass density, smaller SFR surface density, or are located in the outskirts of galaxies. In contrast, our analysis finds no significant correlations between the optical slope of attenuation curves and any of the regional or global properties considered. This discrepancy is likely due to differences in the methods used to derive attenuation curves. The empirical method adopted by \citet{2020ApJ...893...94T} determines attenuation curves by comparing observed SEDs/spectra between dusty galaxies and those with little to no attenuation \citep[e.g.][]{1994ApJ...429..582C,1994ApJ...429..172K,1997AJ....113..162C,2000ApJ...533..682C,2007ApJS..173..392J,2011MNRAS.417.1760W,2016ApJ...818...13B,2017ApJ...851...90B,2017ApJ...840..109B}. As noted by \citet{2011MNRAS.417.1760W}, this approach can introduce bias, as the template or the more attenuated galaxy in a pair will contribute more strongly to the final average curve if the attenuation curve slope depends on the dust attenuation itself, as observed in previous studies. Given the substantial variation in dust content across galaxies and the wide range of attenuation curve slopes, assuming a single attenuation form to rank and compare templates or paired galaxies is no longer appropriate \citep{2020ARA&A..58..529S}. 

Our findings regarding the optical slope are in good agreement with those of \citet{2021ApJ...917...72L}, who also used IFS data from MaNGA but applied the new technique developed by \citet{2020ApJ...896...38L} and Paper I to derive attenuation curves. As demonstrated in these studies, our method allows attenuation curves to be obtained without assuming a predefined functional form, and it derives the attenuation curve before applying our spectral fitting code to derive stellar population properties, thereby significantly reducing the influence of degeneracies between model parameters. Although \citet{2020ApJ...896...38L} did not explicitly examine the correlations between the optical curve slope and stellar population or emission line properties, their results regarding the correlations of stellar and nebular reddening ($E(B-V)_{\text{star}}$ and $E(B-V)_{\text{gas}}$) with $D_n4000$ and stellar age are broadly consistent with our findings in this work.

Numerous previous studies have measured attenuation curves and investigated their correlations with galactic properties at the scale of entire galaxies, covering both the local Universe and high redshifts. A recent review by \citet{2020ARA&A..58..529S} provides a comprehensive summary of these studies. In particular, the anti-correlation between the slope of attenuation curves and sSFR has been well established at galactic scales across a broad redshift range. Our findings from Paper I and the current work demonstrate that these global trends hold at kiloparsec scales in both SF and non-SF regions, regardless of the inclination of host galaxies or the galactocentric distance of regions. This further reinforces the idea that dust attenuation laws are primarily governed by local processes occurring on kpc scales or smaller, rather than by global mechanisms at the scale of entire galaxies.

\section{Summary}
\label{sec:summary}

This is the second paper in a series utilizing the SwiM catalog to investigate dust attenuation properties on kiloparsec scales in nearby galaxies. The SwiM catalog was constructed by cross-matching the Swift/UVOT data archive with the SDSS-IV/MaNGA sample, providing both integral field spectroscopy in the optical and NUV imaging in three bands ({\tt uvw1}, {\tt uvm2} near 2175\AA, and {\tt uvw2}; \citealt{2020ApJS..251...11M}, \citealt{2023ApJS..268...63M}). This dataset is particularly well-suited for studying dust attenuation in kpc-sized regions across optical and NUV wavelengths. In the first paper of this series \citep[][Paper I]{2023ApJ...957...75Z}, based on an earlier version of the SwiM catalog (SwiM\_v3.1) supplemented with Ks-band imaging from 2MASS, we developed a novel method to measure dust attenuation curves. We explored the correlations of optical opacity ($A_V$), the optical and NUV slopes of the attenuation curves ($A_B/A_V$ and $A_{\tt w2}/A_{\tt w1}$), and the strength of the UV bump at 2175\AA\ with the specific star formation rate (sSFR) of kiloparsec-sized regions. In this study, we extend the work of Paper I by applying the same methodology to the latest version of the SwiM catalog, SwiM\_v4.2, which is nearly four times larger than SwiM\_v3.1. Additionally, we classify all regions into two subsets: star-forming (SF) and non-SF regions. Instead of sSFR, we use the specific surface density of H$\alpha$ emission ($\Sigma_{\text{H}\alpha}/\Sigma_\ast$) to quantify the relative strength of H$\alpha$ emission, as sSFR is meaningful only for SF regions. Furthermore, we examine the correlations of $A_V$, $A_B/A_V$ and $A_{\tt w2}/A_{\tt w1}$ with a broad range of stellar population and emission-line properties, including stellar age, stellar metallicity, gas-phase metallicity, and diagnostics of recent star formation history. Finally, we investigate the potential dependence of spatially resolved dust attenuation properties on the inclination of host galaxies and the galactocentric distance of the regions. To ensure reliable measurements of dust attenuation properties, and based on the tests of our method conducted in Paper I, we have restricted our analysis to regions with a continuum S/N greater than 20 and an optical opacity of $A_V > 0.25$.

Our main conclusions can be summarized as follows:
\begin{enumerate}
    \item Overall, when comparing SF and non-SF regions, we find that the optical attenuation properties, characterized by $A_V$ and $A_B/A_V$, exhibit similar correlations with all the stellar population and emission-line properties considered. In contrast, the NUV slopes in SF regions tend to be flatter than those in non-SF regions, and this difference is mainly attributed to the relatively low surface densities of stellar mass in SF regions. 
    \item The optical slope of attenuation curves $A_B/A_V$ shows little to no dependence on any of the stellar population or emission-line properties in both SF and non-SF regions. In contrast, the optical opacity $A_V$ exhibits a positive correlation with specific H$\alpha$ surface brightness (and related parameters), a negative correlation with stellar age (and related parameters), and no clear dependence on stellar or gas-phase metallicity. This is also true in both SF and non-SF regions. 
    \item The NUV slope of attenuation curves, $A_{\tt w2}/A_{\tt w1}$, exhibits an anti-correlation with specific H$\alpha$ surface brightness. This trend is primarily driven by the positive correlation between $A_{\tt w2}/A_{\tt w1}$ and the stellar surface mass density, $\Sigma_\ast$.
    \item The NUV slope of attenuation curves flattens in SF regions that contain young stellar populations and have experienced recent star formation, but it shows no obvious dependence on stellar or gas-phase metallicity.
    \item The spatially resolved dust attenuation properties exhibit no clear correlations with the inclination of host galaxies or the galactocentric distance of the regions. This finding reinforces the conclusion from Paper I that dust attenuation is primarily regulated by local processes on kpc scales or smaller, rather than by global processes at galactic scales.
\end{enumerate}

\normalem
\begin{acknowledgements}
This work is supported by the National Key R\&D Program of China (grant NO. 2022YFA1602902), the National Natural Science Foundation of China (grant Nos. 12433003, 11821303, 11973030), and China Manned Space Program through its Space Application System.

Funding for SDSS-IV has been provided by the Alfred P. Sloan Foundation and Participating Institutions. Additional funding towards SDSS-IV has been provided by the US Department of Energy Office of Science. SDSSIV acknowledges support and resources from the Centre for High-Performance Computing at the University of Utah. The SDSS web site is www.sdss.org.

SDSS-IV is managed by the Astrophysical Research Consortium for the Participating Institutions of the SDSS Collaboration including the Brazilian Participation Group, the Carnegie Institution for Science, Carnegie Mellon University, the Chilean Participation Group, the French Participation Group, Harvard–Smithsonian Center for Astrophysics, Instituto de Astrofsica de Canarias, The Johns Hopkins University, Kavli Institute for the Physics and Mathematics of the Universe (IPMU)/University of Tokyo, Lawrence Berkeley National Laboratory, Leibniz Institut fur Astrophysik Potsdam (AIP), Max-Planck-Institut fur Astronomie (MPIA Hei- delberg), Max-Planck-Institut fur Astrophysik (MPA Garching), Max-Planck-Institut fur Extraterrestrische Physik (MPE), National Astronomical Observatory of China, New Mexico State University, New York University, University of Notre Dame, Observatario Nacional/MCTI, The Ohio State University, Pennsylvania State University, Shanghai Astronomical Observatory, United Kingdom Participation Group, Universidad Nacional Autonoma de Mexico, University of Arizona, University of Colorado Boulder, University of Oxford, University of Portsmouth, University of Utah, University of Virginia, University of Washington, University of Wisconsin, Vanderbilt University and Yale University.

We acknowledge the Tsinghua Astrophysics High-Performance Computing platform at Tsinghua University for providing computational and data storage resources that have contributed to the research results reported within this paper.

\end{acknowledgements}
  
\bibliographystyle{raa}
\bibliography{bibtex,szhou}

\end{document}